This article has been accepted for publication in a future issue of this journal, but has not been fully edited. Content may change prior to final publication. Citation information: DOI 10.1109/ACCESS.2022.3148473, IEEE Access

# Tactile based Intelligence Touch Technology in IoT configured WCN in B5G/6G-A Survey

**Mantisha Gupta[1], Student Member, IEEE, Rakesh Kumar Jha[2], Senior Member, IEEE, Sanjeev Jain[3], Member IEEE**

[1]School of Electronics and Communication Engineering (SoECE), Shri Mata Vaishno Devi University, Katra, Jammu, J&K, India, 182320.
[2]Associate Prof., Department of Electronics and Communication Engineering, Indian Institute of Information Technology, Design and Manufacturing, Jabalpur (IIITDM Jabalpur). (e-mail: jharakesh.45@gmail.com).
[3]Prof. Computer Science Department, Central University Jammu, J&K, India. (e-mail: dr_sanjeevjain@yahoo.com )

Corresponding author: (e-mail: jharakesh.45@gmail.com).

This work was supported by the 5G and IoT Lab, SoECE, TBIC, TEQIP-III at Shri Mata Vaishno Devi University, Katra, Jammu

**ABSTRACT** Touch enabled sensation and actuation is expected to be one of the most promising, straightforward and important uses of the next generation communication networks. In light of the next generation (B5G/6G) system's need for low latency, the infrastructure should be reconfigurable and intelligent in order to be able to work in real time and interoperable with the existing wireless network. It has a drastic impact on the society due to its high precision, accuracy, reliability and efficiency as well as the ability to connect a user from far away or remote areas. Such a touch-enabled interaction is primarily concerned with the real time transmission of the tactile based haptic information over the internet, in addition to the usual audio, visual and data traffic, thus enabling a paradigm shift towards establishing a real time control and steering communication system. Due to the existing system's latency and overhead, it creates delays and limits the usability of the future applications. In light of the aforementioned concerns, this study proposes an intelligent touch-enabled system for B5G/6G and IoT based wireless communication network that incorporates the AR/VR technologies. The tactile internet and network slicing serve as the backbone of the touch technology which incorporates intelligence from techniques such as artificial intelligence and machine/deep learning. The survey also introduces a layered and interfacing architecture complete with its E2E solution for the intelligent touch based wireless communication system. It is anticipated for the next generation system to provide numerous opportunities for various sectors utilizing AR/VR technology in robotics and healthcare facilities, all with the intension of helping in addressing severe problems faced by the society. Conclusively the article presents a few use cases concerning the deployment of touch infrastructure in automation and robotics as well as in intelligent healthcare systems, assisting in the diagnosis and treatment of the prevailing covid-19 cases. The paper concludes with some considerable future research aspects of the proposed system with few of the ongoing projects pertaining to the development in the incorporation of the next generation (6G) system.

## I. INTRODUCTION

Mobile and wireless communications have been playing a decisive role in the current economy with the technologies like 2G,3G,4G,5G,GPRS,EDGE that successfully satisfy the user end with a significant role in the business, education, logistics and other primary industrial applications, effectively connecting the majority of the world's population. These, in the present day are proficient enough to connect to the devices and people for an unprecedented exchange of multimedia and data content, enjoying its fastest growth in the history due to its enabling technologies, encouraging its widespread deployment, further intensifying the communication and the industrial sector[1].

As per the Cisco Visual Networking Index (VNI), there is an effective forecasting of the impact of the visual networking applications on global networks, incrementing from about 11.5 exabytes in 2017 to an expected surge of 77 exabytes towards 2023. The compound annual growth rate (CAGR) is expected to be about 74 percent of the present mobile traffic, 66 percent of the cellular (Wi-Fi) traffic, accompanied by the smart phones, dominating more than 90 percent of the mobile data traffic in the coming few years[2].

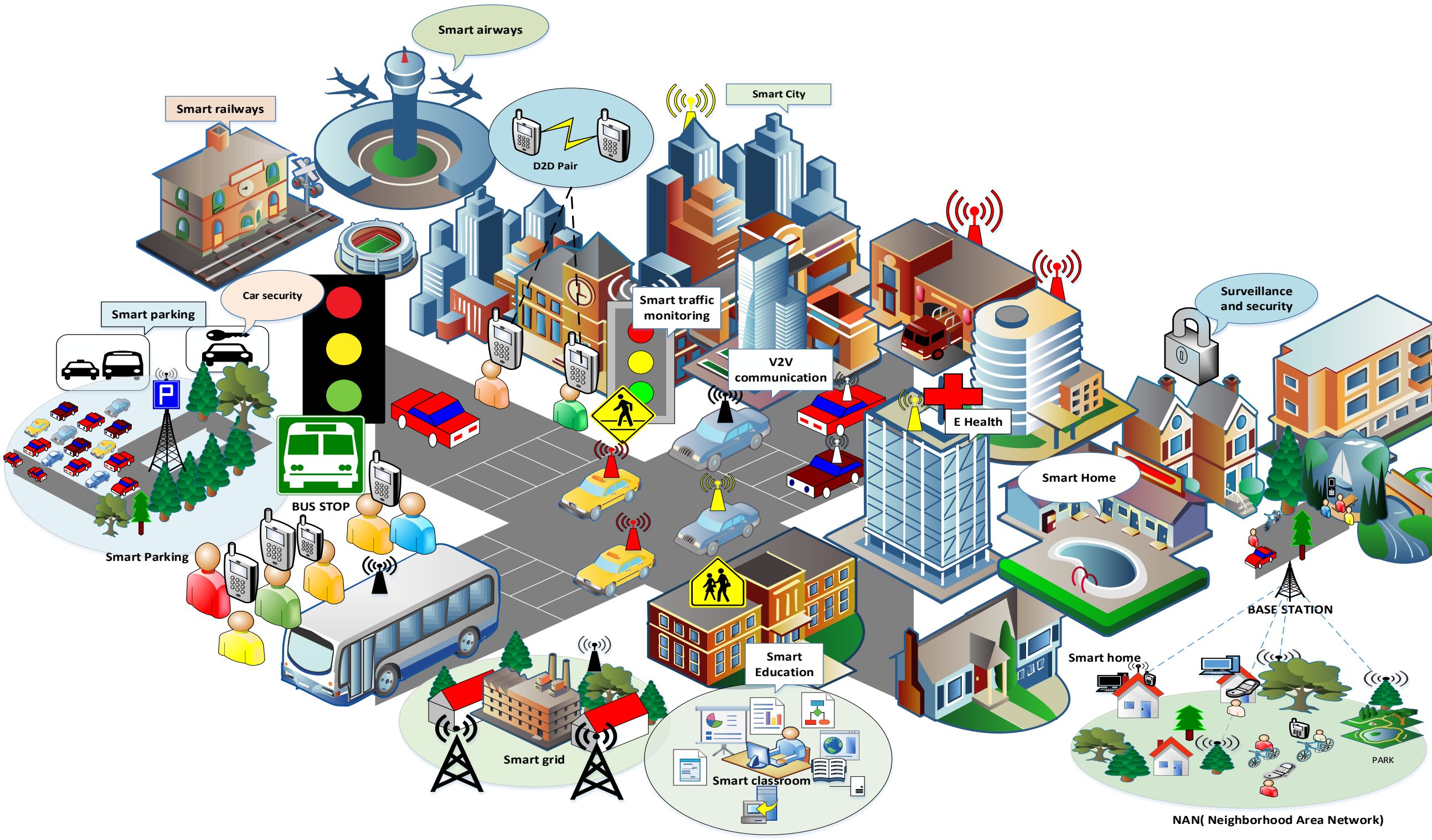

**FIGURE 1: General architecture of existing wireless and IoT based scenario.**

The Internet of Things (IoT) has therefore been a novel paradigm that is swiftly accelerating in the modern wireless communication scenario, connecting billions of devices with a seamless access of internet and applications, altogether forming an internet of everything (IoE). Conclusively the main strength of the IoT lies on the significant impact it has on variety of facets of everyday life and user behavior [3]. Since then, there has been a worldwide increase in the development of cellular network over the past decade. Table I hereby provides a list of all the commonly used acronyms throughout the paper for better understanding.

Many technical challenges have instantiated the designing of a robust wireless network, capable of delivering the necessary performance to support the emerging applications. The previous generations have seen a paradigm shift in the cellular technology and unlike these, the B5G/6G is said to be an integrative structure of the present 5G air interface spectrum. This in combination with the LTE and Wi-Fi provides a seamless user experience, accompanied by the universally high coverage rate. The existing 5G core network inculcates an unprecedented flexibility and intelligence in the upcoming 6G system with an improved spectrum regulation, along with the energy and cost efficiencies.

Along the lines, the system also introduces extreme base stations with high device densities and an unparalleled number of antennas, as well as high carrier frequencies with large bandwidths [4], [5]. From the onset of 2020 onwards and till this date, the existing wireless communication networks have been standardized and deployed globally, with eMBB, MMTC and (URLLC) being the key 5G/B5G communication scenarios [6]–[8].The widely researched, IoT-enabled wireless network architecture, compatible with the B5G/6G system has been effectively illustrated in Fig 1.

The figure represents more or less every possible IoT configured services and applications, requiring high data rate and low latency, interacting with the surrounding environment. All these together constitute a smart system. This smart system entails all the possible smart devices, gadgets and sensors, all in all actuating a smart and automatized infrastructure. This infrastructure therefore highlights the utilization of the D2D communication network[9]–[11], the massive MIMO[12], small cell access points (SCA)[13], the IoT[14] with the network cloud [15], [16], all in all forming a part of the said 5G/B5G cellular framework[17], [18].

All these along with the number of other sensor based interactions, make the existing system automatic enough to ease the human effort and save time, all the while considering the exponential surge in the data traffic, operating the millions of devices that are connected to the internet. The key technologies in this framework incorporates the spectrum sharing [19] with the cognitive radio [20], the interference management [21], the ultra dense network (UDN) [22], the mm-wave [23], 5G/B5G cloud and RAN [24], [25] and SDNs [26]–[28]. The existing 5G network therefore rides on the coattails of an explicit New Radio (NR) interface accompanied by a considerable number of virtualization technologies like the

TABLE I
LIST OF COMMONLY USED ACRONYMS

| Abb. | Definition | Abb. | Definition | Abb. | Definition | Abb. | Definition | Abb. | Definition |
|---|---|---|---|---|---|---|---|---|---|
| 1G,2G, 3G,4G | First, Second, Third, Fourth Generation | DL | Deep Learning | LS | Local Server | NTN | Non Terrestrial Network | UL | Unsupervised Learning |
| 3GPP | 3G Partnership Project | DS | Delay Spread | LTE | Long Term Evolution | OFDMA | Orthogonal Frequency Division Multiple Access | UMTS | Universal Mobile Telecommunication System |
| 5G | Fifth Generation | EDGE | Enhanced Data Rate for GSM Evolution | LTE-A | LTE-Advanced | PD-NOMA | Power Domain NOMA | URLLC | Ultra Reliable and Low Latency Communication |
| 5GPP | 5G Partnership Project | E2E | End to End | MAC | Media Access Control | QoE | Quality of Experience | VR | Virtual Reality |
| 6G | Sixth Generation | eMBB | Enhanced Mobile Broadband | MANO | Management and Orchestration | QoS | Quality of Service | V2V | Vehicle to Vehicle |
| AI | Artificial Intelligence | EVDO | Evolution Data Optimized | MC-CDMA | Multicarrier CDMA | RAN | Radio Access Network | V2X | Vehicle to everything |
| AMPS | Advanced Mobile Phone Service | E-UTRA | Evolved Universal Terrestrial Radio Access | MIoT | Massive IoT | RAS | Robotic Autonomous system | W-CDMA | Wideband Code Division Multiple Access |
| ANN | Artificial Neural Network | FDMA | Frequency Division Multiple Access | MIMO | Multiple Input Multiple Output | RL | Reinforcement Learning | WCN | Wireless Communication Network |
| AR | Augmented Reality | GSM | Global System for Mobile communication | m-MIMO | Massive MIMO | RMS | Root Mean Square | Wi-Max | Worldwide Interoperability for Microwave Access |
| AS | Angle Spread | GPRS | General Packet Radio Service | M2M | Machine to Machine | SC-FDMA | Scalable FDMA | XR | Extended Reality |
| B5G | Beyond 5G | GW | Gateway | ML | Machine Learning | SDN | Software Defined Network | | |
| BS | Base Station | HD | High Definition | MTC | Machine Type Communication | SD-RAN | Software Defined Radio Access Network | | |
| CAGR | Compound Annual Growth Rate | H2M | Human to Machine | mMTC | Massive Machine Type Communication | SL | Supervised Learning | | |
| CDMA | Code Division Multiple Access | HSI | Human System Interface | MR | Mixed Reality | SLA | Service Level Agreement | | |
| CDMA-2000 | Code Division Multiple Access-2000 | HSPA | High Speed Packet Access | NCS | Networked Controlled System | TDMA | Time Division Multiple Access | | |
| CD-NOMA | Code Domain NOMA | IIoT | Industrial IoT | NaaS | Network Slice as a Service | TI | Tactile Internet | | |
| CN | Core Network | IoE | Internet of Everything | NFV | Network Function Virtualization | UDN | Ultra Dense Network | | |
| CNN | Convolutional Neural Networks | IoT | Internet of Things | NOMA | Non Orthogonal Multiple Access | UE | User Equipment | | |
| D2D | Device to Device | ITU | International Telecommunication Union | NR | New Radio | UHD | Ultra High Definition | | |

Network Function Virtualization (NFV), Software Defined Networking (SDN) and Software Defined RAN (SD-RAN)[29]–[32]. The market driven allocation and reallocation of bandwidth are the few efficacious parameters in 5G/B5G system [33], [34].

The smart applications constitute the entire automated smart city, comprising of the smart infrastructures in our day to day lives, ranging from a smart traffic monitoring system, with an efficient V2V/V2X interactions, providing a competent sensor based collision/accident detection system to an equally competent smart parking system. The other applications may vary from the smart health and education facilities like remote health counseling, online smart classroom with the teacher-student interaction, the smart grid system and the smart homes, collectively forming the smart neighborhood area network.

The virtualization facilitates an advanced computation of the network resources and their allocation. These, reasoned with their indispensable applications, facilitate a profitable proposition like network slicing [35]–[37]. The virtualization in the core network (CN) [38]–[40] with the billions of miscellaneous IoT devices [14], [41], improves the integration of the past and present cellular and Wi-Fi standards. It therefore provides a ubiquitous high rate, low latency, giving a smooth experience to all the users in the network.

The prime objective of the existing communication standard has always been to fulfill the demands of increase in capacity, the improvement in the data rate, the reduction in the latency, to provide a better quality of service and experience (QoS and QoE). Therefore to meet these requirements, drastic improvements have been made and are still ongoing, in order to update the existing cellular architecture. Hence the next generation network is said to be agile enough to revert back the intensified network complexity, regardless of handling diverse scenarios.

The existing 5G/B5G networks thus have to rely on the self organization and virtualization approaches, to deal with the disproportionate heterogeneousness and complexity of the network, associated with the massive amount of devices [42]–[44].In the forthcoming years, intelligence in the system is required to yield with such massive connectivity of IoT devices in the existing network with a minimum latency and network complexity. The B5G infrastructure thus has to focus on the considerable scaling and enhancement of the mobile network by incorporating open interfaces to support vertical segments in the network [45].

Such vertical segments are most often the third parties that do not own a particular network infrastructure but require the networking services with their specific requirements, along with their latest business solutions. Automotive manufacturing has been one of the most notable vertical segments in the existing communication system, requiring competent networking capabilities combined with the IoT and edge-cloud services. This in turn helps in the progress of a number of applications like autonomous driving, bird eye view, real-time assessment of road conditions, to name a few[46].

The mobile internet in B5G/6G will thus make provisions for human to human (H2H) interactions with the primary goal of connecting the machines and gadgets to construct an IoT interface which is often built on D2D and H2H interactions. Therefore, to solve the drawbacks such as latency, poor data rate and compatibility, high complexity, privacy and security, the next generation reconfigurable IoT allows for a real time control labeled as the 'Tactile Internet' (TI). According to the ITU[1], it is a network that combines extremely low latency with a high degree of reliability, scalability, and security[47].

TI here provides an improved and virtualized environment which is most likely feasible for the commercial applications like tele-operation with haptic communication like remote surgery. It has therefore been an onset towards revolutionizing every fragment of the society right from the education and healthcare with their prospected future applications varying from inculcating sensations and sentiments using the intelligent robotic applications to the smart health facilities like remote surgery to the virtual shopping experience at the user end.

The applications evolving the modernized WSN like the smart and automated homes and appliances, vehicles, factories, remote sensing and monitoring, augmented and virtual reality(AR/VR) and quantum computing based applications have an IoT as a common backbone, altogether forming an Internet of Everything (IoE)[48], [49]. The touch enabled sensation and actuation is expected to be one of the most fundamental applications of the B5G/6G communication technology due to its potential, simplicity and convenience, taking into consideration the real time scenario.

For this reason, the ultra-responsive internet thus helps enable a real-time control of the physical tactile-based haptic devices, bringing in a paradigm shift toward an intelligent and touch-enabled technology. 'Why do we need a touch technology?' is the most anticipated question here, taking into consideration all the previously-known aspects. The TI although being the most researched domain of the B5G/6G framework, its exploration towards the onset of an intelligent touch based technology is still in the infancy and has thus been emphasized upon in this survey.

### A. SCOPE OF THIS SURVEY

The opening gambit for such technological advancement takes into account the subsequent technical update of the wireless communication network with the contemporarily researched 6G system [50], [51]. It therefore encourages a real time interaction of humans with their environment, with few instances like the actuation of sensors causing the tactile sensation and the real time control/interface in our body system resulting from touching such surfaces.

It therefore defines a new human-machine(H2M) interaction framework enabling a physiological latency of human beings to build a real-time interactive system, with their applications ranging from robotics to healthcare to the autonomous driving including the use of virtual/augmented reality (AR/VR/XR/MR)[52]. For this reason the 'Tactile Internet' is regarded as an impetus and a cornerstone for the deployment of the touch technology and is expected to influence the development, innovation as well as the revolution of the healthcare, education, entertainment, manufacture, automation and smart grids.

This survey therefore paves a concrete path towards the initiation of the intelligent and touch enabled technology in the B5G/6G and IoT based wireless communication network.

### B. MOTIVATION AND CONTRIBUTION OF THIS SURVEY

The main purpose of this paper is to present a comprehensive and the state of art proposal motivated towards deploying an intelligent and touch enabled technology interface in the B5G/6G and IoT based wireless network. We first throw light on the next generation 6G WCN describing its key requirements and applications. The survey then subsequently emphasizes on the role of network slicing in B5G/6G network to sustain a high connectivity of massive number of devices without any latency and buffering at the user end.

[1] ITU: International Telecommunication Union

TABLE II
COMPARATIVE ANALYSIS OF PROPOSED SURVEY WITH THE EXISTING SURVEYS.

| Ref no. | Year | Key contribution | Technology used | Communication system used | Intelligence technique used. | Touch technology | Issues addressed |
|---|---|---|---|---|---|---|---|
| [68] | 2021 | 6G applications like holographic telepresence, e health and in body networks. | 6G | B5G/6G based mobile communication. | AI,ML and edge intelligence | No | An in-depth look at the latest 6G innovations. |
| [67] | 2021 | Space-air-ground-sea integrated communication network. | 6G | Network Slicing and Tactile Internet in 6G | AI | No | Addressing the coverage requirements in terrestrial and NTN[2] like satellite and UAV with high data rates and network security. |
| [66] | 2020 | A novel architecture that employs computing resources in a cross-layered infrastructure to enable network computing | SDN,NFV,TI | Tactile internet, transport and cross-layered protocols. | - | No | Leverages transport and network layers to increase network effectiveness in regards to congestion control and reliability. |
| [65] | 2020 | To provide virtual and logically independent slices for obtaining and deliver slice services from the infrastructure provider to the customers. | ML,DL, 5G network slicing | Network slicing with intelligence and virtualization | ML,DL, SMDP[3], N3AC[4] | No | Performance optimization. |
| [64] | 2020 | Design and operation of B5G wireless network using AI/ML technologies. | AI,ML,B5G | Network slicing and intelligence | AI,ML | No | Overview of ML/AI algorithms with channel modeling and estimation with network management and optimization in B5G wireless network. |
| [63] | 2019 | ML application in the 5G/B5G WCN. | ML,DL | - | ML,DL,DNN[5] | No | MAC layer based resource management, networking and mobility management, and localization in the application layer using ML. |
| [62] | 2019 | Integration of robotics, human-computer interactions and virtual control environment. | AI, Edge Computing | Network slicing with Tactile internet and AI | AI | No | An insight on the Role of Tactile internet in industrial systems as well as an enabling factor of the Industrial Revolution (4.0) with RAS[6] and Virtual control networks. |
| [58] | 2018 | Incorporation of deep reinforcement learning (RL) to handle cognitive smart cities services and improves their performance. | ML, DNN | - | Deep RL | No | Incorporate ML with a high level intelligence in the smart city services. |
| [56] | 2018 | Role of the Network slice Orchestration and Management, Network slice Broker in 5G/B5G network. | 5G/B5G, NFV, SDN, Cloud and Edge Computing. | Network slicing with virtualized fog/edge computing | - | No | 5G network slicing use cases with the E2E slice orchestration and management in eMBB, MIoT, eV2X, URLLC networks. |
| [60] | 2017 | Two types of feedback: Kinesthetic (based on force, torque, velocity, position) and Tactile (based on texture, friction, touch) | AI and predictive analysis | AI based predictive system | AI | No | Enabling touch transmission and actuation in real time using TI, having a control on the real and virtual objects i.e., H2H and MTC interface. |
| [47] | 2016 | The E2E Tactile architecture in real time has 3 domains: Master, Network and Control domain, in TI and 5G network. | HSI, SDN, NFV | Tactile Internet based haptic communication | AI | No | Touch transmission in real time using robotics, haptic equipment, by means of a communication network combining the TI and the 5G network with their applications in industry, automation, healthcare, VR/AR. |
| Our paper 2021 | 2021 | To propose an intelligent touch based system in B5G/IoT system incorporating AR/VR | B5G /6G network slicing, TI, IoT, ML | TI and intelligence based Touch communication system in B5G/6G | AI,ML,DL, Hybrid model | Yes | Layered and interfacing architecture for intelligent touch system for E2E solution. |

[2] NTN: Non Terrestrial Network
[3] SMDP: Semi Markov Decision Process
[4] N3AC: Network slicing Neural Network Admission Control
[5] DNN: Dense Neural Network
[6] RAS: Robotic Autonomous System

The network may be sliced into Cloud and RAN services and applications at the customer end. It therefore enables an easy access, storage facility and virtualization, with each slice serving different applications on the same physical network. As a result the network slicing and the tactile internet in the B5G/6G are therefore the backbone for incorporating a touch enabled infrastructure with an induced intelligence by the AI/ML/DL implementation in the proposed system.

Through this paper we propose here an intelligent touch configured system that would be applicable in the B5G/6G WCN. We hence summarize our contributions as follows:

1) We describe the well researched B5G/6G communication system while throwing light on some of its key parameters and prominent use cases with their applicability in the next generation networks.
2) We provide a discussion on the Tactile Internet along with its applications in the B5G/6G networks as an enabler of intelligent touch based system.
3) We introduce an intelligence in the B5G/6G network with a descriptive mention of few of the intelligence learning techniques in the wireless communication.
4) We further provide a backdrop of the intelligent touch configured technology involving the orchestration of the network slicing with the tactile internet allied with the intelligence and IoT connectivity in the B5G/6G wireless domain.
5) We propose a layered and an end to end (E2E) interfacing architecture enabling the touch technology interface in the B5G/6G domain.
6) We further discuss the research challenges and further explore the research areas of the next generation (6G) deployment and ultra low latency tactile based applications in future.

### C. COMPARATIVE ANALYSIS WITH THE EXISTING SURVEYS

The Table II indicates a complete summary of the existing survey papers on the network slicing and the tactile internet with some of the intelligent technologies like ML/DL implemented in the wireless cellular communication network in 5G/B5G scenarios. In contrast to the other surveys, our survey provides a comprehensive overview of the proposed intelligent touch enabled technology in the B5G/6G and IoT configured WCN.

Farris *et al.* [6], describes the essentiality of the mobile edge computing (MEC) for supporting a wide range of user centric applications. These have an important role in the smart city scenarios presented in Taleb *et al.* [53] where a smart MEC based architecture is significant for reducing the core network traffic while guaranteeing an ultra short latency for the existing network.

Thus MEC here acts as a key factor in enhancing QoS and attaining the *1ms* requisite latency for the 5G/B5G mobile systems. It is accompanied by a considerable number of virtualization technologies like NFV, SDN and SD-RAN. The virtualization in turn promotes an advanced computation and allocation of the network resources. Myriad existing studies have provided a comprehensive overview of the technical challenges and applications associated with the network slicing, TI and the emerging intelligence in B5G/IoT systems.

The works [39], [45], [54] provide comprehensive reviews of literature on network slicing in B5G networks. Richart *et al.*[39] discusses resource slicing and their allocation in virtual networks powered by SDN and NFV, as well as how these can be distributed appropriately to network slices without impairing the efficiency of other slices. While Afolabi *et al.* [45]describes the state of art network slice life cycle architecture operating across the multiple domains thereby enabling an effective network programmability and flexibility with the creation, management and orchestration of the network slices, utilizing the massive IoT and multimedia broadband connectivity.

Foukas *et al.* [54] reviews the concept of network slicing and proposes a generalized layered architecture consisting of an infrastructure layer, a network function layer, and a service layer, along with their associated benefits and challenges. The existing progress in the B5G network slicing with its key trends along with their corresponding potential challenges is presented in [55]. The authors in. [56] throw light on the various concepts of network slicing and softwarization[7] in the B5G technology with their applicability across the RAN and core network, altogether establishing an E2E slicing infrastructure.

The intelligent tools like AI/ML/DL play significant role in outlining automation, deployment and disposition of different applications the existing as well as the next generation networks (B5G/6G) [57]. The work of Mohammadi and Fuqaha [58] provide an intensive facet of the most prevalent deep reinforcement learning in structuring of the cognitive smart cities and its applicability concerning the energy consumption with garnering of water and agricultural utilization.

The Kafle *et al.* [59] successfully highlights the various ways of applying AI/ML techniques for the automation of network functions in different configurations, ranging from development organizations to industrial forums. One such intelligent application of the B5G/6G requiring the *1ms* latency is Tactile Internet (TI) which is quite efficient in establishing a bilateral communication between the humans

[7]Network softwarization is the notion of designing, architecting, deploying and administrating network components, largely based on software programmability properties. It enables flexibility, adaptability, and even total reconfiguration of a network on the spot[37].

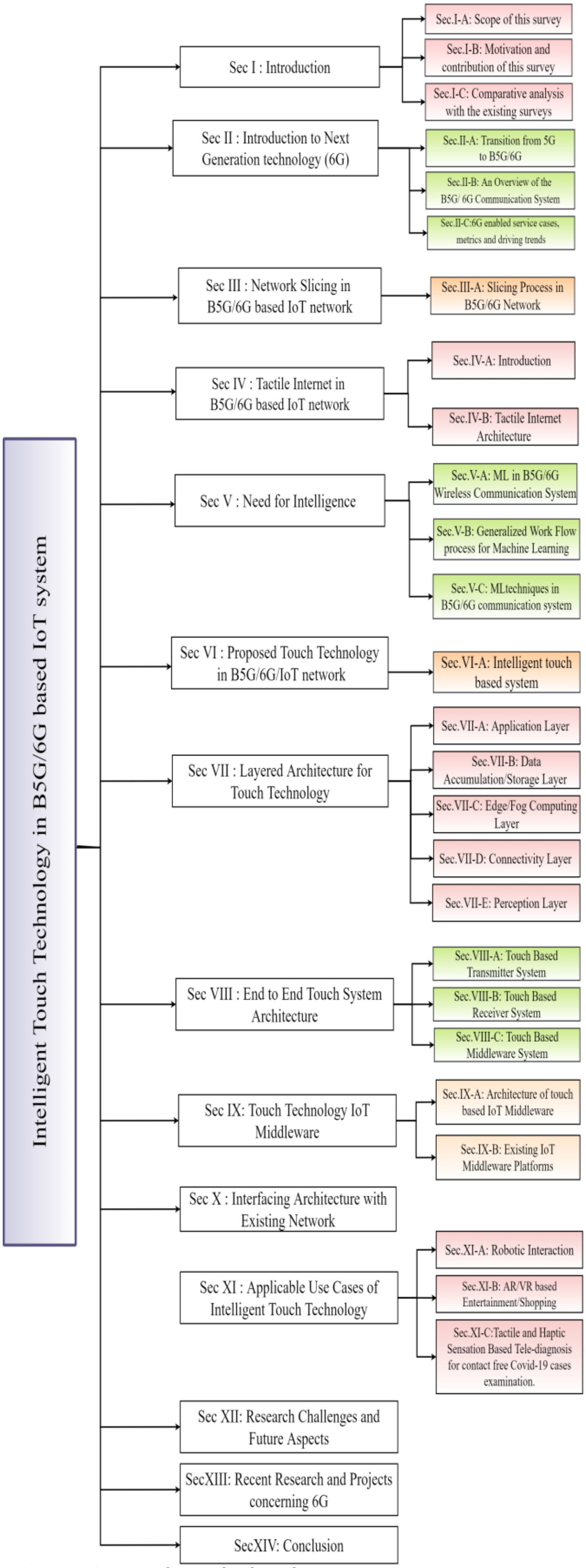

**FIGURE 2: Structural organization of paper.**

and machines, forming a HSI, which ultimately enables a haptic communication system.

The term 'Tactile Internet' coined by Gerhard Fettweis in [46], has been a key enabler in fulfilling the need for a higher data space, essentially resulting in a continuous increase in the storage and computation among various cellular devices connected to the internet. Therefore the TI is centered on H2M interactions with the devices using the B5G and IoT connectivity, enabling haptic and tactile sensations at both the transmitting and receiving end forming a bilateral communication and feedback network. The works of Maier *et al.* [52] highlight the commonalities and various subtle differences between the TI, IoT and the B5G vision.

The authors in Simsek *et al.* [47] highlight the critical requirements and architectural approaches for TI, as well as the technical issues and challenges associated with the resource management, core networking and edge cloud/ AI capabilities. The Aijaz *et al.* [60] helps to examine a number of the stringent design challenges to revolutionize the tactile internet, providing an enhanced haptic perception with a *1ms* round trip delay. Authors in Antonakoglou *et al.* [61] explain the evaluation of methodology and technology assessments for the necessary haptic communication infrastructure.

The examination of the advancements in tele-operation over long distances has an effect on haptic communication, while using the Tactile Internet. Therefore as per [62], the Tactile internet is a key enabler for realizing industrial revolution(4.0) by users and devices in real time. The work by Y.Sun *et al.* [63] explains how machine learning may help with resource control at the MAC layer, network and mobility management in the network layer, and application layer localization.

While C.X Wang *et al.* [64] gives an overview of ML/AI technologies while addressing issues like channel modeling, estimation, network management and optimization in B5G wireless communication network. Following which D.Bega *et al.* [65] encourages exercising ML approach towards the market optimization while maximizing infrastructure provider monetization. Hence to thoroughly optimize the availability of the computing resources, authors in [66] present a novel tactile based flexible next generation internet architecture (FlexNGIA) that capitalizes on the coexistence of transport and network layers to provide better congestion control and reliability services via cross-layered network computing.

You *et al.* [67] gives an insight of the next generation 6G communication highlighting its parameters specifically emphasizing its coverage requirements for the functionality in terrestrial as well as non terrestrial environments. The paradigm aspect of this work is the integration of the space-air-ground and sea based communication network. C.De.Alwis *et al.* [68] discusses several 6G use cases, including holographic telepresence, e-health, and in-body networks that require extremely high data rates, ultra-low latency and high reliability. As a result, the continuous penetration of mobile platforms, robotics, human-computer interaction, and autonomous agents in virtual environments will distinguish future communication and industrial systems.

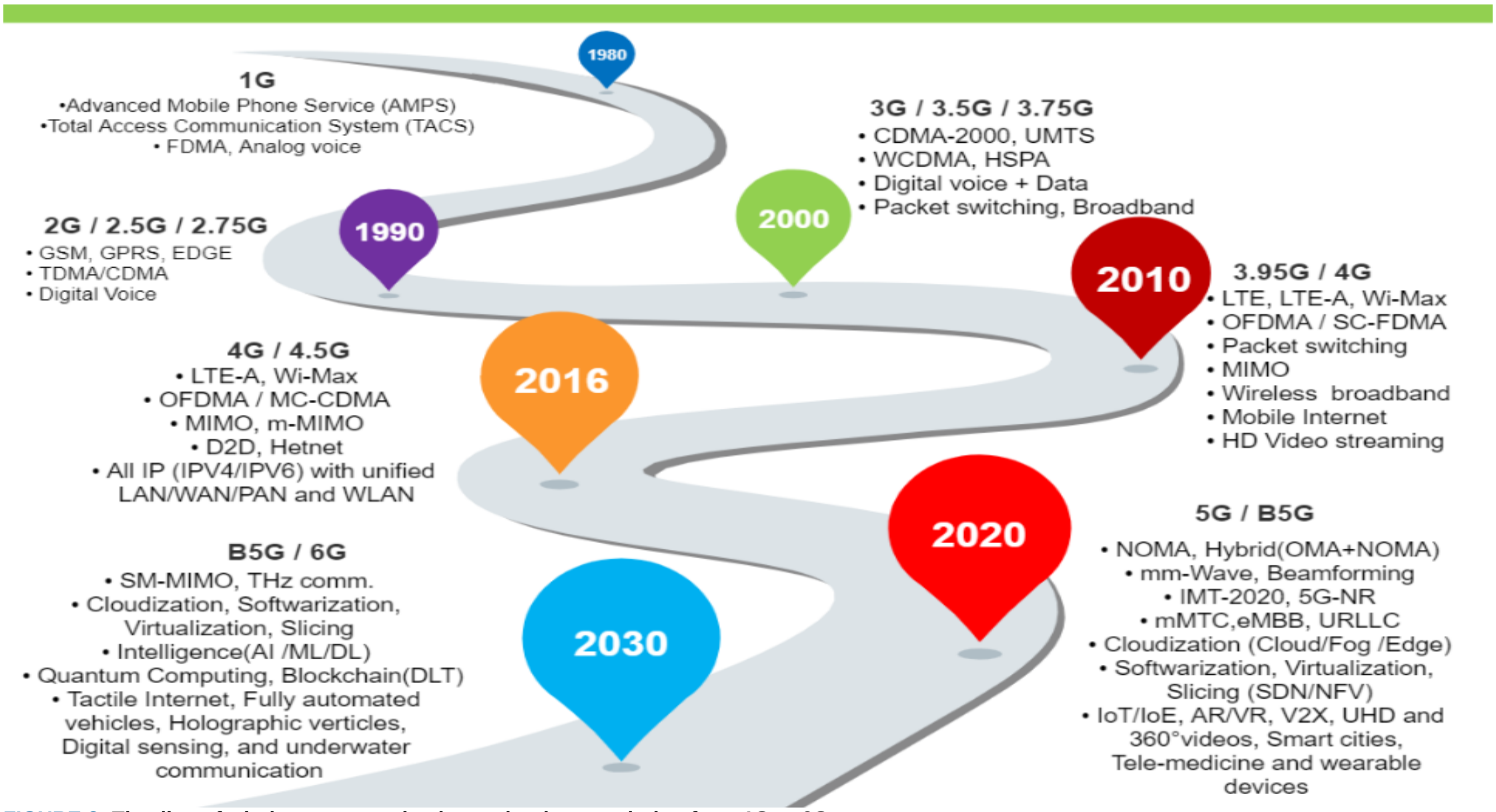

**FIGURE 3**: Timeline of wireless communication technology evolution from 1G to 6G

The remainder of the paper follows the structure depicted in Fig 2. In Section II and III we discuss the next-generation 6G system in detail with its main parameters, service and applications along with network slicing and virtualization concepts in 6G system. Section IV delves into the architecture and state-of-art uses of the tactile internet, which is a crucial component of the touch interface system. Section V introduces the intelligence that is to be incorporated into the next-generation wireless network.

Section VI explains the proposed intelligent touch technology, including its layered and E2E architecture in Section VII and VIII. Section IX discusses some basic major components of the proposed system, while Section X presents an interfacing architecture with the existing network. Section XI contributes some potential and applicable use-cases of the proposed system, while Section XII highlights some of the research challenges and upcoming future aspects. Section XIII summarizes the recent research and few of the ongoing projects on 6G before concluding in Section XIV.

## II. INTRODUCTION TO NEXT GENERATION TECHNOLOGY (6G)

The 6G communication era anticipates how humans will engage with digital virtual worlds beyond 2030, as the projected digital transition with the existing B5G networks have already begun and will continue to evolve over the next decade. Although 5G/B5G is recognized for network cloudification via micro service architecture, the next generation 6G network is strongly linked to the intelligent network orchestration and management. New digital virtual worlds with the connected intelligence must have novel technologies that support these communication and networking challenges beyond 2030.

The $6^{th}$ generation wireless communication network is anticipated to consolidate the terrestrial, aerial, and maritime communication into a robust network that would be more reliable, faster, and capable of supporting a large number of devices with ultra-low latency requirements while remaining cost-effective. Due to the exponential growth in the number of IoT devices, the next generation systems must achieve high spectral and energy efficiency (SEE), low latency, and massive connectivity to provide for services like smart traffic monitoring, VR navigation, telemedicine at the user end, along with digital sensing using a full HD video transmission in connected autonomous devices like drones and robots[69].

The timeline in Fig 3 shows the evolution of the wireless communication network from 1G to the recent 6G. The next section explains why the shift from 5G to 6G is necessary.

### A. TRANSITION FROM 5G/B5G TO 6G

As 5G/B5G networks are consistently deployed, the inherent limitations of this system are being exposed, in comparison to its original assertion as a platform for IoE applications. It is gradually more difficult for the existing multiple access techniques, to cope up with the exponentially growing IoT devices. As a result, the 5G communication systems, already being implemented in the world today are incapable of supporting these many IoT devices. The requirement for faster data rates has fueled the evolution of wireless networks, which has necessitated a continuous *1000*-fold increase in network capacity[70].

TABLE III:
COMPARISON OF PERFORMANCE ATTRIBUTES BETWEEN 5G, B5G AND 6G COMMUNICATION SYSTEMS.

| Performance attributes | 5G | B5G | 6G |
|---|---|---|---|
| Application types | • eMBB<br>• URLLC<br>• mMTC | • Reliable eMBB (ReMBB)<br>• URLLC<br>• mMTC<br>• Hybrid( URLLC + eMBB) | • MBRLLC<br>• mURLLC<br>• HCS<br>• MPS |
| Architecture | • Dense small base stations operating at sub-6GHz in conjunction with umbrella macro base stations.<br>• Mm-wave small cells with a range of approximately 100m for fixed access. | • Denser small cells operating at sub-6 GHz with umbrella macro base stations<br>• Mm-wave cells with a diameter of less than 100m. | • For mobile and fixed access, cell-free smart surfaces with high frequency are supported by mm-wave small cells.<br>• Drone-carried base stations and tethered balloons provide temporary hotspots. |
| Frequency bands | • Sub 6 GHz<br>• Mm-Wave for fixed network accessibility. | • Sub 6 GHz<br>• Mm-Wave for fixed network accessibility. | • Sub 6GHz<br>• Mm-wave for mobile network accessibility.<br>• High frequency and THz bands above 300GHz are investigated.<br>• Non RF technologies like VLC, Optical fiber communication etc |
| Spectral and Energy Efficiency (SEE) | 10x in bps/Hz/$m^2$/Joules | 100x in bps/Hz/$m^2$/Joules | 1000x in bps/Hz/$m^2$/Joules |
| Data rate<br>E2E delay<br>Radio-only delay<br>Processing delay | 1Gb/s<br>5ms<br>100ns<br>100ns | 100Gb/s<br>1ms<br>100ns<br>50ns | 1Tb/s<br><1ms<br>10ns<br>10ns |
| E2E reliability requirement | 99.999% | 99.9999% | 99.99999% |
| Interoperable devices | • Smart phones<br>• Sensors<br>• Drones | • Smart phones<br>• Sensors<br>• Drones<br>• XR equipment. | • Sensors and DLT devices<br>• CRAS<br>• Smart implant system<br>• XR and BCI equipment |

As the demand for wireless capacity continues to surge, the emerging IoT system, which connects millions of people to billions of machines, has resulted in a radical paradigm shift from the rate centric eMBB services from the previous eras towards URLLC and intensified mMTC services, as per the 3GPP, which is working on the implementation of 5G/B5G standard[71]. Although it can be asserted that the evolutionary aspects of the existing 5G supporting the data hungry eMBB services have gained a significant momentum, while the promised revolutionary disposition systems, operating exclusively at high mm-wave frequencies have yet to materialize.

Despite the fact that today's linked 5G systems are easily capable of supporting the most fundamental IoE and URLLC services (such as factory automation), it is still debatable whether or not they will be able to deliver the smart city services based IoE applications in future. Conversely, the initial and the existing B5G deployments are most likely to rely on the low frequencies (sub 6GHz) to support mobile data transmissions. While on the other hand, an enormous influx of new IoT services such as XR (including AR/VR/MR), flying vehicles, and connected autonomous systems would most likely derail 5G's original purpose of supporting small packet and sensing-based URLLC applications[72].

Thus, in order to successfully operate these IoE services, a wireless system must simultaneously provide a high level of reliability, low latency, and a high data rate for a wide range of heterogeneous devices. These new services necessitate the resolution of novel and distinct challenges, unprecedented in terms of their complexity including the tradeoff between latency, throughput and reliability. Not only do these services help entail new approaches for an effective regulation and handling of performance and challenges but also aid in exploration of frequencies beyond 6GHz range in order to create a self sustaining and intelligent wireless network.

This aforementioned network is capable of provisioning and orchestrating communication, computing, control,

TABLE IV
COMPARISON OF VARIOUS 6G SERVICE CASES WITH PARAMETERS AND APPLICATION AREAS.

| S.No | 6G service cases | Performance Attributes | Application Areas |
|---|---|---|---|
| 1. | FeMBB (Further enhanced Mobile Broadband) | • Enhanced broadband in densely populated areas.<br>• Operates in THz communication.<br>• Enhanced multimedia applications like 4D video gaming, mobile TV, connected wearables and sensors. | • Public transportation<br>• High speed trains<br>• Smart cities |
| 2. | umMTC ( Ultra massive Machine Type Communication) | • Reliable connectivity with massive scale (trillions of devices) of connected devices.<br>• Improves connection density.<br>• Enables IoE with ultra dense cellular IoT networks.<br>• SigFox and LoRa as potential technologies for enhanced connectivity and network coverage in 6G. | • Internet of Industrial Smart Things (IIsoT)<br>• Smart buildings<br>• Internet enabled supply chains, logistics, fleet management and water quality monitoring.<br>• Natural/wildlife sensing<br>• Forest monitoring |
| 3. | ERLLC (Enhanced Reliable and Low Latency Communication) | • End to End fast turnout time.<br>• Intelligent framing and coding with efficient resource management.<br>• Intelligent UL/DL communication<br>• Remote robotic surgery using smart surfaces and intelligence. | • Telemedicine<br>• Internet of Healthcare (IoH)<br>• Remote Robotic Surgery<br>• XR |
| 4. | ELPC (Extremely Low Power Communication) | • Uses Intelligent Reflecting Surfaces (IRS) known as Reconfigurable Intelligent Surfaces (RIS).<br>• Reduces hardware dependency and Tx-Rx complexities.<br>• Reduced energy consumption with passive array transmission. | • Smart homes<br>• Smart cars<br>• UAVs |
| 5. | LDHMC (Long Distance and High Mobility Communication) | • High mobility and seamless communication for long distances (>1000km).<br>• Accurate channel estimation.<br>• Use of FBMC and UFMC as alternative to OFDM. | • Deep sea tourism<br>• High speed transportation<br>• Space sightseeing |
| 6. | MBBLL (Mobile Broad Bandwidth and Low latency) | • MEC to attain end to end low latency.<br>• Low complexity mechanism for VR experience by user. | • Mobile AR,VR |
| 7. | MLLMT (Massive Low Latency Machine Type) | • Data availability, ultra scalability and low latency.<br>• Time critical applications where decision making takes fraction of seconds.<br>• Automation, controlling and monitoring of industrial 4.0 use cases. | • Home and building automation<br>• UAVs<br>• IoT enabled Healthcare |
| 8. | MBBMT (Massive Broadband Machine Type) | • Touch based experience with high data rates.<br>• Massive IoT connectivity in densely populated areas.<br>• Tactile sensations captured by sensors/devices converts to digital data. | • Tactile Internet |

localization and sensing of the scenarios that are best suited for IoT needs[73]. So to address these issues, a game-changing 6G wireless system is required, with a design that is organically tuned to the performance requirements of IoE applications and associated technical advancements.

### B. AN OVERVIEW OF THE B5G/ 6G COMMUNICATION SYSTEM

The 6G wireless communication network, which is currently being researched, is expected to integrate terrestrial, aerial, and maritime communication systems into a robust network that is more reliable, fast, and capable of supporting a large number of devices with ultra-reliable and low-latency requirements. The AI, ML/FL, quantum communication, blockchain/DLT, beyond 6GHz and towards Terahertz communication, TI, swarm UAVs, Zero touch network and service management (ZSM), large intelligent surfaces (LIS), NTN and 3D networking, VLC, compressive sensing with an efficient energy transfer and harvesting are just few of the currently proposed by the ongoing researches worldwide[74].

Owing to a massive growth in the number of IoT devices, realization of the advanced services like smart traffic monitoring, VR based navigation, smart medical facilities like tele-medicine and HD video transmission in drones and robots is possible. Hence the B5G/6G communication systems aim to achieve high SEE, low latency, and massive connectivity. The ever-increasing number of IoT devices makes it difficult for the existing multiple access strategies to handle such a huge number of devices, therefore requiring a more extensive network in order to make use of the massive bandwidth capabilities offered by B5G/6G communication systems.

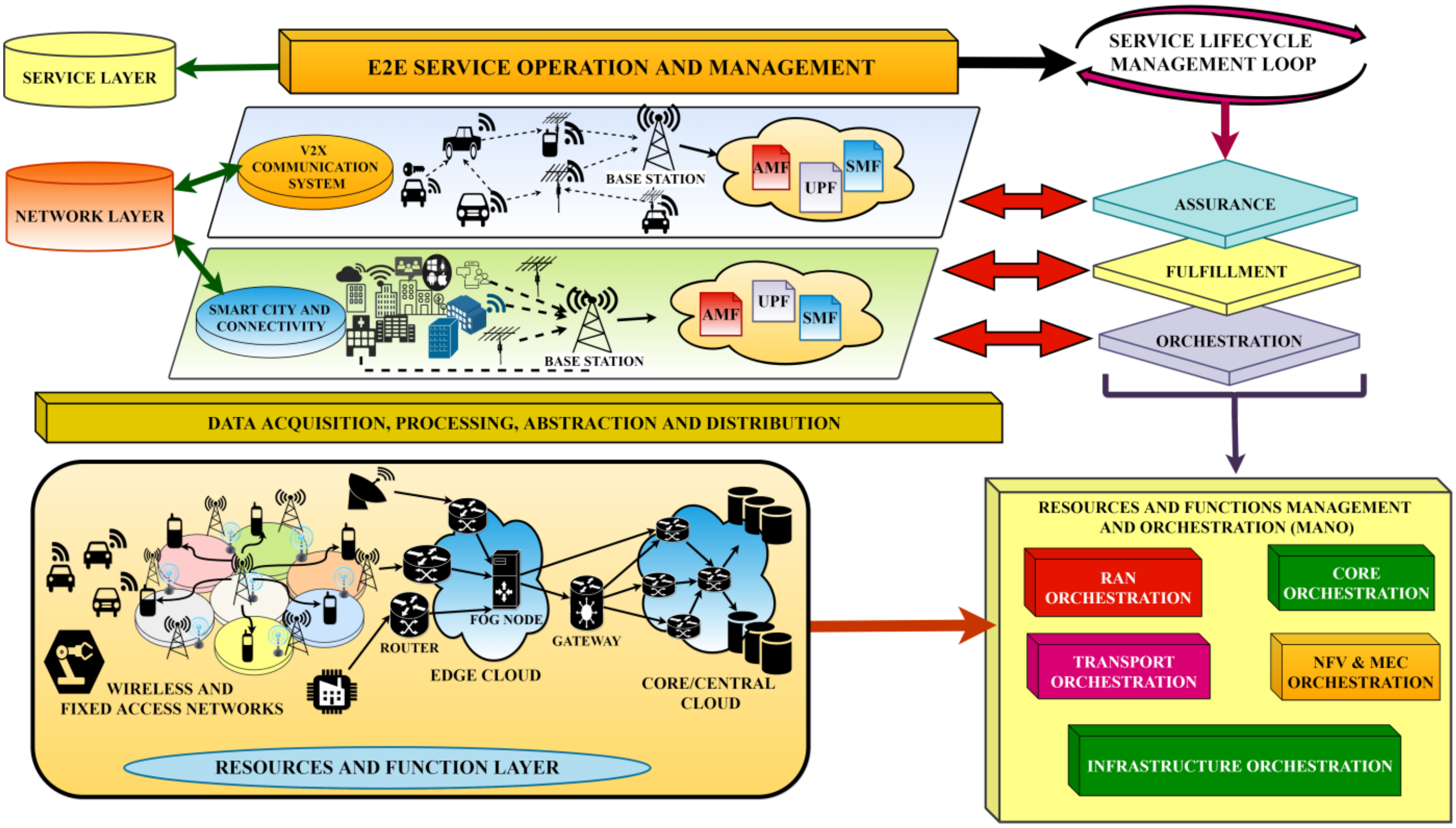

**FIGURE 4: Network slice functionality in B5G/6G network**

According to the work by Zhang *et al.* [75], a speculated vision and functionality of 6G networking scenario provides a technological framework and requirements for industries in the future generation communication system with cell less architecture, decentralized networking, and resource allocation with 3D radio interoperability. The next generation wireless network comprises of a large number of linked devices with numerous base stations (BSs) and access points (APs), each of which will serve multiple devices at the same time, forming a coordinated multipoint (CoMP)[76].

Hence more data would be transmitted via future wireless communication networks, since most of the value-added apps and services rely significantly on the data exchanges. Therefore large devices will generate a massive quantity of data, which will need high-performance processing units and backhauling connections. The following subsections go through the new 6G enabled driving trends, metrics, and use service cases.

### C. 6G ENABLED SERVICE CASES, METRICS AND DRIVING TRENDS.

With the new performance metrics, new technological trends step up to redefine the prevalent B5G applications by morphing the classical URLLC, eMBB, and mMTC into something entirely new and innovative. As a result, Table III gives a comparison of some of the key performance attributes of the 5G, B5G, and 6G wireless communication systems[73]. Various countries have initiated projects aiming at the research and deployment of B5G/6G communication networks, as discussed in Section XIII. The research on 6G is accelerating, and has been documented in recent works like [77]–[79].

According to the recent research, a variety of possible 6G applications have been classified as mobile broad bandwidth and low latency (MBBLL)[80], massive broad bandwidth machine type (mBBMT)[78], massive low latency machine type (mLLMT)[81], further-enhanced mobile broadband (FeMBB)[82], extremely reliable and low-latency communications (ERLLC)[83], ultra-massive machine-type communications (umMTC)[44], long-distance and high-mobility communications (LDHMC)[84] and extremely low power communications (ELPC)[85]. Detailed descriptions of each of these are provided in the preceding Table IV, along with their parameters and application areas.

## III. NETWORK SLICING IN B5G/6G BASED IOT NETWORKS

The forthcoming B5G/6G networks aim to serve a wide range of applications, thus recognizing the 5G epoch as the century of mobile telecom networks, all the while promoting dedicated use-cases and endowing unequivocal services, to meet diverse user requirements. Hence the technologies like UHD, multimedia, AR/VR/XR therefore needs a faster speed and a relatively higher capacity and connectivity, compared to the mission-critical applications like IoT/MIoT and autonomous systems require an ultra-low latency and ultra-reliable operation.

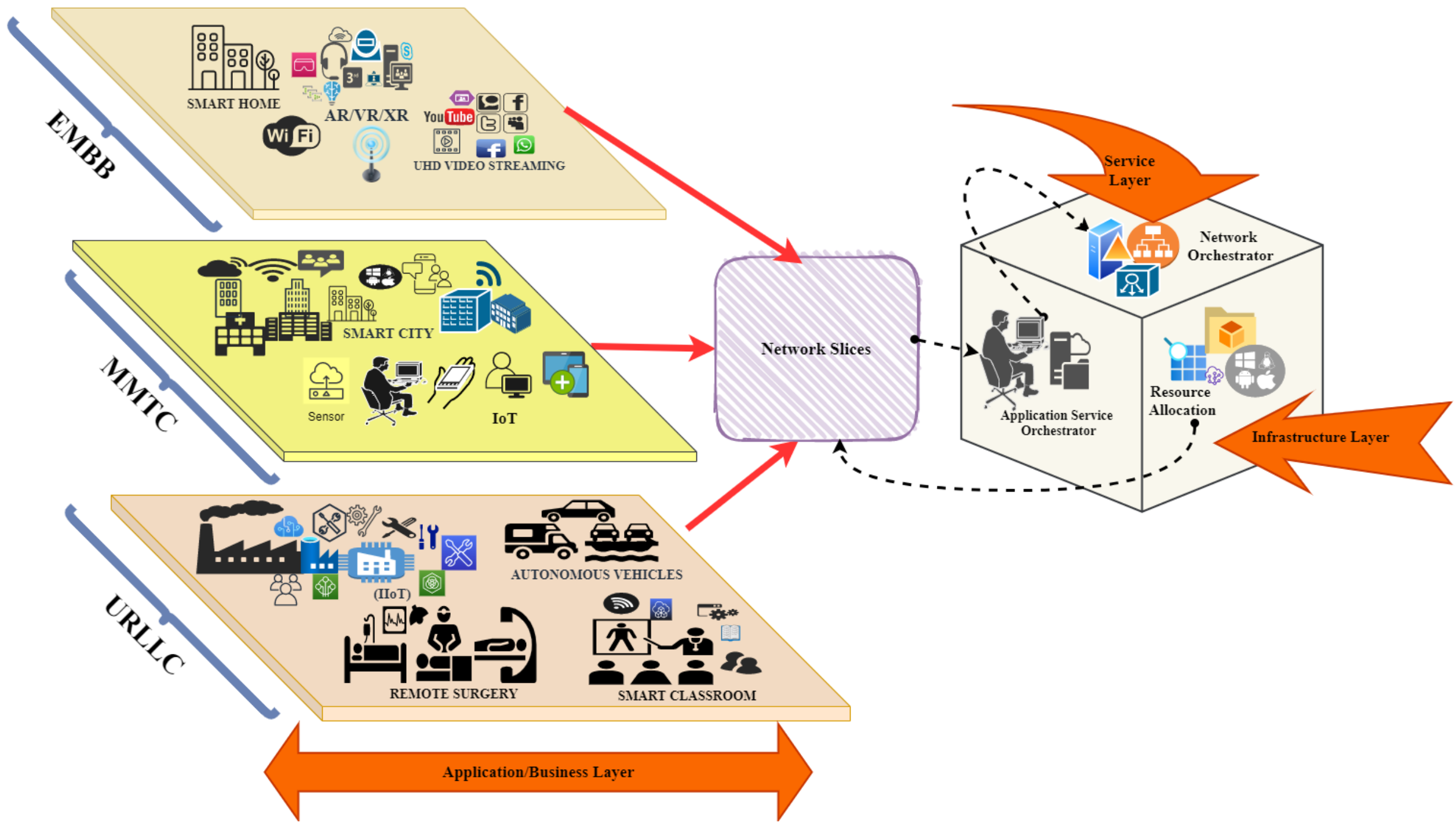

**FIGURE 5: Network slice orchestration and management**

The 6G cellular framework formulation anticipates to be accomplished on the existing and researched 5G/B5G technology, thus supporting a surplus of network services with miscellaneous performance requirements. The advancement of cellular networks and resulting generation-wide improvements is motivated primarily by the desire to enable better data-based services. A variety of aspects render 5G important, including the mm wave spectrum distribution and reallocation of bandwidth. Virtualization with a billion individual networks in the CN and IoT facilitates the convergence between previous and present cellular and Wi-Fi requirements[83].

This in turn offers a pervasive high rate and low latency experience for network customers. The most significant part of the B5G/6G infrastructure comprises of the network service and its development platform. It is highly capable of improving the network scalability while fulfilling the user requirements by utilizing existing services. The network slicing functionality in B5G/6G domain has been depicted in Fig.4. The virtualized infrastructure here has provisions for slice instances, and collectively functions with the infrastructure resources from another slice instance[37].

A set of homogenous APIs are made available for creating an abstraction layer to facilitate with the slice management while controlling its virtual resources during its operation. These slices can therefore be accessed by different tenants or third parties using these APIs. Here SLA act as the slice blueprints, using which the tenant specifies its requisite slice characteristics ranging from topology, management, control and so on. The slice lifecycle is regulated by the service lifecycle management loop, openly accessed by all the functioning slices[45]. The management and orchestration (MANO) offers an integrated and a holistic approach towards the regulation of network slicing and the NFV management.

It offers a standardized level of data abstraction followed by the adapt specification of its network infrastructure together with its service management and implementation process[31]. This section discusses the concept of network slicing in B5G/6G communication networks with its functionality, management and orchestration in RAN and CN and finally its application in the 6G with the proposed architecture.

### A. SLICING PROCESS IN THE B5G/6G NETWORK.

The existing B5G is most likely to consider a variety of business and service quality requirements like the enhanced capacity coupled with the intelligent traffic and offloading techniques accompanied with a highly complex and heterogeneous network. All of these fulfill the required performance criterion together with an autonomous network management. A high data rate guarantees a high level of end user service quality with an unlimited mobile broadband connectivity in the jam-packed areas like stadiums, concerts and shopping centers, by means of the terminals having the AI capabilities [86].

The reduced latency with high data rates are capable of supporting the UHD streaming from the cloud technology and improved VR devices and other wearable computing gadgets. It therefore provides a faster web downloading while enabling a premium user experience in services like YouTube streaming, Netflix and so on with a high video resolution. The network slicing is a fundamental key for the B5G/6G technology. Thus the network slices here are an end to end concept of the next generation technology where the slice operator supports a massive amount of customers. Here each of them in the long run requires a multiple end to end individual and logical networks referred as network slices. They are categorized into three components, each of them governing the RAN, core and transport domain[56].

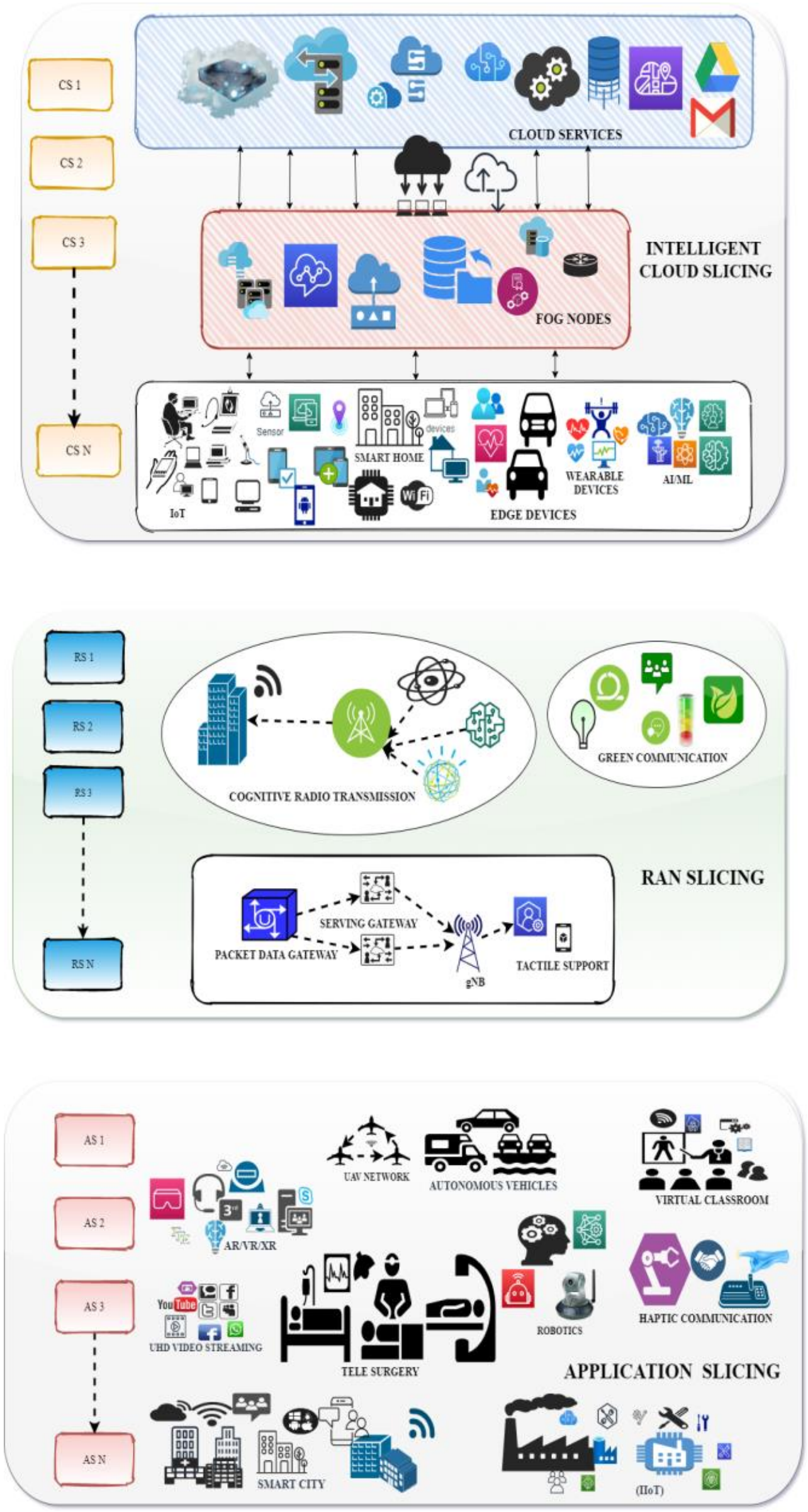

FIGURE 6: **Network slicing in next generation networks (B5G/6G).**

The RAN and core slices consist of the application context and personalities respectively with the transport slices being connectivity between the RAN and core. Each individual domain has a controller, i.e. the RAN controller, core controller as well as the transport controller, all of these supported by an end to end orchestrator. To realize the B5G/6G networks, 5G network slicing plays a crucial role in guiding about slice utilization for automation, assurance and optimization of transport slices involving various low latency and high reliability applications.

These applications may range from automated vehicles, tactile applications, smart devices and so forth[87]. The Fig.5 diagrammatically illustrates the network slice orchestration process applicable in B5G/6G networks. At the back end, the resource allocation takes place in the infrastructure layer where resources are provided to the individual slices. The slicing applications are managed by the network orchestrator in the service layer, so as to enable an effective application slice management and orchestration. It can be undoubtedly claimed that the most defining feature and in other words the 'secret sauce', for the 5G/B5G success is the E2E network slicing, which will be applicable in 6G networks as well.

Hence the network slicing is responsible for the optimal resource efficiency and flexibility in the network. It therefore enables the implementation of new business models as NaaS, supporting various mission critical use cases including the industrial automation(4.0) and availment of remote health facilities[50], [88]. The network slicing architecture pertaining to the B5G/6G system has been illustrated in Fig 6. The slicing process is therefore described as the three sub processes, interlinked with each other: the intelligent cloud slicing, the RAN slicing and the application slicing, all of which are functional in the B5G/IoT enabled networks.

Commencing from the lowermost stratum we have various applications that furnished at the consumer end ranging from the real time online gaming with a UHD streaming to the live online classroom teaching sessions, efficiently making use of the AR/VR technology to deliver the required information. The other applications may also involve the use of robots in real time actuations and tactile and touch based haptic communications, which may be put to a practical use in industrial operations, automation in the robotics and machinery, vehicles and UAV fleet to accomplish the services requested by the users at the customer end in the form of application slices.

The succeeding layer is that of RAN slicing, taking place in between the CN and RAN, i.e., at the backend of the network, routing the clients with the final applications. It is capable of enabling an effective resource and spectrum allocation with power and energy efficient cognitive radio network system. All of this is in the form of RAN slices and successfully connects the edge devices with the cloud network. The cloud computing and storage enables intelligent cloud slicing technique, where the cloud enabled applications are accessed in the form of cloud slices. These support different edge devices at the user end while facilitating the requested services.

The following section describes the tactile mode of communication in B5G/6G system that is to be incorporated with the aforementioned slicing techniques, so as to deploy and configure the proposed touch interfacing wireless system with incorporated intelligence which is discussed in detail in the subsequent sections.

## IV. TACTILE INTERNET IN B5G/6G SYSTEM

The internet initially was designed and indented to be a reliable and interoperable means of communication across the globe. With time, not only has it evolved to convey a large amount of content, but it also helps enhance the real world experience of the users. The tactile internet thus improves remote real-time physical engagement with existing and virtual items. Whereas integrating TI with touch interface will provide a two way interactive experience that blurs the boundaries between the real and the virtual world. The works in [89]–[92] examine the technical requirements of TI and its ability to support future smart city applications. It therefore provisions for a real-time control with the physical and haptic sensations to be experienced by the users remotely.

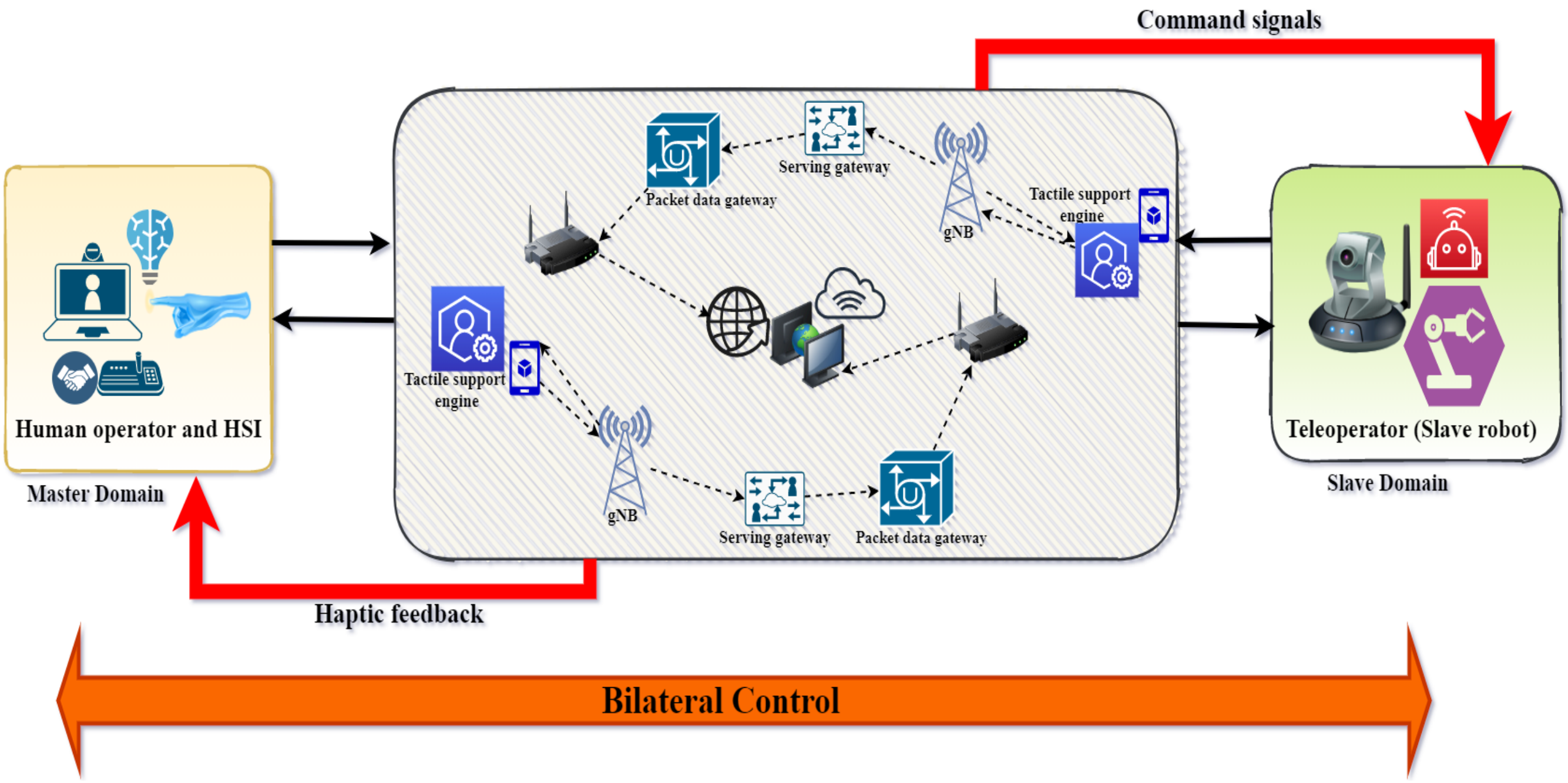

**FIGURE 7:** Tactile internet E2E architecture with bilateral feedback system

Unlike the traditional internet applications, the new tactile internet (TI) intends to serve as a medium for remote, real-time physical interactions with actual, physical objects, including humans, machines, and processors. The tactile internet enables enhanced virtualized remote classroom instruction with the participants in collaboration with the remote environment collectively amid the presence of remote and virtual resources. Thus, the data rates of wireless communications have been increasing, which is primarily due to innovation in electronics and the latest communications technology, including text messaging, video streaming, emails, and file sharing.

The TI is centered on the H2M interactions with the haptic devices. To facilitate haptic communication, the transmission of data via tactile internet creates a network that is both extremely reliable and extremely responsive. Haptic interaction is a type of interaction that includes the use of remote touch. It refers to the kinesthetic perception of information conveyed by the muscles and joints of the body via force, torque, position, and velocity, as well as the tactile perception of information conveyed by the mechanoreceptors of the human skin via surface texture and friction. Sharing information through kinesthetic mode facilitates a global control loop with stringent latency requirements, while containing a feedback of generally audio/video type[93].

Now with the enabling of haptic data, the TI enables a networked control system (NCS) supporting the connected sensors and actuators while controlling highly dynamic processes. Therefore it helps in digitally transmitting the sense of touch from one place to another, facilitating the URLLC network in the B5G/6G system. In other words, the tactile internet's purpose is to provide a remote and dynamic way for people to experience physical haptic or the touch based control, while exchanging closed-loop information between the virtual and physical worlds[90].

Wireless communications can thus be a medium for controlling and directing real and virtual objects using such a platform. This revolutionary technology continues to transform healthcare, transportation, education, logistics, smart grid systems and many more, hence covering a major portion of the economy sector in the society. It thus provides sub-millisecond connectivity for the healthcare applications like remote surgery.

This section draws attention to one of the most popular applications of the B5G/6G communication system: 'Tactile Internet' and helps review its parameters, applications and the basic architecture, while focusing on its application in the intelligent 6G and IoT systems.

### A. INTRODUCTION

The 'Tactile Internet', as coined by Gerhard P. Fettweis, has been a catalyst for the economic development and creativity and in bringing a new stage of maturity in adapting technologies for a changing global environment. [46]. Given that cellular communications connects a vast majority of people worldwide, it is therefore imperative to connect the technology as well. According to IEEE P1918.1 working group, the Tactile Internet may be defined as a network of networks that can be accessed, perceived, and manipulated by people or machines on a remote, real or virtual basis in real-time perceptions[94].

The tactile internet application offers the standard required latency needed to guide and control real and virtual objects without causing cyber sickness, revolutionizing education, accessibility and traffic, healthcare, athletics, culture, games, and the smart grids, thereby profoundly shaping our community. The i-phone for instance has an astounding haptic interface, provided by the gyroscope and the modern touch screen technology, which has been a welcome step that could drastically transform the way we connect.

Additionally, we have an instinctive (or innate) sense of our surroundings, which has a tactile understanding of the real time connection with our world. The tactile feedback and a phone give our hands the sensations, which in turn enable our whole system to be modulated with its proximity. Whereas inside the vehicle it modulates several sensors and controls elucidating a real time human-machine communication interfaces. Thus all TI needs is a highly responsive, smart, and reliable connectivity in order to provide a medium for intelligent, real-time touch, control, sensing and operation.

With high availability of TI, accompanied with its very fast reaction time and reliability, the human interaction with the machines enables a new dimension by creating an interactive, real-time system, which revolutionizes almost every segment of the society[94]. Taking into account the industrial dimension, TI is an interconnected system of specialized components and applications used in industrial environments to monitor and operate the physical equipment. Hence the works in [62], [89], [92] contribute to the role of TI in industrial systems by examining its potential in emerging and future industrial sectors.

Against this backdrop, the goal of this study is to identify and address the cutting-edge challenges to implement an intelligent touch enabled system via tactile internet in the perspectives of B5G/6G based wireless and IoT communications networks. Thus, tactile internet serves as a key to realizing the vision of "Touch Technology". The TI has an expected potential and future scope through the increasing penetration of mobile and cell platforms with the robotic-human and computer interactions in the virtual control environments, for interconnecting people, machines, appliances and processes in real time.

In order to accomplish this, we present our discussion with an overview of haptic communication via tactile internet architecture, with an emphasis on its potential development in the touch-enabled framework in 6G and IoT mobile networks.

### B. TACTILE INTERNET ARCHITECTURE

The haptic or the touch sensation helps ascertain a connection between the humans and peripheral environment in a way analogous to the auditory and visual senses. It therefore occurs bilaterally as a touch, sensed by imposing a motion or a movement in an environment by some reacting force. The haptic communication thus provides an additional dimension and advantage over the traditional audio visual communication for a real time control and accessibility in the distant and remote environment. The tactile internet architecture consists of a radio access network (RAN) and a core network (CN), both of which are expected to meet critical requirements for TI functionality realization.

Each of the three domains in the tactile E2E architecture can be separated into three sections: the master domain, the network domain, and the controlled (slave) domain. The master framework comprises of an operator, either human or machine, and an operating system interface. Using various coding techniques, this interface acts as a master robot or as a controlling device, converting the operator's input into tactile input. If the controlling device is haptic in nature, it lets human interact with objects in virtual or real environments through physical means such as touching, feeling, manipulating, or controlling.

It is primarily responsible for the controlled domain's operation. In case of a network controlled system, the master domain includes a controller that issues command signals to the sensor or actuator system. A domain, where both robotic machines and other objects at distant locations in a controlled environment, and directly accessed by the master domain via command signals is referred to as a controlled domain. When remote operations are carried out via haptic feedback, energy is transferred between the master and the controlled domains and the global control loop is completed.

As apparent from Fig.7, the components in the E2E tactile internet architecture are explained as below:

1) MASTER DOMAIN: Generally, an HSI/HMI[8] is a robot where the user may touch, feel, and move virtual and real-world items while directing the actions in the slave domain.
2) SLAVE DOMAIN:
The slave domain is controlled by a tele-operator, through different command signals from the master domain, interacting with its surroundings.
3) THE NETWORK OR CONTROLLER:
The network domain kinesthetically integrates the person with its surroundings and distant environment.
4) TACTILE SUPPORT ENGINE: Being on the edge network, it effectively offers AI capabilities that are crucial in system stabilization.
5) HAPTIC DEVICE: Enables the tactile communication, which means a user may touch, feel, and engage with virtual or real-world things.

As a result, the most common design for a tactile haptic device is a linkage-based system that consists of a robotic arm connected to a stylus and capable of applying force on its tip. Thus, the growing degree of freedom (DOF) is an essential phenomenon for the envisaged applications that integrate the network interface in a direct and indirect connection with a cellular network. Since they both include sensations rather than conventional multimedia, it is essential to differentiate between the tactile internet and the haptic communication, which are comparable to traditional multimedia like speech, data, and video.

[8] HSI/HMI: Human System Interface or Human Machine Interface

To summarize, haptic communication networks include communications across the wired and mobile internet, as well as applications that operate on the tactile internet. This means that the haptic communications and the tactile internet have a service and medium connection with each other, respectively.

## V. NEED FOR INTELLIGENCE

The existing and the future B5G/6G wireless network are expected to endow the users with an improved coverage and high data rates with a better cost efficiency, resource utilization, scalability, adaptability and security. Hence the B5G/6G wireless communication system is anticipated to be a backbone of the digital revolution in the next generation network providing a ubiquitous reliable and practically an instantaneous connectivity for the humans and machines. The Artificial Intelligence (AI) may be regarded as the 'processing and simulation of the human intelligence by machines' and therefore has a great potential in working out several intractable and unstructured problems containing large amount of data[84].

In other words AI may be defined as a science of constructing computers that are capable of performing tasks requiring human-like cognitive intellect[95]. All the while AI has therefore been a widened approach for the machines to be able to smartly carry out assigned tasks. ML on the other hand is presently the widely accepted application of AI, empowering the machines to train and learn from large datasets and perform tasks without any need for explicit programming. Next in order is deep learning (DL), a subset of ML that analyses the artificial neural networks (ANNs). These have more number of hidden layers to replicate the human brain, making it one of the most widely used ML techniques. DL therefore has a lucrative application in fields like computer vision, bioinformatics and speech recognition.

Such induced intelligence in the wireless communication network not only reduces the manual effort in network deployment, configuring and management but also helps in an improved system performance. It also asserts the adaptability and reliability of the communication network by taking robust decisions in real time according to the prediction and behavior of the users and network. Hence due to the recent advances and research in the intelligence (AI/ML), a wide range of novel technologies like self driving cars, voice assistants, holographic telepresence, e-health and wellness applications, pervasive connectivity in smart environments, industry 4.0 applications, massive robotics with the unmanned mobility in 3D, AR/VR have become possible.

6G wireless networks thus offer a broadband network, which is fast, instantaneous and safe, in order to enable mass data exchange at various frequencies using a wide range of technologies. In addition, these technologies are moving towards intelligent devices in IoT that will demand a more reliable, stable, efficient, and a secure connectivity. Hence the complex connected devices therefore require a dynamic communication network to address their inherent complexity. The future wireless networks will eventually need a self-organizing and configuring capability alongside their cooperation and coordination between the different nodes and communication layers.

It enables us to effectively meet challenges like coverage, spectrum, and energy efficiency. The continuous acceleration of the machine type communication (MTC) devices adds to the existing ultra dense network's complexity. Therefore many such applications supported by B5G network must attain a short transit time and low latency with high reliability, availability and security. The majority of these are resource constrained and unable to rely on their bounded resources and thus call for an uninterrupted and safe operation as its main concern. Consequently these applications owing to their delay and bandwidth constraints cannot be moved to the cloud or network controller[96].

Furthermore these devices generate diverse range of datasets with a large scale of erroneous or missing values. Many wearables with VR/AR, intelligent products and support systems, and other data hungry use cases have a built-in end infrastructure to afford and deliver content based services. Thus in order to incorporate an intelligent system, an intelligent and content aware approach must be implemented for the planning, design, analysis and optimization of such network. This necessitates integration of the network systems with their data sources, decision making and cyber physical infrastructures, as well as sensing and communication networks [97]–[99].

Conclusively the favorable conditions for the implementation of the intelligent learning techniques in 6G wireless networks range from:

1) Network interoperability with the distribution, affordability and accessibility of computing resources.
2) The predictive nature of the network characteristics.
3) Accessibility of a considerable amount of data.

Therefore a densely integrated wireless network may be engineered by adopting artificial intelligence (AI) principles combined with the incorporation of machine learning (ML) techniques with reasoning and decision making mechanisms. Accordingly the development of an intelligent touch based system calls for a promising development in facilitating the efficient resource distribution in the cloud, fog and edge nodes, aiming to put together system intelligence and data processing abilities in close proximity with the original data source.

### A. MACHINE LEARNING IN B5G/6G WCN

ML as a member of the quintessential AI technology has nowadays become a key component of 6G wireless communication network. ML is said to be of a plausible advantage in the communication system owing to fact that the dynamic nature of the wireless communication channels complicates the channel interference models in the B5G scenarios. Therefore ML techniques are capable of extracting information from the unknown channel by learning from the communication data while taking into account previous learning experience. Furthermore, the rapidly growing number of wireless access points necessitates a global optimization of communication resources as well as a fine-tuning of the system design[73].

However the existing approaches together with the massive amount of resources complicate the tasks concerning the optimization and correlation of the system parameters. In contrast, advanced ML techniques like as deep learning and probabilistic learning can represent highly nonlinear relationships and aid in the determination of optimal system parameters. Sequentially, ML aids in the realization and instillation of learning-based adaptive network configurations by identifying and evaluating their behavioral patterns in advance rather than reacting to unanticipated outcomes[63].

As a result, the current cellular networks that were built and managed based on the preceding premise may be unable to keep up with the growing complexity of data produced and therefore fail to provide the necessary capacity, dependability, and flexibility. Now as response, the network may be unable to respond fast enough to expected occurrences, thereby compromising real-time communication services. Because the majority of AI/ML algorithms are not purpose-built for the wireless communication networks, it is difficult to apply them directly to the B5G/6G networks.

All the above arguments call for an intelligent communication interface in the real time, facilitating a stable and efficient connectivity within the network. The 6G wireless communication system guarantees a wide range of frequency bands, including sub-6GHz, mm-wave, THz, and optical bands, while also increasing the computational complexity in the channel model. As a consequence, comprehending new channel characteristics for modeling new channel scenarios is a lengthy process.

Owing to the significantly high computational channel complexity in many situations, traditional techniques may aid in certain approximations and assumptions to help simplify the channel modeling and processing methodology. This ensures the balance between accuracy and complexity tradeoffs of both channel modeling and processing methods is beneficial.

### B. GENERALIZED WORK FLOW PROCESS FOR MACHINE LEARNING

For most prevailing and functional machine learning algorithms, the generalized work flow process in basic steps is described and shown in Fig. 8 [100]:

1) PROBLEM FORMULATION:
Since the ML training process is time consuming, it is critical that the problem be appropriately formulated at the start of the process. Moreover there should be a strong correlation between the problem and the information gathered. Classification, clustering, and decision making are three important types of machine learning. The proposed model should also be considered within these three categories to aid in the identification of learning model as well as data collection. Improper problem formulation results in an unsatisfactory learning model and performance.

2) DATA COLLECTION:
There are two types of data collection: offline and online data collection, where data collected in real-time is used as model feedback in online data collection and is also used for re-training of models. In contrast, offline data may be retrieved from the source without an Internet connection [101]. By utilizing monitoring and measurement tools, online and offline data can be collected efficiently, securely and stored for model adaptation. Data collection marks the beginning of training and learning. Validation and testing are set in motion after that.

3) DATA ANALYSIS:
It is divided into two types: preprocessing and feature extraction where preprocessing is used to reduce noise from the gathered data. The data's features are then extracted, which is a prerequisite for learning and training [102]. The types of characteristics that can be extracted from the network include packet level features and flow level features. The packet size, mean, root, and variance are extracted at the packet level and mean flow length and mean number of packet flow features extracted at the flow level.

4) MODEL CONSTRUCTION:
While iterative process selection, training, and tuning are all necessary aspects of the model selection process, they are applied differently and a suitable learning model must be chosen depending on the dataset size. The training stage entails training the model with the dataset that will be collected at the start of the stage, whereas the tuning stage have the model learn for itself by comparing it to the trained data.

5) MODEL VALIDATION:
Cross validation of the testing process is used to check the model's accuracy, which aids in optimizing the model and preserving the overall efficiency of the system.

6) DEPLOYMENT AND INFERENCE:
Throughout the deployment and inference stages, the model's trade-offs and stability are monitored to ensure accuracy and to determine the optimal sequence of steps to be taken.

TABLE V
MACHINE LEARNING TECHNIQUES AND THEIR APPLICATIONS IN B5G/6G SYSTEM.

| ML Techniques | Definition | Type | Principle | Applications in 5G /B5G communication system | | | | | | | | | | | | | | | | |
|---|---|---|---|---|---|---|---|---|---|---|---|---|---|---|---|---|---|---|---|---|
| | | | | m-MIMO | Small Cells | D2D | Hetnets | Small cell,D2D, Hetnet clustering | Spectrum sensing and allocation | Resource allocation | Anomaly/ fault detection | QoS requirement | Outage mgmt. | SINR improvement | Channel estimation/ detection | CSI | Behavioral learning | Cognitive radio | Energy harvesting | Smart grid |
| SVM[9] | Data point separation using a hyper plane or kernel functions. | SL | Classifier function: Linear /non linear. | ✓ | ✗ | ✗ | ✗ | ✗ | ✓ | ✗ | ✗ | ✓ | ✗ | ✗ | ✓ | ✓ | ✓ | ✓ | ✗ | ✗ |
| KNN[10] | Test point decision by voting of the k nearest neighbors. | SL | Non parametric lazy learning algorithm for classification and regression. | ✓ | ✗ | ✗ | ✗ | ✗ | ✓ | ✗ | ✗ | ✓ | ✗ | ✗ | ✓ | ✓ | ✓ | ✓ | ✓ | ✗ |
| K-Means Clustering | Segregation of n data points into K clusters, each associating to cluster with nearest mean. | UL | Iterative refinement with cluster allocation to the data point with least ED[11] from it. | ✗ | ✓ | ✓ | ✓ | ✓ | ✓ | ✗ | ✗ | ✗ | ✗ | ✗ | ✗ | ✗ | ✗ | ✗ | ✗ | ✗ |
| Bayesian Learning | Data points trained by GM[12], EM[13], HMM [14] models. | SL | A posteriori probability distribution. | ✓ | ✗ | ✗ | ✗ | ✗ | ✓ | ✗ | ✗ | ✗ | ✗ | ✗ | ✓ | ✗ | ✗ | ✓ | ✗ | ✗ |
| PCA[15] | Relevant information extraction from large data sets using orthogonal transformation. | UL | Data sets reduction into principal components. | ✗ | ✗ | ✓ | ✓ | ✗ | ✓ | ✗ | ✓ | ✗ | ✗ | ✗ | ✗ | ✗ | ✗ | ✗ | ✓ | ✓ |
| Q Learning | A model free RL where an agent has an access to a set of possible states and environment, with no concern of rewards or transition between them. | RL | Off policy RL to get best action for the current state. | ✗ | ✓ | ✓ | ✓ | ✗ | ✓ | ✓ | ✗ | ✓ | ✓ | ✓ | ✗ | ✗ | ✓ | ✗ | ✓ | ✓ |
| MAB[16] | Multiple agents, sequentially taking actions receive random reward to achieve a steady state. | RL | Trade off between the best action and information to achieve a larger reward in future. | ✗ | ✓ | ✓ | ✓ | ✗ | ✓ | ✓ | ✗ | ✗ | ✗ | ✗ | ✓ | ✗ | ✓ | ✗ | ✓ | ✓ |
| MDP[17] | A discrete time stochastic control state transitioning process. | RL | A single agent with partly random and partly controlled states. | ✗ | ✓ | ✓ | ✓ | ✗ | ✓ | ✓ | ✗ | ✗ | ✗ | ✗ | ✓ | ✓ | ✓ | ✓ | ✓ | ✓ |

[9] SVM: Support Vector Machine
[10] KNN: K Nearest Neighbor
[11] ED: Euclidean Distance
[12] GM: Gaussian Mixture
[13] EM: Expectation Maximization
[14] HMM: Hidden Markov Model
[15] PCA: Principal Component Analysis
[16] MAB: Multi Armed Bandits
[17] MDP: Markov Decision Process

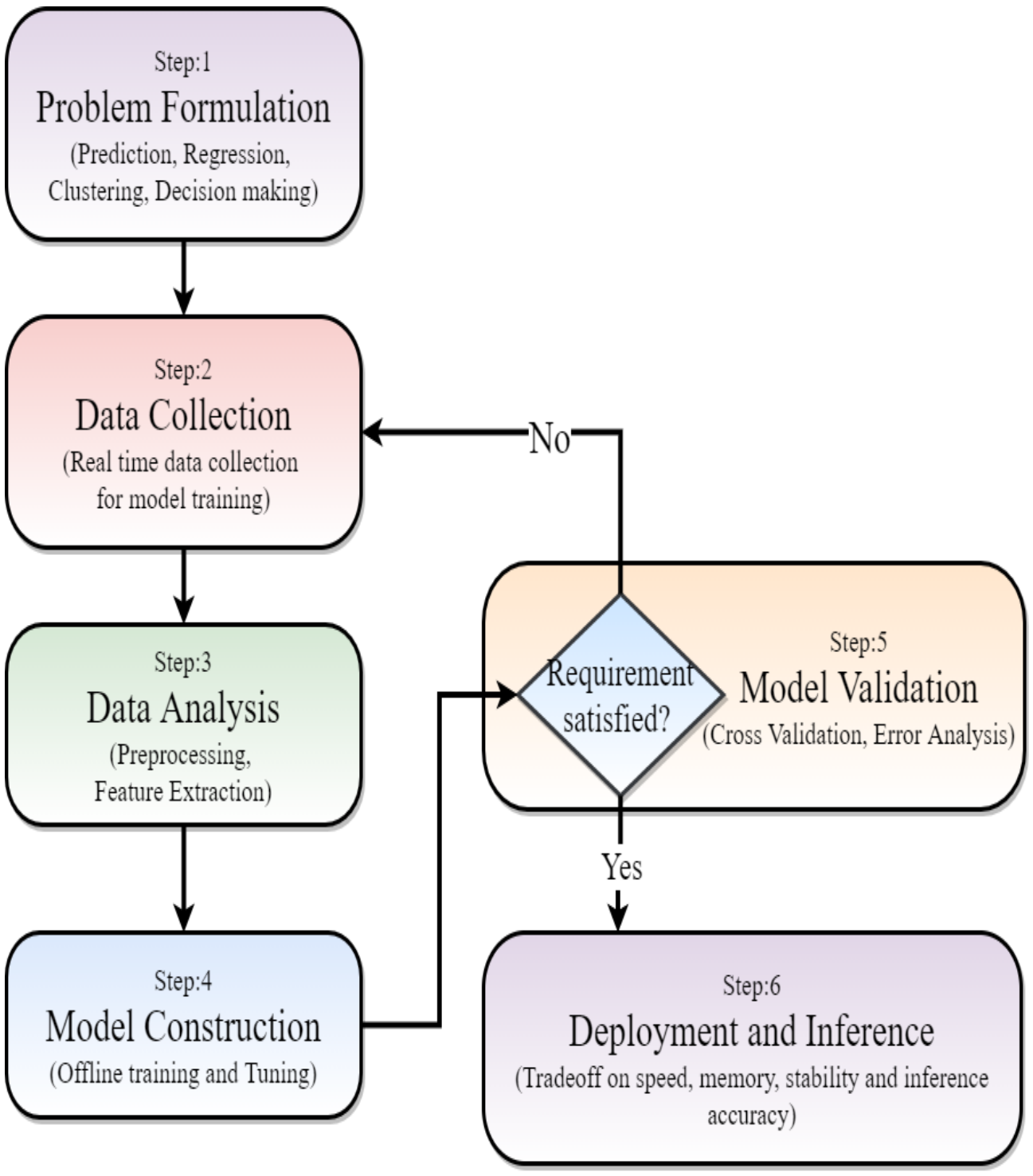

FIGURE 8: Machine learning basic workflow process[100]

### C. MACHINE LEARNING TECHNIQUES IN B5G/6G WIRELESS COMMUNICATION SYSTEM.

A few existing works containing a comprehensive overview of the existing intelligent techniques with their application in the B5G/IoT wireless communication systems are discussed here. Authors in [99], [101] talk about different learning techniques in IoT applicable scenarios while taking into consideration their input information and computational complexity. While the works in [103], [104] provide fundamental concepts about the state of art AI based technologies applied in the existing network, making the system competent enough to accomplish self configuration, self optimization and self healing of the concerned system.

Additionally authors in [58] discuss framework and application of the popular deep reinforcement learning (RL) technique that is vital in the engineering of the cognitive smart cities. Furthermore the articles in [105], [106] further provide a comprehensive survey of the existing RL algorithms and their stability and behavioral adaptability with the other learning agents in the implemented network. Speech recognition, bioinformatics, and computer vision are just a few of the applications for machine learning that make efficient use of the technology.

Machine learning is primarily used for prediction and classification, but it also plays a role in performance prediction and intrusion detection in networking systems. Therefore to make decisions directly, machine learning constructs models that can learn from data without adhering to a set of rules[107]. ML hence allows a model to enter a self-learning mode without having to be explicitly trained. To learn system characteristics that cannot be represented by an explicit mathematical model, ML models therefore are used as computing systems.

These models are employed in tasks including categorization, regression, and intelligent agent-environment interactions. Using basic arithmetic calculations, the model can efficiently complete the task once the system characteristics are learned[108]. Models may be trained by providing them with data sets, and when they are exposed to fresh data, they are able to learn, forecast, and develop. There are three types of machine learning algorithms: supervised learning, unsupervised learning, and reinforcement learning.

In supervised learning, a model is trained on labeled datasets and then learns on its own by comparing the training dataset to the predicted output. This method is commonly used for classification and regression issues. Unsupervised learning is a type of machine learning that uses an unlabeled dataset to detect patterns and relationships in the data. It is mostly used to solve clustering and association problems. During reinforcement learning, an agent interacts with a system set and learns how to map all information about action, without any training data[109].

This section provides a brief overview of some of the most commonly used artificial intelligence and machine learning techniques, with Table V illustratively defining and highlighting various machine learning techniques and their functionality in B5G and 6G wireless communication systems. The works in [110]–[113] encompass few of the researched investigations concerning the ML application on the B5G/6G wireless communication channel. However most of the existing works exercise intelligence on a limited part of the existing wireless communication channel. Some of the wireless channel characteristics influenced by intelligent learning techniques are discussed below:

1) CHANNEL MODELING USING ML:
   The ML helps deal with the channel modeling problem by implementing the model based approach, extracting the wireless channel features from the existing data. ML has an efficacy in predicting channel feature, estimating channel parameter, CIR[18] modeling, MPC[19] and classification of scenarios pertaining to the concerning channel environment, derived out from the above cited works.
2) CHANNEL MEASUREMENT:
   A channel model based on a feed forward neural network(FNN[20]) and RBFNN[21] is shown in [114] which is functional in predicting channel properties like received power, RMS delay and angle spread(DS/AS).Therefore in addition to the transmitter and the receiver coordinates, their input

[18] CIR: Channel Impulse Response
[19] MPC: Multipath Component Clustering
[20] FNN: Feed Forward Neural Network
[21] RBFNN: Radial Basis Function Neural Network

parameters also influence the distance and frequency of the Tx-Rx link.

3) NOISE :
ML in [115] makes use of ANN to remove noise from CIR, while [116] makes use of CNN to identify the relevant wireless channel. In CNN and wireless communication channels, MPC characteristics like as amplitude, latency, and Doppler frequency serve as input and output parameters.

4) CHANNEL ESTIMATION:
In order to obtain accurate channel estimation, the work in [117] uses a 2D non linear complex support vector regression (SVR) in a rapidly fading and time varying multipath channel. On the other hand, the work in [118] refers to a deep learning-based channel estimation method for beam space mm-wave massive MIMO systems that can learn the channel structure from a huge number of training datasets.

5) MASSIVE RADIO INTERFACE:
ML algorithms help in analyzing the enormous amount of data produced by the massive MIMO arrays, where the conventional channel estimation and detection algorithms are rendered incapable. Deep learning methods and techniques, particularly image processing and video analytics, provide the most exciting algorithmic approaches[119].

6) WIRELESS NETWORK LOCALIZATION:
The continuous development and updating of wireless channel locations has been made feasible via automated learning from crowd-sourced data employing a large number of mobile devices, yielding precise localization results. In exchange, it enables consumers to benefit from improved location-based services.

7) NETWORK MANAGEMENT:
Machine learning and artificial intelligence has the ability to optimize a variety of tasks such as fault detection and user tracking over a wirelessly linked network.

8) RESOURCE MANAGEMENT:
The resource management mechanism is only able to function once the system has memorized the states and conditions of the network users and their real time wireless environment. This therefore helps improve the system performance with time and in turn helps the system incorporate an intelligent and dynamic decision making phenomenon.

Following consideration of all relevant factors, we intend to incorporate intelligence into the tactile internet infrastructure in order to achieve a complete automation of the systems that surround us, taking into account the feasibility, interoperability, and functionality of the 6G and IoT-based wireless communication networks. An intelligent touch-infused technological framework is proposed in the next section, which incorporates intelligence from existing smart IoT infrastructure and interfaces it with next-generation systems by creating a tactile/haptic feeling as we interact.

## VI. PROPOSED TOUCH TECHNOLOGY IN B5G/6G AND IOT NETWORK

The proposed and futuristic anticipated system infrastructure is expected to encompass an intelligent and a reconfigurable touch enabled system that is pertinent in an IoT interfaced B5G/6G systems. The intelligent Touch-based IoT paradigm can be made up of a variety of functional elements that help smart objects perform various functions such as sensing, actuation, identification, management, and communication. The touch based IoT system's functional elements[3] can be summarized as follows:

1) SMART DEVICES:
The primary components of the IoT based Touch System, performing sensing, actuation, and control functions are capable of sharing data with other applications and servers. To connect to other smart devices, each IoT device has to be prepared with numerous interfaces including the internet access, I/O interfaces for sensors, audio and video, storage and memory interfaces.

2) FUNCTIONALITY:
The device functionality ranges from smart-watches, wearable sensors, automatic cars, industrial machines, LED lights and so on. From office automation and household appliances to manufacturing lines and commodities tracking, intelligent IoT techniques are used in a wide range of applications. As a result, IoT services must be used to improve IoT application development and accelerate installation.

3) SERVICES:
These are dedicated to identity and device modeling and are commonly grouped under the umbrella term identity services. Additional subcategories include information aggregation, discovery, control, collaborative awareness, ubiquitous services, and data analytics and publishing.

4) REMOTE ACCESS:
As opposed to devices that use mechanical switches or buttons to remotely manage, IoT devices have either no human involvement or can be remotely managed without the need for human intervention.

5) SECURITY:
Taking into consideration the security aspect, as far as the data on wireless networks is concerned, especially with regards to denial of service (DoS), spoofing, and eavesdropping, the information is vulnerable to an array of attacks. Thus, IoT systems use many security features, such as privacy, authorization, authentication, data security, content integrity, and message integrity, to attempt to thwart these attacks.

6) IOT APPLICATION:
In order to provide IoT users with interfaces, the application layer supplies IoT users with various interfaces that enable them to monitor and control the various aspects of IoT applications.

Thus the projected system is to be most likely based on the IoT functionality, and is to be expected to be implemented in the B5G/6G networks, where the system intelligence is of an utmost need so as to be compatible with the accelerating high data rate and in turn satisfy the low latency requirements of the next generation system[120], [121].

### A. INTELLIGENT TOUCH BASED SYSTEM

The problem statement here describes the need for the intelligence in B5G system, and therefore can be elaborated as per the works in [122]–[125] stating that the ever since the evolution of the wireless network from 1G to the existing 5G/B5G and subsequently the 6G networks, have led to a tremendous ascend in the billions of connected devices bound together by IoT, forming an integrated IoE network. The increased traffic, due to growing number of devices, therefore requires high energy efficiency and lower latency.

The growing new use cases in the evolving B5G network incorporates the AR/VR/XR based smart systems including the smart road monitoring, the smart cities, consisting of the smart homes and IoT governed smart appliances. These systems are externally controlled and therefore greatly lack in intelligence. Hence for the efficient functioning of these devices in the next generation 6G network, an intelligent system is required to govern the existing AR/VR based 5G/B5G network effectively.

It therefore requires an interfacing mechanism with the existing network infrastructure. To implement this in real time, the B5G/6G enabled tactile internet proves to be a promising technology, which in turn can provide a new dimension to human to machine interaction by enabling haptic sensations and therefore a touch enabled communication interface, in real time. The touch-governed IoT system is expected to permeate many facets of the contemporary daily living, including the ability to sense, process, analyze, and infer environmental parameters from natural resources and ecologies to human environments.

The main aspect of this touch-based IoT network is its ability to intelligently connect to the other existing and future networks, to all the resources that are utilized by these networks, and to help accomplish that through the advancement of the prior networks and communication protocols. For this vision to succeed, we will need to advance beyond conventional mobile computing technologies and design an IoT system in which everything we can touch is connected and capable of acting as a smart and intelligent extension of ourselves.

An in-depth knowledge of IoT issues span from an awareness and a better understanding of the IoT concerns and the complexities involving the size and depth of the pervasive communication network, software architectures, and data transfer and processing. This knowledge is used to create autonomous and intelligent devices in IoT systems. The primary goal is to set up a global network of connected smart objects and devices, all of which can connect to each other without human intervention.

Each object that has been embedded with a smart interface and connected to the user platforms and digital environments is assigned a virtual identity and interfacing, allowing it to connect and communicate with other embedded objects[3]. As we build our IoT network, our physical and virtual entities turn into virtual things in a cyber world, each with specific abilities that all IoT entities can use to realize personalization, specialization, and autonomy based on the communications protocols used to make the smart entities unique and provide them with virtual personalities.

The combination of B5G/6G network slicing technology and the TI application will thus prove to be a driving factor in the realization of this suggested reconfigurable and intelligent touch enabled framework, whose research is still in its infancy. Consequently, it can be concluded that the main goal of the IoT-based Touch technology system is to improve the lives of people, where all objects around us have the ability to figure out what we want, what we require, and what we like, as well as serve it accordingly without us having to explicitly command them.

The Fig.9 presents the layered architecture of the proposed intelligent touch technology in B5G/IoT networks. Hence it requires an AI/ML technology, coupled with the B5G/6G based Network Slicing and Tactile Internet; to implement and interface the intelligent touch based system. Hence the network slicing and tactile internet explained in the previous sections proves to be the cornerstone in the incorporation of a reconfigurable intelligent touch system.

## VII. LAYERED ARCHITECTURE FOR TOUCH TECHNOLOGY

The Internet of Things connects millions of smart objects, increasing data traffic and necessitating the use of large processors and storage systems. This emergence of IoT system together with the rising demand for the wireless capacity ultimately paves way for the futuristic intelligent Touch Technology system. Many factors, including as interoperability, scalability, QoS, and reliability, must be considered while designing such an IoT based intelligent infrastructure. As a result, the required intelligent IoT architecture based on touch technology should have the following characteristics:

1) DISTRIBUTIVENESS:
The IoT model for the proposed system should enable data collection from various sources and their processing by various smart entities in a distributed manner.

2) INTEROPERABILITY:
IoT devices from different vendors must communicate to achieve common goals.

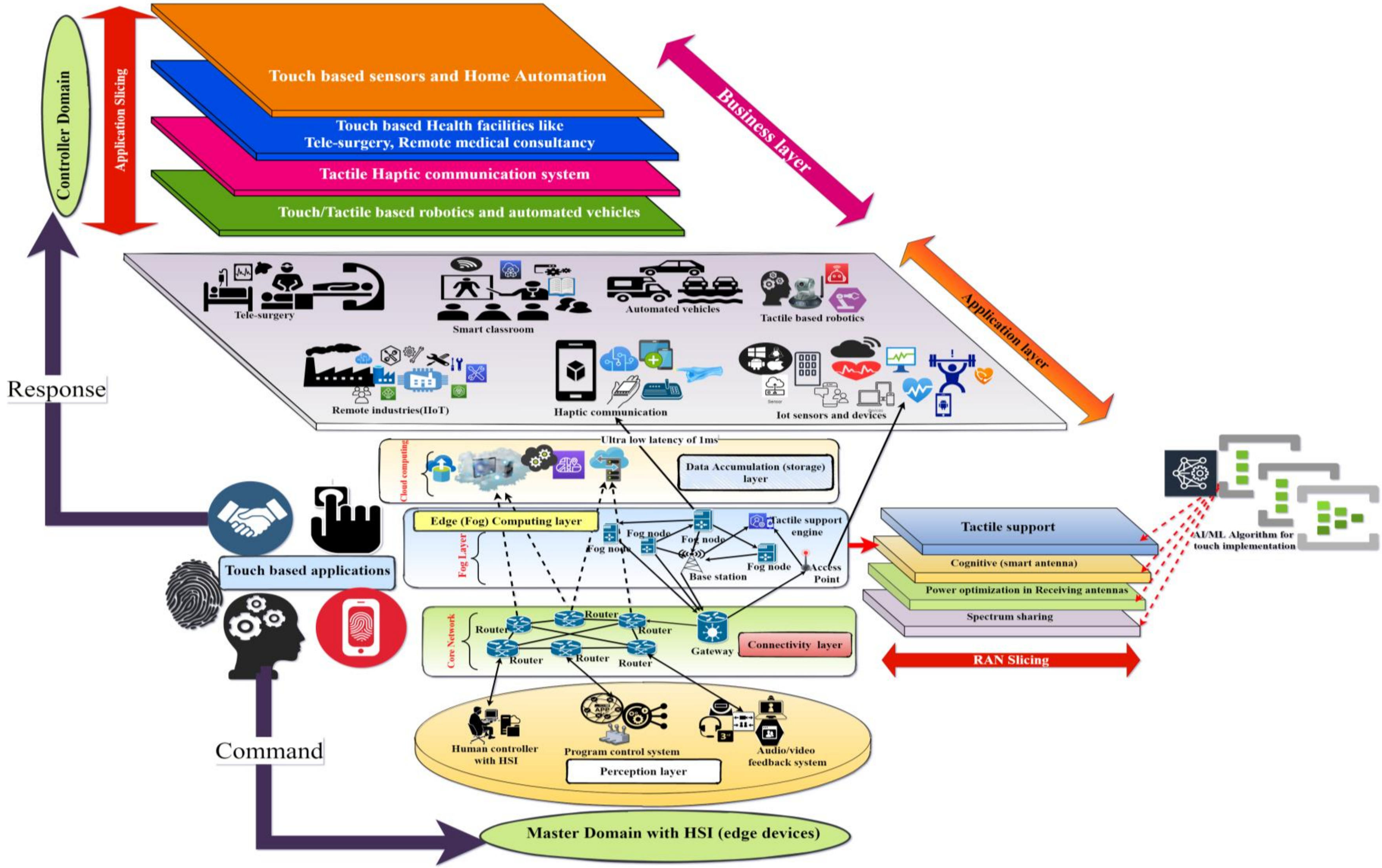

**FIGURE 9: Proposed layered architecture for intelligent touch technology**[22]

[22]The architecture presents the IoT Layered Infrastructure comprising of : Perception, Connectivity, Edge computing(network), Cloud Storage(transport) and Application layers, providing tactile-based communication and device interaction, as well as efficient network slicing between the connectivity and application levels in order to transfer and completely support requested services and user requirements.

Additionally, protocols and systems must be designed in such a way that smart devices from a variety of manufacturers can exchange sensed data in an interoperable manner.

3) SCALABILITY:
Any such IoT network is expected to have billions of objects connected at any time. Because these platforms run many applications and systems, hence such applications should be able to handle and process a huge amount of data.

4) LIMITED RESOURCES:
Both computing units and energy are considered highly rare units in regards with the limited resource availability.

5) SECURITY:
The feelings of helplessness and vulnerability that users have when they are under the control and dominance of an unknown external device could be detrimental to the given system deployment.

This section presents the proposed layered architecture for the intelligent and touch enabled technology which is to be functionally implemented in the B5G/6G and IoT configured wireless communication network. This architecture, which is defined by the tactile internet and network slicing as its cornerstone components, can thus be broadly classified into master, controller, and network domains.

The master domain is comprised of user-facing edge devices such as a robotic arm, gesture-based glove, or a gaming console for real-time online gaming, which collectively form the HSI network. Thus, this layer contains various end devices and sensors, which aid in the touch-based sensory data transmission. All of this is located on the perception layer as per the layered architecture presented before. The perception layer is responsible for transmitting the generated sensory data to its required destination, over the network.

The reliable and timely transmission of data from the perception layer devices to the edge computing layer is ensured by the efficient connectivity between the layers. Connectivity layer is therefore responsible for processing and communication between the master devices accompanied with their effective routing and switching, enabling a reliable delivery of information across the network. This layer also ensures the security of the network.

The third layer is the edge computing or the Fog layer, where the data evaluation is done and is processed at higher levels. The information is then processed and transferred by the tactile support from the end user to the base station. Thus the fog and edge network therefore facilitates a distributed computing, storage; control and communication of the network functions with reduction in the latency, system response and cloud workload of the transmitted data. It is further connected to the RAN layer which is responsible for introducing intelligence in the system through radio network.

The B5G/6G network slicing, virtualization and fragmentation takes place in this layer, the details of which are explained in the earlier sections. Hence to further avail the tactile facilities, the RAN slicing is accomplished, fragmenting it into the isolated frameworks, each of which is further meant to furnish various high data rate applications simultaneously, while maintaining the latency of the requisite ≤1ms. The routed network endows with a provision for network slicing and storage of data, enabling access to the various cloud services, for effective data storage and acquisition.

The data acquisition process, before further furnishing the requisite applications, follows the data accumulation and abstraction. Data accumulation involves the data capture and storage, to be used by the applications later. It also deals with the query based data processing. Data abstraction further virtualizes and consolidates the data at a place before its slicing and cloud storage. The setup of the signaling procedure and protocol stack along with the physical layer optimization allows the accessible devices to direct and control the power of the sensors, edge, fog, and cloud-based platforms.

The application layer further furnishes the applications due to application slicing. Here the applications required are entirely based on the TI and are intelligent and touch enabled. These may vary from the haptic communication by the haptic and IoT sensors, the automated remote industrial operation and monitoring, which is an example of the industrial revolution *4.0*. It has a great potential in the evolution of healthcare, education, entertainment as well as edutainment. Since the application layer is in charge of monitoring, controlling, and analyzing the data, it must be at the heart of all systems.

The business layer in the end provides these services at the consumer end i.e. at the controller domain at the receiver at the destination. For an easy understanding, let's go over each layer of the architecture with their associated protocols and functionality. Following the bottom up approach, the five layered IoT configured architecture of the proposed touch technology comprises of the following layers:

1) Perception layer.
2) Connectivity (Data Link) layer
3) Fog/Edge computing (Network) layer
4) Data Accumulation (Transport) layer
5) Application layer

As a result, we will go over each of these layers one by one, explaining their functionality and interoperability as we go. This section describes each layer one by one with its functionality and interoperability beginning with the application layer upto the perception layer following a top down approach [126] in Fig 9.

### *A. APPLICATION LAYER*

As the front end of the IoT architecture, the application layer is where the majority of the IoT potential will be realized. This is because the application layer provides IoT developers with the user interfaces, platforms, and tools that are required to implement IoT applications such as smart homes, intelligent transportation, smart health, and smart cities, among others. Furthermore, it is in charge of receiving the data that has been processed from the network layer.

The IoT application layer provides appropriate protocols and services needed to transmit messages at application level. When choosing a communication protocol for a certain application, many elements should be considered, including power consumption, necessary BW, transfer and connection time, delivery guarantee, data security, and packet size. A description of a few protocols that are commonly used at the application layer is provided below and summarized in the Table VI:

1) MQTT: It uses middleware and apps to enable communication between IoT devices and the network in a variety of ways, including M2M, server to server, and machine to server, and it runs on the TCP/IP protocol. Additionally, it supports communication over low-bandwidth and unreliable links and is used for publishing and subscribing to lightweight message exchanges[127].
2) XMPP: It allows for low latency communication and minimal message delivery, making it ideal for video calls, instant messaging, and chats, as well as publish/subscribe systems, gaming, and IoT applications. Because of its simplicity and adaptability, it makes it possible to communicate between heterogeneous systems [127].
3) REPRESENTATIONAL STATE TRANSFER (RESTFUL): The REST protocol is a collection of best practices, rules, and constraints for developing web services that enable data exchange and communication between various devices, as well as for developing distributed hypermedia systems with desirable properties such as modifiability and scalability. RESTful is a request-response and client-server architecture that allows clients to access server resources in IoT contexts. It is based on the HTTP protocol. Because they are lightweight and straightforward protocols, RESTful APIs are considered to be a good choice for a variety of IoT applications[128].
4) CONSTRAINED APPLICATION PROTOCOL (COAP): In IoT applications, it enables resource-constrained and unsynchronized devices to communicate while providing flow control, reliable delivery, and simple congestion control. It uses multicast and unicast requests to enable publish-subscribe communication strategy. CoAP uses UDP due to its small message size, low code footprint, and lack of TCP handshake overhead before transmission[129].
5) AMQP: It is widely used in commercial and business domains because it offers reliable and secure communication between heterogeneous devices and supports publish subscribe architecture based on an efficient and reliable messaging queue. In addition, it uses the TCP protocol for increased reliability. The message queue and exchange queue models are used to transfer data over AMQP. In the message queue model, messages are retained until they are transmitted to the receiver, whereas in the exchange queue model, messages are routed in a proper order[129].

TABLE VI:
SOME COMMONLY USED APPLICATION LAYER PROTOCOLS

| Protocol | MQTT | XMPP | REST ful | CoAP | AMQP |
|---|---|---|---|---|---|
| Standard | OASIS[23], Eclipse Foundations | IETF | IETF | IETF, Eclipse Foundation | OASIS, ISO/IEC[24] |
| TCP/UDP | TCP | TCP | TCP | UDP | TCP |
| Architecture | Publish/Subscribe | Publish/Subscribe; Request/Response | Request/Response | Publish/Subscribe | Publish/Subscribe |
| QoS: | ✓ | × | ✓ | Confirmable, Non-confirmable, Acknowledge, Reset | ✓ |
| Semantics | Connect, Disconnect, Publish, Subscribe, Unsubscribe, Close | Get, Post, Put, Set, Result | Post, Put, Delete, Cut | Post, Put. Delete, Get-CON (Confirmable), Non (non-confirmable). ACK (acknowledgement),RST(Reset) | Consume, Deliver, Publish, Get, Select, Ack, Delete, Nack, Recover, Reject, Open, Close |
| Security Features | TLS/SSL[25]<br>• For high latency and low BW networks.<br>• For resource constrained devices | TLS/SASL<br>• Decentralization by server as there is no central master server.<br>• Flexibility<br>• Open standards<br>• High network overhead | TLS/SSL<br>• Scalability<br>• Easy implementation and interaction<br>• Flexibility<br>• Unsuitable for distributed networks. | DTLS/IPSec<br>• Reliability<br>• Lost packets retransmission<br>• Multicast support<br>• Resource monitoring<br>• Low overhead | TLS/SSL, IPSec, SASL<br>• Scalability<br>• Reliability<br>• Performance security<br>• Heavy protocol as it requires memory and power resources. |

[23] OASIS: Organization for Advancement of Structured Information Standards
[24] ISO/IEC: International Organization for Standardization/ International Electrochemical Commission
[25] SSL: Secure Sockets Layer

TABLE VII
SOME COMMONLY USED DATA ACCUMULATION LAYER PROTOCOLS

| Protocol | TCP | UDP | DCCP | STCP | TLS | DTLS | RSVP |
|---|---|---|---|---|---|---|---|
| Packet size | 20-40 bytes header | 8 bytes header | 12-16 bytes header | 12 bytes header | 5 byte header | $2^{24}$-1 bytes (handshake message) | 16 bytes header |
| Packet transport form | Segment | Datagram | Datagram | Datagram | Segment | Datagram | Datagram |
| Flow control | ✓ | × | × | ✓ | × | × | ✓ |
| Congestion control | ✓ | × | ✓ | ✓ | × | × | ✓ |
| Error detection | ✓ | × | ✓ | ✓ | ✓ | ✓ | ✓ |
| Reliability | ✓ | × | × | ✓ | ✓ | ✓ | ✓ |
| Features | • Performance, robustness and network capacity improvement.<br>• Orderly data delivery between hosts.<br>• Transmission not possible if sequenced packet is not acknowledged. | • No guarantee on packet delivery<br>• High packet loss<br>• Packets may arrive out of order.<br>• Retransmission needed when data is corrupted. | • Eliminates delay in out of order waiting data packets.<br>• Supports strict, partial and unordered delivery modes.<br>• Multi-homing support<br>• Congestion control. | • Flexibility for VoIP applications<br>• Multi-homing method<br>• Minimizes DoS attacks<br>• Dynamic IP addressing | • Prevent tampering by intruders<br>• Prevent passive listening by attackers<br>• Adds more latency | • Security and reliability for handshake message transmission.<br>• Unordered message queuing<br>• Retransmission timer to reduce packet loss probability.<br>• Slower than STCP | • Data integrity.<br>• Error reporting<br>• Multicast comm. among heterogeneous devices.<br>• QoS routing<br>• Scalability issue. |

### B. DATA ACCUMULATION/STORAGE LAYER:

It interacts with the application layer to send and receive data without mistakes in the typical way. The transmitting side here is in charge of breaking down messages received from the application layer into segments and sending them to the network layer. Following that, the segmented messages received will be reassembled into messages that will be passed directly into the application layer by the receiving side of the communication channel.

This layer is responsible for ensuring the integrity and reliability of transmitted data by providing features such as packet delivery order, congestion avoidance, multiplexing, byte orientation, and data integrity. Hence known as the routing layer because it is in charge of routing data packets through the network area, where its protocols are in charge of packet ordering, error detection, and correction. This section provides an overview of a few protocols that are commonly used at the data accumulation layer. The protocols are summarized in the Table VII:

1) TRANSPORT CONTROL PROTOCOL (TCP): It is a heavyweight, connection-oriented protocol, where the connection is not established until all the required data has been sent between each end device. TCP is suited for reliable communication since it requires acknowledgement messages to ensure each sending and receiving procedure, as well as support for retransmission of lost or corrupted packets and a flow control mechanism. Hence, this protocol's packet overhead is extremely high, resulting in increased power consumption from devices and thus making it impossible to operate on power-constrained devices. TCP divides the data packet into multiple packets, each with an ordering number and source and destination IPs[130].
2) USER DATAGRAM PROTOCOL (UDP): It is a connectionless protocol that attempts to give protocols and applications that run over IP, an unreliable, minimum message queuing, message passing, and best effort transport. There is no requirement for end-to-end connectivity between communicating entities, which enables efficient communication for some applications that require real-time performance with low latency, such as video and voice. UDP does not provide a port with any attribute for addressing the source and destination functions and provides a data integrity check [130].
3) DATAGRAM CONGESTION CONTROL PROTOCOL (DCCP): It establishes unicast bidirectional connections for datagrams with unreliable dynamic congestion control. These characteristics make DCCP ideal for applications that transmit large amounts of data and require a trade-off between reliability and timeliness, such as VoIP and media streaming. Due to its unreliability and absence of a receiving window, the flow rate of DCCP can be progressively increased[130].
4) STREAM CONTROL TRANSMISSION PROTOCOL (SCTP): It is a connectionless, message-oriented, IP transport layer protocol

similar to UDP that enables SCTP-based peer-to-peer (P2P) communication and reliable transmission for applications communicating over an IP network. As a result, it inherits the majority of TCP's functionality, such as packet recovery and congestion management[130].

5) TRANSPORT LAYER SECURITY (TLS): It was created to provide security channels among communicating peers and to give authentication, data secrecy, data integrity, and encryption to applications by preventing eavesdropping, message forging, and tampering. It runs on top of several transport layer protocols. It is composed of two components: the handshaking protocol, which is responsible for authenticating communication ends, agreeing on shared keys, and negotiating cryptographic parameters and modes; and the record protocol, which divides the traffic into multiple records and protects them using the traffic keys[131].
6) DATAGRAM TRANSPORT LAYER SECURITY (DTLS): It was created to secure datagram applications that do not require or provide dependable data delivery, such as datagram online gaming, internet telephony, and media streaming, which are deemed delay sensitive. DTLS is an enhancement of the TLS protocol that prevents message forgery, tampering, and eavesdropping when transmitting data streams. Therefore, it should be able to cope with and resolve a variety of datagram difficulties, such as packet loss, packet reordering, and delay[132].
7) RESOURCE RESERVATION PROTOCOL (RSVP): It is a multicast and unicast control transmission protocol that was created to enable data stream transmission with a flexible, robust, scalable, and heterogeneous resource reservation setup at each router. There is also support for resource reservations in each node along the data path, multipoint to multipoint communication paradigm, cache management routers, and receiver initiated reservation[133].

### C. EDGE/FOG COMPUTING LAYER:

It is responsible for providing data with routing paths so that the packets can be transmitted across the network area. This layer creates logical connections, sends out error messages, and maintains the data transmission routing path. Furthermore, this layer contains all network devices such as switches, firewalls, bridges, and routers that are required to work with appropriate communication protocols such as 3G-4G-5G, Wi-Fi, infrared, ZigBee, and Fibre to the X.

This layer is in charge of forming, addressing, and routing data packets, as it receives datagram packets from the transport layer and converts them to data packets before transmitting them to the destination side. This section gives an overview of some protocols commonly used in the edge/fog computing layer with the protocols summarized in Table VIII:

1) ROUTING PROTOCOL FOR LOW POWER AND LOSSY NETWORK (RPL): It is a tree-based, IPv6 proactive distance vector routing protocol developed by the routing-over-low-power and lossy networks working group to run commercial appliance networks with insecure connectivity, poor data rates, and substantial losses. It has a storing and non-storing mode to reduce memory requirements and eliminate loops in low-resource applications. It is prone to high packet loss owing to congestion, has a long delay, and is vulnerable to assaults since it lacks end-to-end encryption. Overhead packets for control are flooded into the networks as a result[134].
2) COGNITIVE ROUTING PROTOCOL FOR LOW POWER NETWORK (CORPL): It is an extension of the RPL protocol, which was designed to be compatible with cognitive networks in order to improve performance. This feature ensures high packet delivery ratio, and keeps the network from colliding. The CORPL routing mechanism takes advantage of an opportunity to select the most effective forwarding next hop from a pool of eligible neighboring nodes. There are minimum collisions and duplication of data packets[135].
3) CHANNEL AWARE ROUTING PROTOCOL (CARP): It is a distributed protocol with a lightweight data package that was created for underwater and IoT applications. Network initialization and data forwarding are two steps in the routing operation performed by CARP. The receiving node updates its distance to the sink node, broadcasting the welcome messages with their ID and hop count. In data transmission, the sender sends a ping message to its neighbors, determining the best relaying node based on the link quality and information received from their ping messages, and then forwards data. When selecting a relaying node, residual energy, network quality, and buffer space are all taken into account[136].
4) COLLECTION TREE PROTOCOL (CTP): A tree-based routing system that was created to give the greatest effort for any cast communication in networks with low energy demands. An early form of networking is where a source node (sink node) announces itself as the root node, where minimal cost is paid to deliver data to the root. Other nodes connect to the root tree via lightning ads and then send their collected data into the sink node with the minimum amount available. CTP, on the other hand, does not permit reverse routing from the sink to the sensors[137].
5) LIGHTWEIGHT ON-DEMAND AD HOC DISTANCE VECTOR ROUTING PROTOCOL-NEXT GENERATION (LOAD-NG): When compared to the on demand distance vector (AoDV) protocol, it is a more lightweight distance vector and reactive protocol that is designed to provide a secure, scalable, and efficient routing in lossy and low power networks. As a reactive protocol, LOAD-ng generates on-demand route requests to discover a path to the target node and when data has to be delivered, the receiving unicast replies hop by hop from the destination node back to the sender node. If a route is found to be broken, attempts to fix it are made, or an error message is sent to the requested node[138].

6) AD-HOC ON DEMAND MULTIPATH VECTOR FOR IOT (AOMDV-IOT): It seeks to discover and establish connections between nodes and the internet. For each node, AoMDV-IoT generates two routing tables: an internet connecting table (ICT) and a routing table. In addition, it converts IP addresses into internet linking addresses (ILA) and when a node wants to be connected to the internet, the IP associated with the desired internet is connected to ILA so that the search function can be utilized. If there is no internet node in ICT, the source node will broadcast the requested packet to update both tables until an optimal route to an internet node is discovered[139].

### D. CONNECTIVITY LAYER:

The IoT connectivity layer in the touch system architecture is comprised of a variety of communication protocols that are primarily responsible for providing services to the network layer. Hence, it is in charge of connecting and transmitting signals from end devices to higher layers via routers and gateways.

The IoT connectivity layer consists of a variety of communication protocols that, depending on the transmission range and coverage area, provide services to the network layer. Some of the most commonly used protocols are reviewed further down this section.

1) NFC PROTOCOL: A short-range protocol that allows mobile objects to communicate with one another over a few cm of distance and allows data to be transferred in seconds between the connected NFC devices that are in close proximity to one another. It is RFID-based and thus uses an alternate magnetic field to connect devices that are either active or passive. In active mode, all of the connecting devices generate a magnetic field, whereas in passive mode, one device generates a magnetic field while the others use load modulation to transmit data. Passive mode is energy saving and is widely used in today's smart phones[140], [141].

TABLE VIII:
SOME COMMONLY USED EDGE/FOG COMPUTING LAYER PROTOCOLS

| Protocol | RPL | CORPL | CARP | CTP | LOAD-ng | AOMDV-IoT |
|---|---|---|---|---|---|---|
| Network Topology | Mesh, Hierarchical | Cognitive M2M networks, Mesh | - | Tree-based topology, Mesh | Grid | Dynamic IoT network |
| Routing metrics | Connectivity, Link quality, BW, Reliability. | Reliability, Collision risk | End to end packet latency, energy consumption per bit, buffer spaces, packet delivery ratio. | ETX[26] of neighbors | Hop-count | Lifetime Hop count |
| Network metrics | MP2P,P2MP[27] communication | P2P, MP2P, P2MP | MP2P, P2MP,P2P | MP2P,P2MP | P2P | P2P,P2MP |
| Data rates | Low data rates | Low data rates | Low data rates | Low traffic rates | - | - |
| Latency | High Latency | Delay of sensitive applications | Low Latency | High Latency | High Latency | Low Latency |
| Algorithm | Distance vector | Distance vector | Link state | Distance vector | Distance vector | Distance vector |
| Scalability | ✓ | ✓ | ✓ | ✓ | ✓ | ✓ |
| Security | × | × | × | × | Uses integrity check values. | × |
| Network mobility | × | × | ✓ | ✓ | ✓ | ✓ |
| Applications | • Home automation. • Industrial automation • Building automation • Smart grid and Smart cities. | • Smart Grid. | • Underwater WSNs application | • Commercial products • Industrial WSNs • Teaching • Research | • Home applications • Industrial application | • Mobile IoT applications |

[26] ETX: Expected Transmitted Count
[27] MP2P: Multipoint to Point Communication ;P2MP: Point to Multipoint Communication

TABLE IX:
SOME COMMONLY USED SHORT RANGED CONNECTIVITY LAYER PROTOCOLS.

| Protocol | NFC | 6LowPAN | BLE | Zigbee | Z-Wave |
|---|---|---|---|---|---|
| Network type | P2P | Star, Mesh | Star | Star, Tree, Mesh, Cluster, Hybrid | Mesh |
| Frequency band | 13.56MHz | 2.4GHz | (2.402-2.481) GHz | 2.4GHz, 915MHz, 868MHz | 868(Europe), 908(USA), 900ISM |
| Transmission range | 10cm | (10-100)m | 100m | (10-100)m | 30m |
| Data rate | 106Kbps-424Kbps | (20,40,60 )kbps | 125kbps (12 Mbps) | 250Kbps | (9.6, 40, 200)kbps |
| No. of nodes | 2 | 65000 | 65535 | 65000 | 232 |
| Power consumption | 15mA | - | 15mA | 30mA | 5mW |
| Routing protocols | NFC possesses routing features | RPL, AoDV | RPL, 6LoWPAN | Zigbee, RPL, AoDV, Zigbee network routing (ZBR) | AoDV, DSR |
| Applications | • P2P data transfer<br>• Payment and ticketing applications | • Smart home<br>• Smart agriculture<br>• Industrial IoT<br>• Healthcare applications | • Mobile phones<br>• Smart homes<br>• Wearables and PC<br>• Security and privacy<br>• Healthcare<br>• Sports and fitness etc | • Smart home<br>• Medical monitoring<br>• AI sensors<br>• Consumer electronics. | • Home automation<br>• Smart lighting |

2) 6LOWPAN: It allows smart devices to connect to the internet via the IPv6 protocol while also taking into account the nature of wireless IoT networks by creating a very small header message format. It also removes obstacles to using IPv6 addressing protocol in IoT devices with limited processing power, data rate, and power[142]–[144].

3) BLUETOOTH LOW ENERGY (BLE) PROTOCOL: BLE is a low-power alternative to short-range wireless communication developed by the Bluetooth Special Interest Group. Additionally, it allows for fast data packets to be transmitted at speeds up to 2Mbps in the ISM band[145].

4) ZIGBEE: The objective is to develop a low-cost, scalable and power-sipping wireless connectivity that is suitable for a wide range of controlling and monitoring purposes. With intelligent routing and setup procedures, this protocol builds on IEEE *802.15.4's* features to enable high failure resilience and easy installation. It also works well with other wireless communication technologies due to its strict security and listening techniques[146].

5) Z-WAVE: Smart light controllers and other sensors in home devices use this low-power wireless communication technology. With low latency transmissions and data rates of up to *200kbps*, this technology operates over *900 MHz* ISM bands[147], [148].

6) LOW POWER WIDE AREA NETWORKS (LPWAN) PROTOCOLS: LPWAN protocols are low-power, low-bandwidth, and low-cost protocols that are particularly useful for long-distance communications. Furthermore, the devices that implement these protocols have transmission ranges ranging from *1m to 50 km.* The general characteristics of LPWAN protocols are as follows, followed by a brief discussion of each protocol:

   a. These LPWAN protocols are implemented by low-power devices.
   b. These protocols are limited to the transmission of small smart packets, typically *100 bytes* or less.
   c. Devices that implement LPWAN protocols are composed of extremely low-cost components, typically costing less than a few dollars.
   d. Within and outside their domains, these devices are designed to provide good coverage and reliability.

7) LONG RANGE WIDE AREA NETWORK (LORAWAN): It is a physical layer communication protocol that uses very less power and has a battery life of up to ten years. LoRaWAN specializes in M2M, industrial and smart city applications, which require long-range communication, ranging from *2 to 5 km* in urban areas and up to *15km* in suburban areas. The process of communication through large networks, which contain billions of intelligent devices, also promotes the data rates of this protocol in the complete duplex wireless medium, from *0.3 to 50Kbps*[149].

TABLE X:
COMPARISON OF LPWAN PROTOCOLS

| Protocols | LoRaWAN | SigFox | NB-IoT |
|---|---|---|---|
| Topology | Star of stars, Mesh | Star | Star |
| Modulation | CSS[28] | BPSK | QPSK |
| Frequency | Unlicensed ISM bands: 868 MHz (Europe), 915 MHz (N. America), 433 MHz (Asia). | Unlicensed ISM bands: 868 MHz (Europe), 915MHz (N. America), 433 MHz (Asia) | Licensed LTE frequency bands |
| Transmission range | 5 km (urban), 20 km (rural) | 10 km (urban), 40 km (rural) | 1 km (urban),10 km (rural) |
| Data rate | 250bps-50kbps | 100bps | 200kbps |
| No. of nodes | 1000 | 100 | 5500 |
| Power/ Current consumption | 50mW | (3-50)µA | 500mW-4W/ (19-49)mA |
| Handover | No end devices will be associated with a single base station. | No end devices will be associated with a single base station. | End devices are associated with a single base station. |
| Applications | • Smart city<br>• Smart logistical and transportation<br>• Industrial application<br>• Real time monitoring<br>• Video surveillance | • Smart farming<br>• Status monitoring<br>• Smart building<br>• Asset tracking and logistics | • Electric metering<br>• Manufacturing<br>• Automation<br>• Smart city |

8) NB-IOT: It is a narrowband radio technology that was developed and standardized by the 3GPP to support IoT applications with low data rates and high complexity. It proposes a new radio access method based on LTE standards but with less capabilities in order to lower the power consumption of IoT devices with limited resources[150].

9) SIGFOX: It is a technology of narrowband and ultra narrowband for connecting a large number of power-controlled devices. In order to operate, the protocol must operate on a frequency band of *868MHz*, where the spectrum is split into *400* channels of *100Hz*. Rural areas can receive signals from IoT devices that can transmit up to *140* packets per day at a data rate of up to *100bps*, and urban areas can receive signals that can reach distances of *(30-50)km* in rural areas and *(3-10)km* in urban areas[149].

The Table X compares the characteristics of the LPWAN protocols discussed in this section[149]:

### E. PERCEPTION LAYER

The major goal of this layer is to feel the physical characteristics of the entities that surround us and within the dominating IoT network, which relies on sensing technologies like RFID, WSN, and GPS. It's also in charge of translating sensory data into digital signals that may be transmitted via a network. Indeed, embedded intelligence and nanotechnology play an important role in this layer, as they improve the processing capabilities of any object by inserting small chips or microcontrollers into everyday smart devices. This layer consists of all user end devices (smart devices, wearables, sensors, actuators etc.) that are connected to the IoT and cellular network and capable of accessing and transmitting tactile sensations over the network.

Additionally, there are some fundamental attributes that are an integral part of the end devices, as detailed in the Table XI [151]. For the system to enable an intelligent and reconfigurable touch based system, both the master domain and the controller domain, need to have a bilateral connectivity with the intelligent RAN network. It is further governed by the AI/ML algorithms, responsible for intelligent resource allocation and data transmission. Hence motivation for the proposed model lies entirely in the fact that to incorporate an entirely intelligent system in the 6G and IoT framework requires the AI/ML technology implemented on the B5G network slicing and its application, TI as the building blocks of the touch system.

TABLE XI:
FUNDAMENTAL ATTRIBUTES OF CONNECTED USER END DEVICES.

| S.No | Attributes | Description |
|---|---|---|
| 1. | Identification (ID) | Each of the connected objects is assigned an ID based on conventional parameters such as universal product code, Media Access Control (MAC-ID), and IPv6 ID. |
| 2. | Meta information | It contains device model, ID, revision, hardware, serial number, and manufacture data for each IoT device. |
| 3. | Security controls | Allows the device owner to employ security settings to limit the types of devices that can connect to the user's device. |
| 4. | Relationship management | It enables each IoT device to establish, update, and terminate relationships with other devices. |
| 5. | Service composition | This component allows smart objects to interact, with the goal of providing the optimum integrated services to users and is also in charge of processing the data obtained from different objects to provide user with best solution. |

[28] CSS: Chirp Spread Spectrum

One of the most promising aspects of this proposal is that it is based on the assumption that 6G applications will have a high data rate and latency requirement of less than *1ms* in the majority of cases. The application of the TI based intelligent systems are still in the nascent stage and therefore require a well defined technology to deploy the touch enabled technology in the next generation networks, combining the network slicing with the TI.

Thus an IoT enabled interface acts as a backbone of the proposed reconfigurable and intelligent tactile based touch communication wireless system. Summarizing the important requisites for the proposed touch based technology can therefore be listed below as:

1) An interfacing architecture with the existing B5G/IoT framework is required that can link the present B5G network with the next generation 6G network through IoT based intelligent and sensory devices and sensors.
2) The network slicing and TI prove to be a strong backbone in implementing the given framework. Also the intelligence may be incorporated in the system using the ML/AI based algorithms, to increase the system computational capacity.
3) The 6G applications are largely based on the AR/VR/MR/XR, having large power consumption and real time simulation and therefore need proper energy efficiency and power optimization techniques for its effective implementation. It can therefore prove to be an important future aspect that needs to be considered.
4) Also the minimum infrastructure costs, with an energy efficient performance are important parameters that need to be considered while implementing the proposed system.

Hence it can be concluded by saying that, all the above enlisted parameters, will eventually pave a way towards the establishment of intelligent and a configurable touch enabled system, interfacing with the B5G/6G Wireless Communication Network. We have further proposed an end to end touch interfacing architecture to be implemented in the real time B5G/IoT based wireless communication network as described ahead in this section.

The IoT infrastructure includes physical objects integrated in the WCN, which are designed to provide intelligent in-house service to users. Thus the IoT system comprises of five layers as the perception, connectivity, network, transport and the application layers starting from the device end. The network layer in turn includes the connectivity with the edge/fog computing as well, as the data storage functionality within itself as given in Fig 9, altogether forming the middleware component of the system.

The proposed Touch based intelligent and IoT configured system may therefore be realized as an end to end architecture in the Fig.10. This end to end architecture is therefore to be realized in three phases which include:

1) The TI enabled real time touch communication network establishments at the transmitting end complete with its bilateral feedback system.
2) Similar TI enabled real time touch reception system at the receiver end that is operated by the remote industrial machinery and robotic system.
3) Lastly it requires an intelligent interfacing algorithm to simultaneously operate both the ends and a suitable slicing mechanism to fulfill almost all the URLLC and IoT enabled intelligent Touch based tactile application.

## VIII. END TO END TOUCH SYSTEM ARCHITECTURAL MODEL

The E2E system architectural model of the proposed intelligent touch technology has been represented in Fig.10. The real time applications at the user ends satisfying the desired output at the destination may range from AR/VR based transmission and communication system, real time online gaming, remote medical service accessibility like tele-surgery and remote health consultation and remote industrial operation involving the major role of robots and automatic machinery. The system model has been divided into three components and therefore may be analyzed in three phases. These being categorized as:

1) PHASE1: The bilateral communication at the transmitter comprising of the tactile based user devices like the robotic arm, AR/VR gear and gesture based haptic devices, real time online gaming console and many end, more, interacting with the BS, the local server (LS), transmitting the data to the middleware through the gateway.
2) PHASE2: It represents the bilateral communication at the receiver end, accomplishing the intelligent, touch based applications ranging from intelligent healthcare facilities like remote surgery, remote industrial operations, real time online classrooms and many more.
3) PHASE3: It acts as the mediating controller, connecting the above phases and is therefore responsible for implementing intelligence in the proposed system. The AI/ML algorithms are therefore routed in this phase. The aforementioned phases may be separately analyzed as the transmitter end, the receiver end and the processing end.

### A. TOUCH BASED TRANSMITTER SYSTEM

The transmitter end of this end to end architecture of the intelligent touch interfaced system may be represented as phase 1, comprising of the tactile user devices like the robotic arm, augmented or virtual reality gear, gesture operated tactile devices, real time online gaming modules. The Fig.11 presents the touch enabled flow process at the transmitter end. The process may begin with the abovementioned devices at the user end requesting for diverse high data rate and IoT based URLLC applications varying from the remote medical consultancy or tele-surgical operation to the assembling and modeling of the machinery pertaining to the remote industrial function. The user may request for a particular service by the means of a touch sensor or an AR/VR device, virtually furnishing the data towards the server, to be routed further.

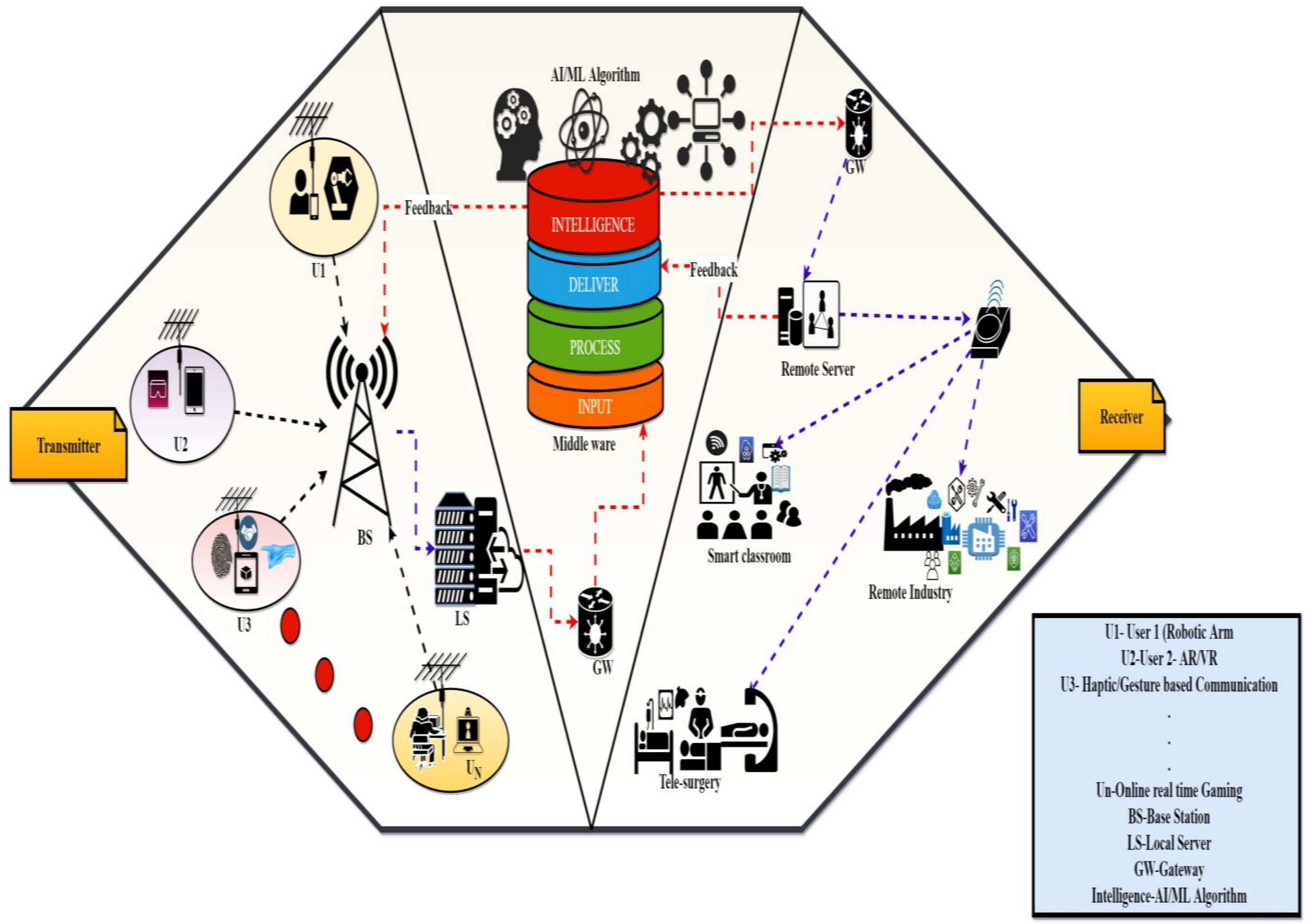

**FIGURE 10: Touch interfaced end to end architecture.**

The further the convenience may be at the individual viewpoint, making use of tactile enabled robotic system for household function. In addition it may have a future application as in virtual e-commerce or virtualized or holographic shopping where the customer may virtually access or try the product by tactile sensing and touch enabled interface between the user end and the linking server forming a bilateral communication feedback system at the user/customer end.

The TI based applications place a request to the local server in order to check for the channel availability. The request is forwarded when the channel is available and if the channel appears to be busy, the entire process starts all over again from the first step, where the user must ask for a service. Following the channel access granted to the user, and prior to sending the data it tries to establish connection with the server. This is followed by the handshaking request-response procedure between the channel and server.

The server gives acknowledgement on the channel availability and thus firmly establishes the connection between the customer and the server, virtually accessing, forwarding and routing the data further to the gateway. Here ‘A’ represents the data to be forwarded to the gateway. A similar kind of procedure is conceded at the receiver end explained in the next section.

This middleware is to be positioned between the edge and the cloud computing layers, successfully participating in the network slicing and providing the requisite virtual platform for supporting different tactile applications. The slices may be controlled, scheduled and allocated using different ML/DL based algorithms, successfully implementing the proposed touch based network.

### B. TOUCH BASED RECEIVER SYSTEM

The flow process at the receiver end of the touch interfaced system is presented in Fig.12. The point ‘A’ which represents the user end devices at the transmitter end, further routes the information signal in the channel via BS, router and gateway. The gateway connects the transmitter with the middleware. The process therefore briefly illustrates the role of the middleware infrastructure with its connectivity at both the transmitter and receiver end devices.

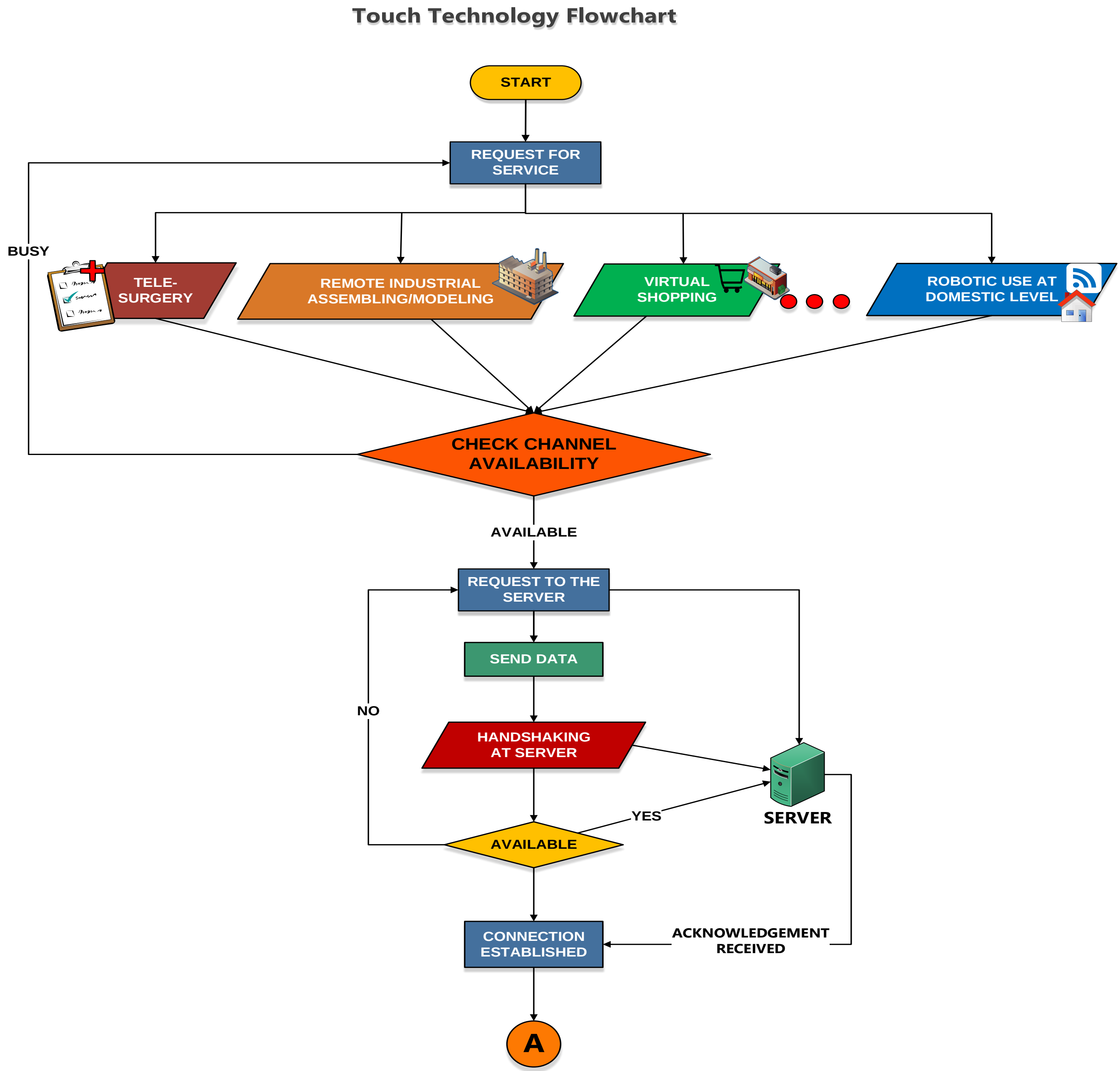

**FIGURE 11: Touch enabled flow process at the transmitter end.**

This middleware exists at the junction of edge and cloud computing layer. Middleware performs the virtualization process in conjunction with network slice abstraction and allocation, fragmenting the existing physical network into multiple virtual networks. The techniques and algorithms required for the proficient implementation of the slices bearing n number of virtual networks may perhaps be applied in the middleware layer. The middleware further connects to the LS on the receiver side and establishes connection.

The process similar to that in the transmitter and is followed with the middleware forwarding the information data to the server after establishing connection with the server via handshaking mode. The LS at the receiver end in turn indicates whether or not the channel is available for signal transmission by means of request and acknowledgement process. The signal is processed as the output after the channel availability is declared at the LS. The receiving end of the proposed system may perhaps be represented by point 'B'.

This point 'B' may further endow the requested TI enabled touch based applications at the destination end. Hence the user/customer successfully gets the desired output. It may be the successful tele-surgical operation or it may the remote industrial operation. It may also consist of the successful modeling and assembling of robotic and machinery components. Additional applications vary from the smart and virtual classroom to the virtual holographic shopping and e-commerce, with successfully accessing the desired product without actually having to lay a hand on it. It is possible to virtually access and analyze the product through holographic imagery and the transmission and reception of the information may be established by implementing the TI based touch configured system, which will prove to be an impacting factor in 6G system.

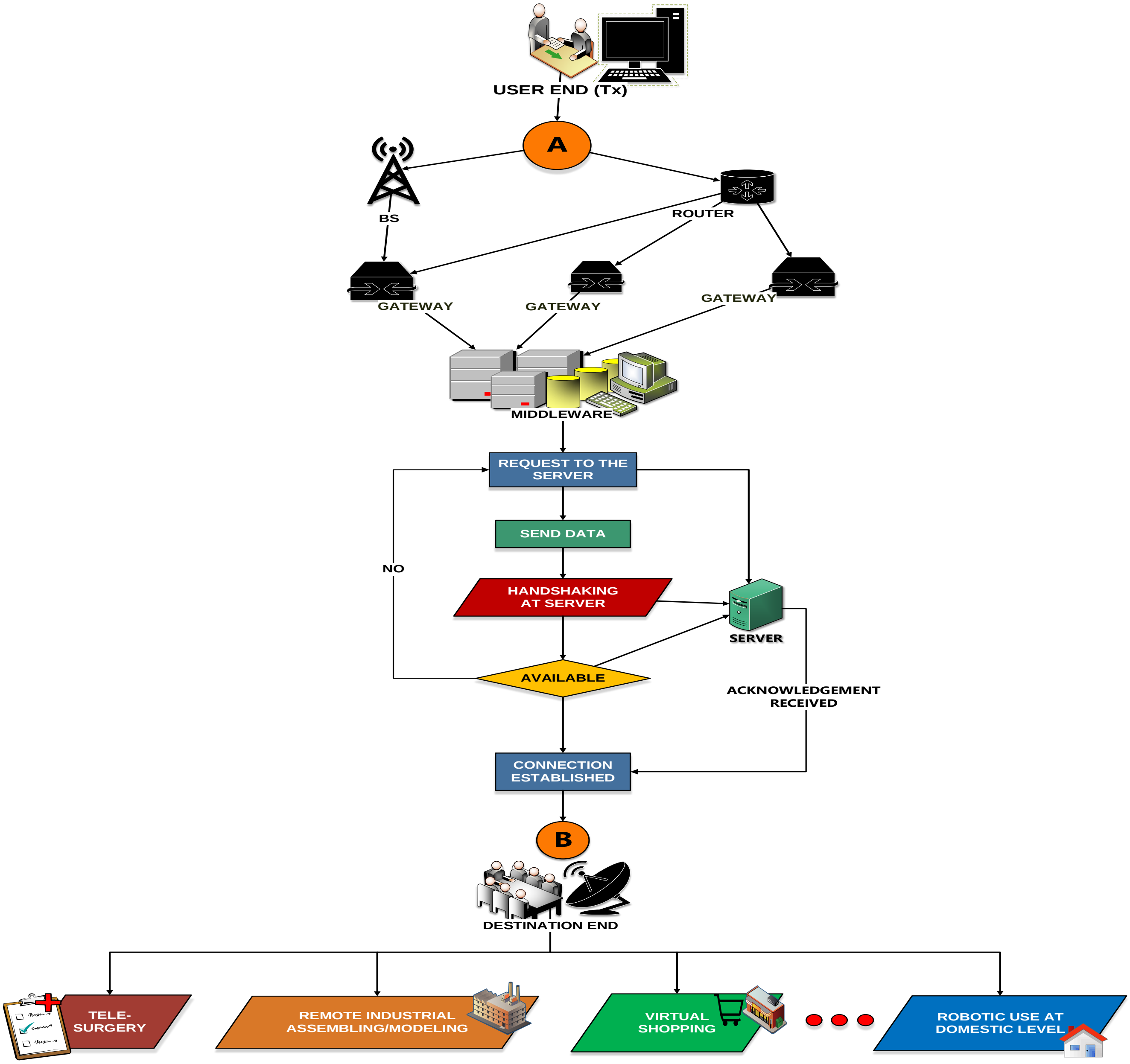

**FIGURE 12: Touch enabled flow process at the destination end.**

### C. TOUCH BASED MIDDLEWARE SYSTEM

The middleware system is an integral and the most important component of the proposed intelligent touch configured system. It therefore helps in integrating and configuring both the transmitter and the receiving user/application ends. Hence it is imperative that both its connecting ends have the same configuration and dimensionality for the system to act smoothly and without any undesired latency. The LS at the user end routes the requested information via BS, router and gateway to the middleware infrastructure. At this juncture the information requested is virtualized in numerous network functions enclosed in several network slices.

This system too comprises of a layered infrastructure, and we have considered a 4-layered centralized middleware network in Fig.13. These layers may be categorized as: input layer, processing layer, delivery layer and the output service layer. The input layer forwards the input data from the transmitter end; the processing layer helps analyze the data from the transmitter end; the processing layer helps analyze the data using the virtualization and network slicing process. The delivery layer categorizes the slices as per their use cases and data rate and bandwidth requirements, whereas the output service layer arranges the final application slices to be delivered at the destination end.

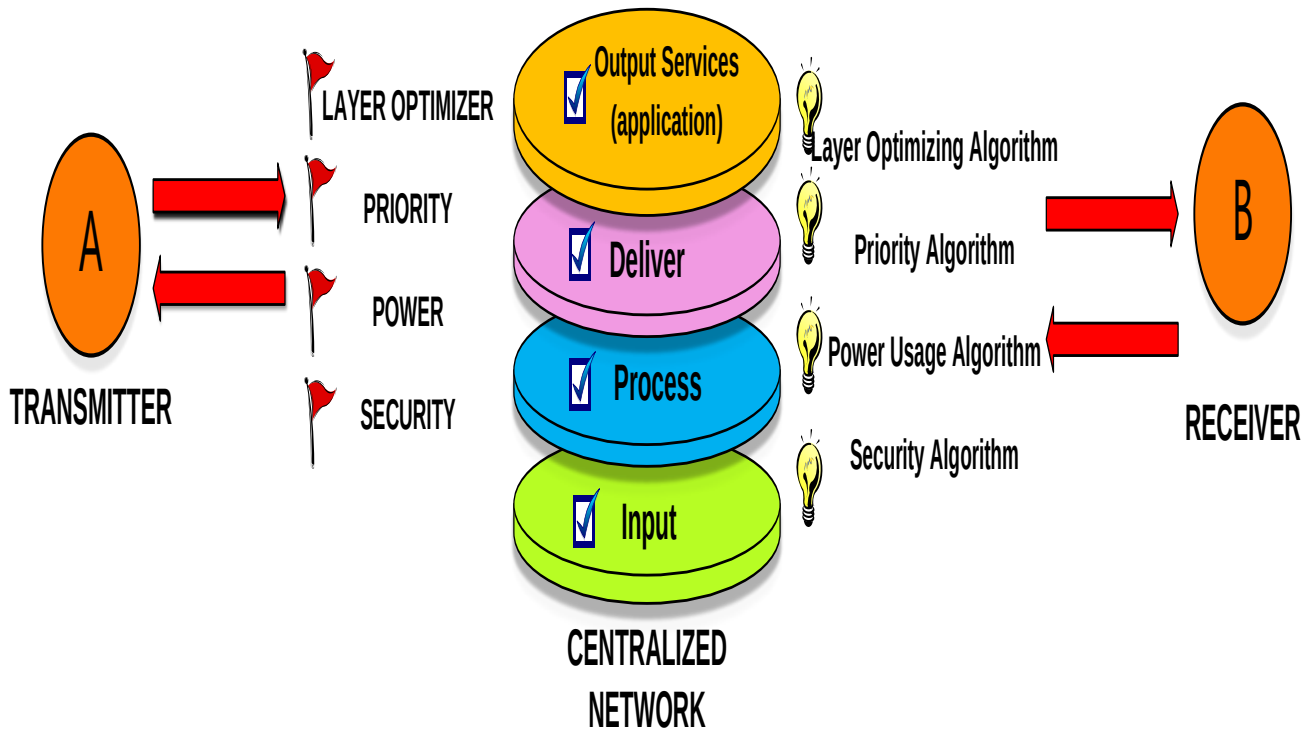

**FIGURE 13: Middleware component system**

Therefore the proposed requisite intelligence in the system is to be established via this middleware infrastructure. There is a bilateral feedback system at both the transmitter and receiver end which helps in transmitting the feedback and incorporating the intelligent techniques/algorithms at both the connecting ends. The input to this system is all the way through the transmitter user end represented by 'A'. Consequently the input data fed to the middleware may be arranged in a sequence considered and governed by factors like optimization, priority, power consumption and security.

The next subsection further elaborates the architecture and functioning of the middleware to be used in the proposed touch based system. Hence at the output end these factors may be implemented as their corresponding techniques and algorithms like the layer optimizing algorithm, the priority based algorithm, the power usage algorithm and the security based algorithm, so as to obtain the desired output result. The AI/ML/DL algorithmic implementation of the above listed factors, in wireless IoT and B5G/6G systems has been a research interest of many scientists and researchers.

Thus for the real time implementation of the intelligent touch enabled system we call for an interfacing architecture that integrates the existing B5G technology with the 6G technological interface, enabling the TI enabled touch connectivity through intelligent network slicing. Consequently the subsequent sections describe the interfacing architecture of the present scenario with the next generation system along with its future scope and applications towards the implementation of the intelligent touch system.

## IX. TOUCH TECHNOLOGY IOT MIDDLEWARE

The IoT is a vast network of connected smart devices that aim to make the surrounding environment intelligent and autonomous. Thus most vendors do not care about compatibility of their products with other competitive brands, which is one of the major challenges that IoT paradigms face in machine to machine communication. This problem has been addressed in several ways, one of which is the enforcement of universal standards, which is extremely difficult to implement. Another approach has been suggested, which is the implementation of middleware software to facilitate communication between these devices.

Middleware can be defined as the software that offers interoperability between incompatible applications and devices; as well as shielding customers from the smart object's complexity. Hence, it serves as a software bridge between applications and things, allowing IoT systems to communicate and collaborate more effectively with one another. There are a plethora of middleware solutions available, whether proprietary or open source, that are provided by various companies, with the majority of these solutions being very similar to one another.

However, there are no benchmarks, performance indicators, or performance measurements that allow us to compare various systems. Following our examination of several articles[152]–[154], we have come to the following conclusions about some of the issues faced by IoT middleware:

1) ABSTRACTION AND INTEROPERABILITY: IoT middleware aids in allowing various smart devices to interface easily with each other in order to facilitate collaboration and data exchange among heterogeneous devices.
2) DEVICE DISCOVERY AND MANAGEMENT: This attribute allows IoT devices and services to be located in their network domain where the IoT environment infrastructure is primarily dynamic because all newly joined devices must announce their existence and services.
3) SCALABILITY: The IoT middleware must be scalable and must provide APIs in order to list all IoT devices, their capabilities, and their services, among other things.
4) DEVICE CATEGORIZATION: APIs must also allow users to categorize devices based on capabilities, manage devices based on remaining energy, report IoT device problems to users, and perform problem load balancing procedures.
5) BIG DATA AND ANALYTICS: Because of the fragile nature of wireless sensor networks, part of the detected data may be incomplete, requiring the middleware to take this into account and extrapolate incomplete data using a suitable machine learning method.
6) PRIVACY: Since this majority of data coming from IoT applications and services is related to human daily life, security and privacy issues must be considered when transferring and processing it, necessitating the development of middleware that addresses these issues.
7) CLOUD SERVICES: Cloud computing is the vital layer of any IoT system. The data captured through sensors will be stored and analyzed in a

centralized cloud, and, as a result, IoT middleware should run well in various types of clouds as shown in Fig. 14.

8) CONTEXT DETECTION: Ambient data collection applications and real-time reactive applications are the two types of IoT applications. In the first, sensors collect data that will be processed offline later to obtain reasonable information that will be used for similar scenarios in the future, while in the second, systems must make a real-time decision based on the sensed data.

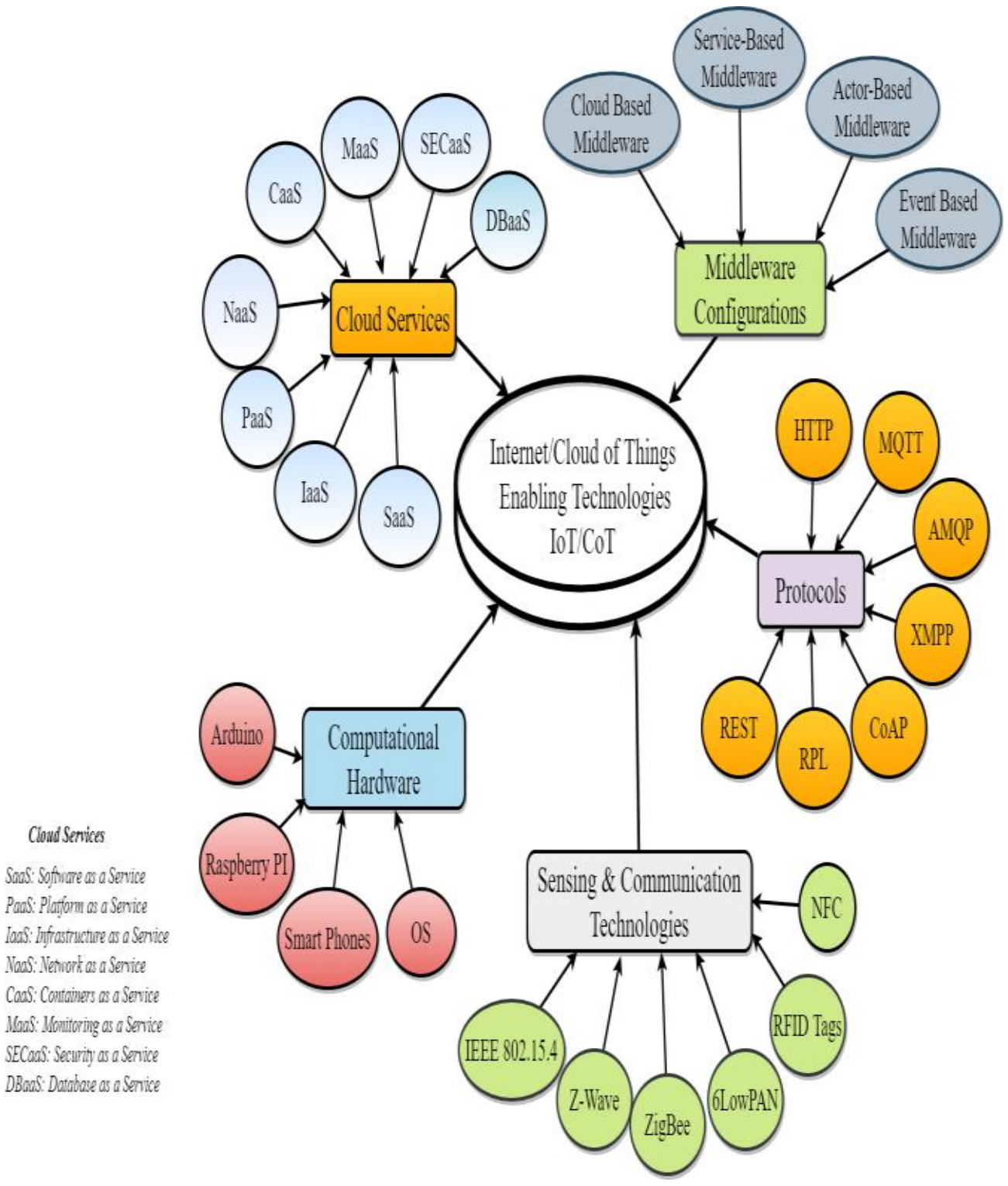

FIGURE 14: **Cloud and IoT enabling technologies.**

### A. ARCHITECTURE OF TOUCH BASED IOT MIDDLEWARE

Following are types of IoT middleware architectures that are currently available, which are classified based on the services that they provide[155]:

1) SERVICE ORIENTED ARCHITECTURE (SOA) OR SERVICE BASED SOLUTION:

Users and developers are given the ability to employ or add different types of IoT devices that can be used as services in the service oriented middleware architecture (SOA). Three layers make up SOA architecture: the physical layer, which contains actuators and sensors, the virtualized layer, which contains cloud and infrastructure servers that are responsible for performing various computational operations, and the Applications layer, which contains all services and utilities as shown in Fig.15. Access control, storage management, and event management are just a few of the general intermediate layer services accessible.

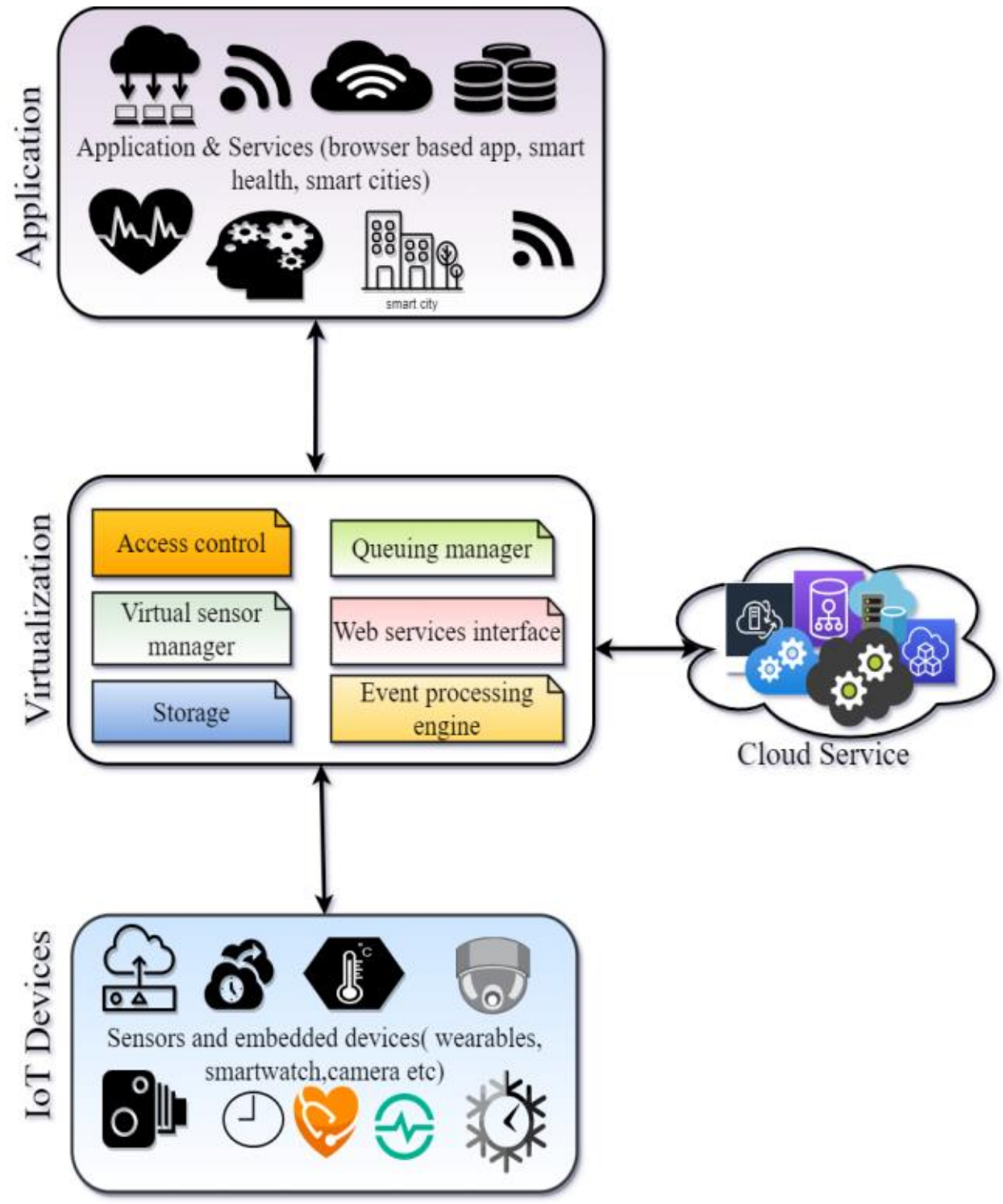

FIGURE 15: **Service oriented IoT Middleware architecture.**

SOA is a powerful middleware that may be deployed on nodes that communicate with cloud servers or on a powerful gateway that sits between the cloud levels and the IoT layer. As a result, this sort of middleware is incompatible with resource-constrained devices and does not allow for device-to-device communication. The most widely used service-based middleware is discussed here and summarized in the Table XII.

a) **Link Smart (Hydra):** A web service platform that intends to reduce the heterogeneity of different devices and entities in the IoT ecosystem, as well as control all types of smart devices regardless of their communication protocols, such as Zigbee, RFID, Bluetooth, Wi-Fi, and so on. This middleware is unique and it allows IoT devices to be used as services by embedding the required services. Health care, agriculture, and home automation are few of the IoT applications that can be managed with Link Smart [156].

b) **Kaa:** It is an open-source platform that enables the creation of IoT solutions and is administered by Cyber vision Inc and Kaa IoT technologies. By utilizing a web-based GUI built on the Apache platform, it is possible to create data delivery schemes, support multi-tenancy on servers, and generate endpoint software development kits (SDK). Kaa enables direct or indirect communication with endpoint devices, while encrypting their data using AES[29] and RSA [30]encryption methods[154].

[29] AES: Advanced Encryption Standard
[30] RSA: Rivest Shamir Adleman

TABLE XII
COMPARISON OF COMMONLY USED SERVICE-BASED MIDDLEWARE ATTRIBUTES.

| Service Based Middleware | Link smart (Hydra) | KAA | GSN | Thing speak | Aura |
|---|---|---|---|---|---|
| **Deployment type** | PaaS, SaaS | IaaS | PaaS | PaaS | IaaS, SaaS |
| **Network connectivity** | HTTP, REST,MQTT | MQTT, CoAP | HTTP | MQTT, REST API | MQTT, HTML |
| **Data format supported** | JSON | REST,JSON,API | JSON, Sen ML | XML, Thingspeak, API, JSON | REST ful API,JSON |
| **Programming Language** | C#, .NET, Java, Python, PHP, JavaScript | C,C++, Java | C, Java, Ruby | MATLAB | Python, PHP, C++, JavaScript |
| **Session persistence** | Machine learning algorithm | - | - | Using MQTT | Aura session |
| **Stream processing** | CEP queries, Esper EPL | - | SQL queries | MATLAB | Aura library |
| **Applications** | • Device abstraction<br>• Stream mining<br>• Live data management<br>• Data storage<br>• Online ML | • Analytics<br>• ML<br>• Event reporting<br>• Visualization | • Network structure visualization<br>• Data stream processing<br>• Data plotting | • Real-time analytics<br>• Event reporting<br>• Visualization | • Real-time applications connecting to a GUI<br>• Online video services<br>• Billing system<br>• Consoles and mobile devices and Smart TVs |
| **Service discovery** | REST API | MQTT with Kaa protocol V1 | REST HTTP query, Sbt 0.13+, Java JDK 1.7, Scala 2.11 | - | Environment management |
| **Security and Privacy** | • Encryption<br>• Authorization<br>• Authentication | • Encryption | • Authentication<br>• Access control mode | • Encryption | • Authentication<br>• Authorization |

c) **Global Sensor Network (GSN):** Its objective is to provide a standardized platform that enables adaptable deployments, sharing, and integration of heterogeneous Internet of Things (IoT) objects. This platform is built to meet the requirements of physical and virtual sensors and actuators, whether they are connected via a wired connection or wirelessly. GSN is a Java platform that can be placed on IoT cloud or servers, and it allows a series of wrappers to feed the system with collected line data that is later processed using XML specification files[157].

d) **Thing Speak IoT:** It is a MATLAB-developed analytic open-source platform service that allows people and things to communicate. In order to collect, visualize, and analyze real-time data from devices and sensors, Thing Speak provides users with tools that allow them to use the HTTP protocol over the internet to store and retrieve data from them[158].

e) **Aura:** This middleware facilitates the development of pervasive mobile IoT applications by abstracting device differences and allowing them to communicate freely. Aura makes an effort to optimize the screen backlight and CPU in order to improve performance while also reducing power consumption. In engaging with events, Aura uses two concepts: proactive and reactive. In a proactive concept, system layers respond immediately to the higher layer, whereas in a reactive concept, all layers adapt their resource and performance based on demand[159].

2) CLOUD BASED MIDDLEWARE:

User options are limited in the cloud-based middleware framework due to the limited number and variety of smart devices connected to IoT applications. In addition, because different use cases can be programmed and then determined in advance, the sensed data can be collected and interpreted with relative ease and accuracy. The operational component of this middleware is restricted by the resources available in the cloud computing environment.

Although IoT functions have a general presence in the IT architecture, like storage systems and computation engines, these functions are represented and controlled by APIs where IoT services are accessed by either cloud-based RESTful APIs or by vendor-provided applications as shown in Fig.16. The most widely used cloud-based IoT middleware is listed ahead and summarized the Table XIII.

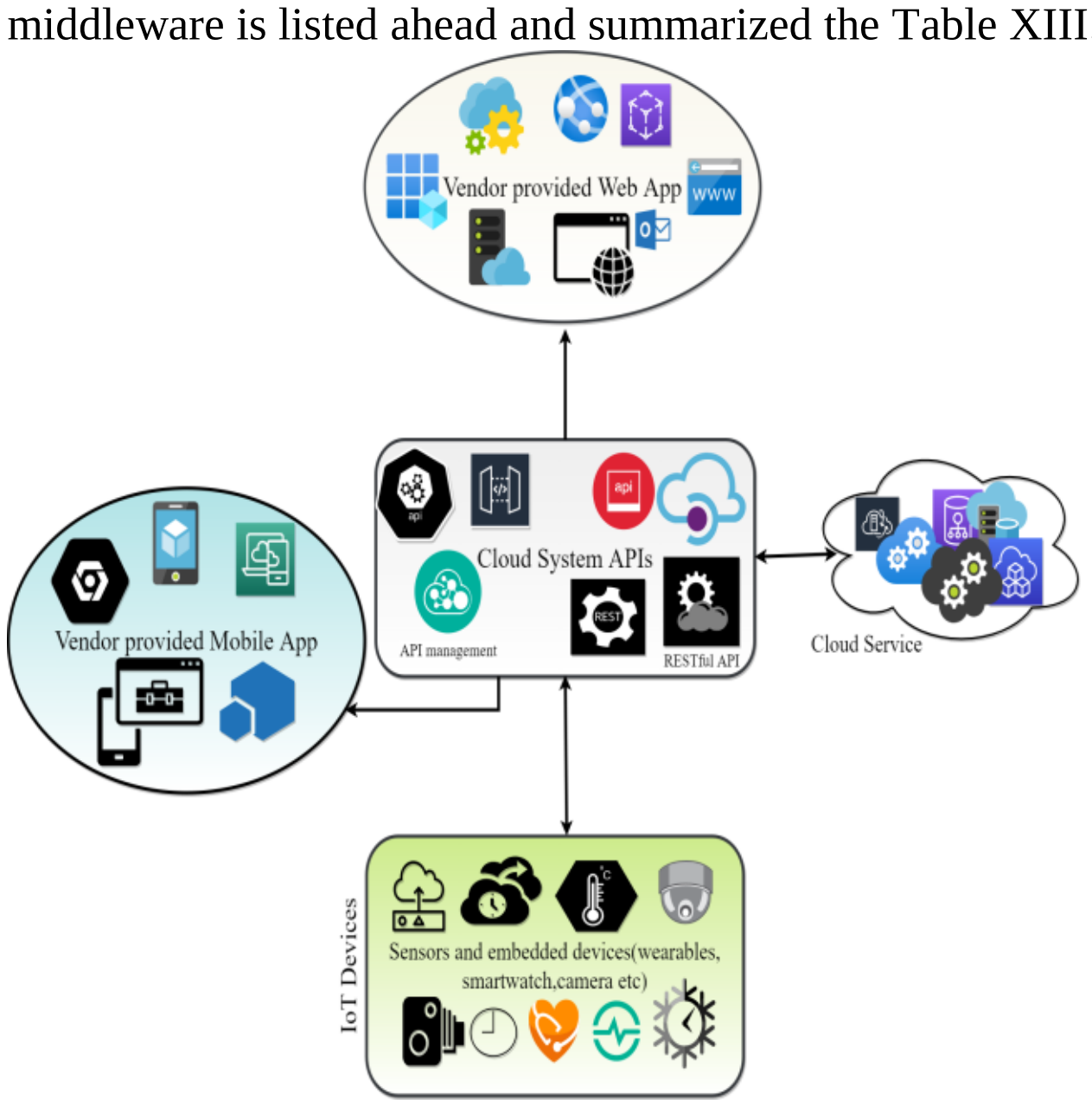

FIGURE 16: **Cloud based IoT Middleware architecture.**

TABLE XIII:
COMPARISON OF COMMONLY USED CLOUD-BASED MIDDLEWARE ATTRIBUTES.

| Cloud Based Middleware | AWS IoT | Azure IoT Hub | IBM Watson IoT | Google Cloud IoT | Oracle IoT |
|---|---|---|---|---|---|
| Deployment type | IaaS, PaaS | IaaS | IaaS, PaaS | IaaS, PaaS | PaaS |
| Interoperability | Web services | Azure IoT SDK | MQTT | REST APIs | Oracle Service bus |
| Network connectivity | MQTT, HTTP, Web-socket | HTTP, AMQP, MQTT over Web-socket | MQTT, HTTP | MQTT, HTTP | MQTT, HTTPs |
| Applications | • Real-time analytics<br>• AI/ML<br>• Event reporting<br>• Visualization | • Real-time analytics<br>• ML<br>• Event reporting<br>• Visualization | • Real-time analytics<br>• ML<br>• Event reporting<br>• Visualization | • Real-time analytics<br>• ML<br>• Event reporting<br>• Visualization | • Real-time analytics<br>• Event reporting<br>• Visualization |
| Technologies for application development | • AWS Cloud Trial<br>• Amazon Cloud Watch<br>• Kenisis<br>• Amazon ML | • SQL database<br>• Azure tables<br>• Azure Cosmos DB | • Cloudant<br>• No SQL Database | • Firebase Google's Big Data tool<br>• Big Query | • No SQL Database |
| Service discovery | • Discovery API<br>• AWS Lambda<br>• AWS App Mesh<br>• Amazon Route 53<br>• Netflix eureka | • Azure container services with Kubernetes<br>• Eureka<br>• Consul | • Discovery knowledge graph<br>• Watson discovery | • Consul<br>• Zookeeper | • JAVA WSDP |
| Security and Privacy | • Encryption<br>• Authorization<br>• Authentication<br>• Auditing | • Authorization<br>• Authentication<br>• Encryption | • Authorization<br>• Authentication | • Authentication | • Authentication<br>• Authorization |

a) **AWS IoT:** In order to manage cloud services, such as allowing millions of connected devices to communicate securely and easily with other devices and cloud applications, Amazon developed this platform. AWS IoT enables customers to build IoT applications that collect, process, analyze, and sense data in order to make appropriate decisions without the need for infrastructure management by utilizing AWS services such as Amazon Kinesis and Amazon Cloud Watch. AWS IoT customers can also always monitor all devices that communicate with their applications[160].

b) **Azure IoT Hub:** It is a central platform developed by Microsoft for managing bidirectional communication between IoT applications and the devices to which they are linked. Because of Azure's extensive capabilities, it enables clients to develop full-featured, scalable IoT solutions that provide secure and reliable communication between the hosted cloud and connected devices. Azure is a Microsoft product. When it comes to controlling IoT connected devices, Azure IoT Hub supports a variety of messaging patterns, including request-response, file upload from devices, and device to cloud telemetry[161].

c) **IBM Watson IoT:** This platform, which is built on top of the IBM Cloud, allows users to connect and control a variety of IoT appliances, sensors, industries, and home appliances. Using IBM Watson, its clients can create and manage their own IoT applications and appliances. They can also extract KPIs from their data and use them to control their tools and applications, as well as process their collected data using historical and real-time analytics. IBM Watson also offers a block chain service[162].

d) **Google Cloud IoT**: Essentially, it is a fully managed device that is composed of a set of tools that provide a comprehensive solution for secure and easily connecting and processing of data generated, whether they are located in the cloud or at the network edge. Google cloud IoT aspires to develop models capable of optimizing client business, anticipating problems, and increasing operational efficiency[154].

e) **Oracle IoT:** A cloud-based service platform that lets users create a real-time IoT solution to be linked with enterprise applications while leveraging rigorous security cloud capabilities and cutting-edge edge analytics. Furthermore, it integrates IoT data quickly and easily into customer business. Oracle IoT enables clients to connect their devices to the cloud, which will aid them in making critical strategies and decisions in their businesses[163].

3) ACTOR BASED MIDDLEWARE FRAMEWORK:

In terms of functionality, it is a lightweight middleware that can be implemented at the sensory, gateway, and cloud computing layers. Unlike other middleware, the computational operations of this middleware are distributed across multiple layers, including the sensory layer and mobile access layer. The sensory swarm is the outermost layer, made up of sensors and actuators, while the mobile access layer, made up of gateways, smart phones, Raspberry Pi, and laptops, is the intermediate layer, and the cloud is the innermost layer. The middleware which is also the actor host is designed to be lightweight and can be embedded in any layer of the application stack as shown in Fig.17.

TABLE XIV:
COMPARISON OF COMMONLY USED ACTOR-BASED MIDDLEWARE ATTRIBUTES.

| Actor Based Middleware | Calvin | NODE-RED | Ptolemy Accessor Host | Akka |
|---|---|---|---|---|
| **Deployment type** | IaaS | PaaS, SaaS | - | - |
| **Interoperability** | Actor model (event driven) | Web services | Accessor | Aggregate programming |
| **Network connectivity** | MQTT | HTTP, MQTT | HTTP,HTML | HTTP,HTML |
| **Data format supported** | JSON | JSON | JSON, XML | JSON |
| **Programming Language** | C, Python | JavaScript, Node.js | C, C++, JavaScript | Java, Scala |
| **Session persistence** | Distributed hash table | MQTT | Local file system | Akka persistence library. |
| **Stream processing** | Data flow processing | Node-red-contrib-cep | Discrete event director | Akka hop, Akka stream library and Apache Flink |
| **Applications** | • Distributed applications<br>• Runtime applications | • Connecting to IoT<br>• Connecting and binding databases<br>• Collecting and storing IoT data in event driven applications. | • Finite state machine applications<br>• Web applications | • Real-time streaming<br>• Real time applications<br>• Building powerful and concurrent Web applications |
| **Service discovery** | • Calvin control APIs | • Bonjour /Avahi | • Discovery.js<br>• Discovery function | • Akka discovery method<br>• Kubernetes API<br>• AWS<br>• Consul<br>• Marathon API |
| **Security and Privacy** | • Authorization<br>• Authentication | • Authentication<br>• Encryption | • Authentication<br>• Encryption | • Authentication<br>• Authorization<br>• Encryption |

A storage device, for example, may not be included in the actor-based middleware used on a smart watch. If the storage device is provided by an actor, it can be downloaded from a cloud repository whenever it is required. The most widely used actor-based middleware is discussed here and summarized in the Table XIV.

a) **Calvin:** It is an open source IoT platform developed by Ericsson to be used on energy-constrained smart devices because it offers a portable and lightweight unified programming architecture with input and output ports that define the interfaces. Furthermore Calvin can also be used at the edge of IoT ecosystems to reduce long-distance connections, lowering latency and reducing power consumption of IoT devices. Their main advantage is its ability to move from one environment to another[164].

b) **Node-RED**: It is an open source IoT platform developed by IBM and built on the node.js[31] programming language. Because of its small footprint, this platform can be used at the edge of an IoT network, while on the server side, a JavaScript platform with an event-driven module and non-blocking I/O is used. It allows users to build IoT applications by dragging and dropping connected blocks that represent IoT components. The platform drawbacks include the fact that it does not support service discovery and only allows for password authentication for security[165].

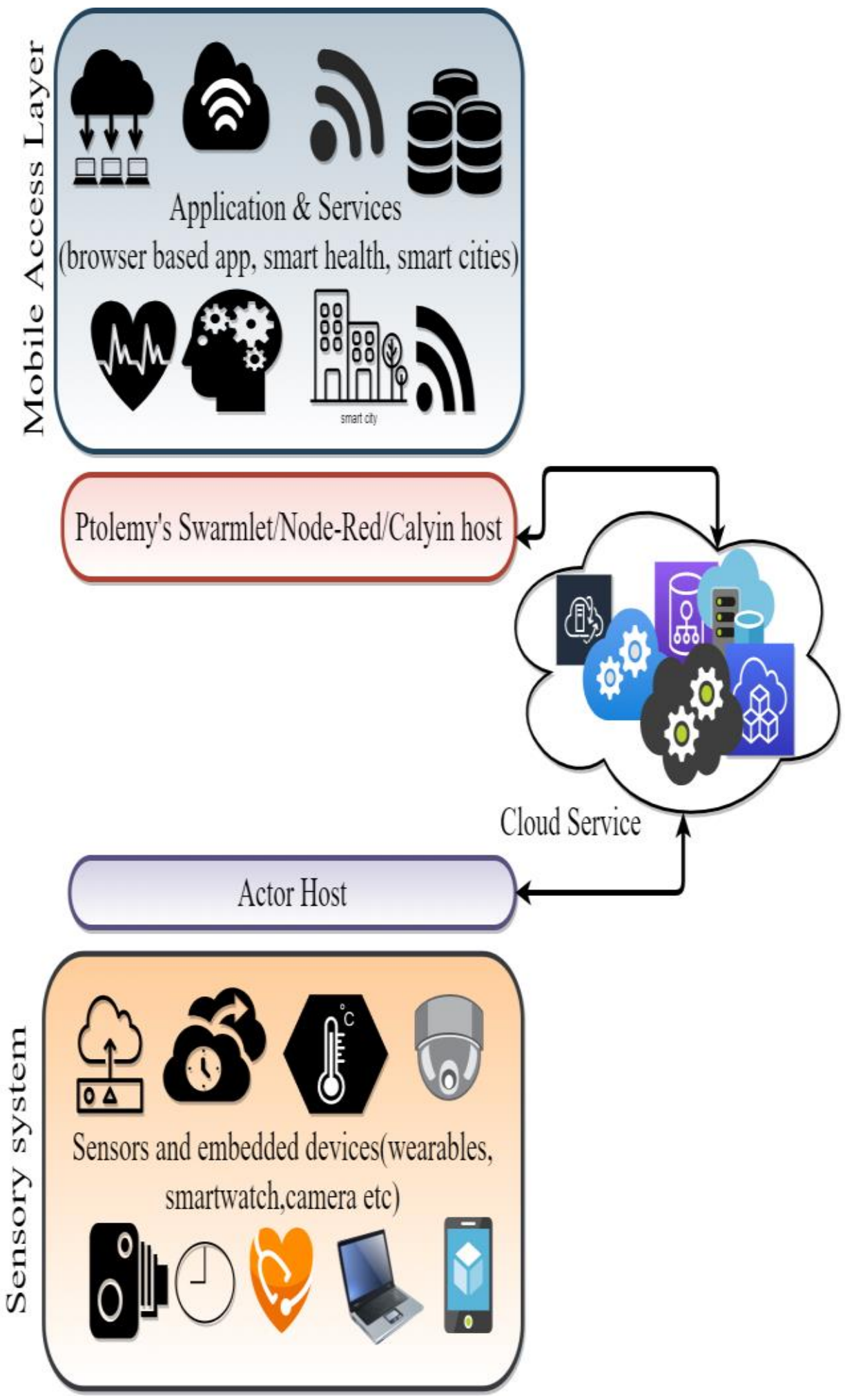

FIGURE 17: Actor based IoT Middleware architecture

[31] Node.js: It is an open source, cross platform and run time environment for executing JavaScript code on the server side.

TABLE XV:
COMPARISON OF COMMONLY USED EVENT-BASED MIDDLEWARE ATTRIBUTES.

| Event Based Middleware | Hermes | Gryphon | Rebeca | Fiware |
|---|---|---|---|---|
| Deployment type | PaaS | SaaS | PaaS | PaaS |
| Interoperability | Active message abstraction, 5 layered architecture by Fenix, Pegasus | Information flow graph between devices and broker | HTTP,SNMP,RMI | IoT Agent Framework Library |
| Network connectivity | ACL, HTML,XML | HTML, HTTP | HTTP, SNMP, Java RMI | MQTT, HTTP web socket |
| Data format supported | JSON, Hermes XML | JSON, NYSE, NASDAQ | XML | HTTP,JSON |
| Programming Language | C, Java, Python | Python, Java | .NET,C#, Java | C++, Java |
| Session persistence | Java persistence | Buffered stream, JMS persistent events | Fault tolerance plug-ins, sliding window scheme | Apache scheme, My SQL, Postgre SQL. |
| Stream processing | Open RTSP | Relational Subscription Scheme | - | Fiware Kurento, Web RTC |
| Applications | • Internet based distributed applications<br>• Large scale ubiquitous applications<br>• Web service | • Exchange connections<br>• Ledger accuracy guarantee<br>• Fault tolerance<br>• Monitoring<br>• ML<br>• Quantitative analysis | • Monitoring and management<br>• Fault tolerance<br>• Publishing methods | • Collecting and processing data<br>• Visualization and data analysis<br>• Data access control<br>• Monetization<br>• Publication<br>• Communication<br>• Publish/subscribe |
| Technologies for application development | • Type base routing algorithm<br>• Service agents | • Java message service (JMS)<br>• BKS+99<br>• Information flow graph<br>• Publisher-hosting broker | • Java management extension<br>• Object oriented API | • Fiware Content broker |
| Service discovery | • Service agents<br>• Yellow page service<br>• Discovery component<br>• Matchmaker service agent | - | Publish/subscribe mechanism | • Selection component Fiware NGSI<br>• REST ful API |
| Security and Privacy | • Encryption<br>• Authentication | • Authentication<br>• Auditing | • Authorization<br>• Authentication<br>• Encryption | • Authentication<br>• Authorization |

c) **Ptolemy Accessor Host:** Professor Edward Lee created this open source platform in 1996 to design, simulate, and model embedded and real-time devices. The underlying concept of this platform is that an IoT system is constructed from software components that interact and communicate with one another through messages sent through interconnected ports on a computer network[166].

c) **Akka:** It is a collection of open source libraries and a free actor-based platform that was created to allow developers to create distributed and run-time applications in either the Java or Scala programming languages. It enables users to meet business requirements without having to write large low-level code, resulting in high performance, fault tolerance, and reliability. Akka also allows multi-threading, isolates communication between applications and their devices, and provides a clustered architecture with excellent availability[167].

4) EVENT BASED MIDDLEWARE FRAMEWORK:

In order to improve the development of distributed systems, middleware that supports the publish/subscribe paradigm is being developed and implemented. According to this definition, this paradigm is a communication infrastructure that aims to provide clients with general-purpose services by assisting them in dealing with the heterogeneity and complexity of large-scale and distributed environments as shown in Fig.18. The event-based middleware hides some of the complexity of distributed applications from the programmer, which will make it easier to develop and program much different functionality in the future. The most significant differences between these architectures are their openness to supporting new IoT device types, the types of middleware services or computational units they support, and the locations where the IoT middleware can be embedded or deployed.

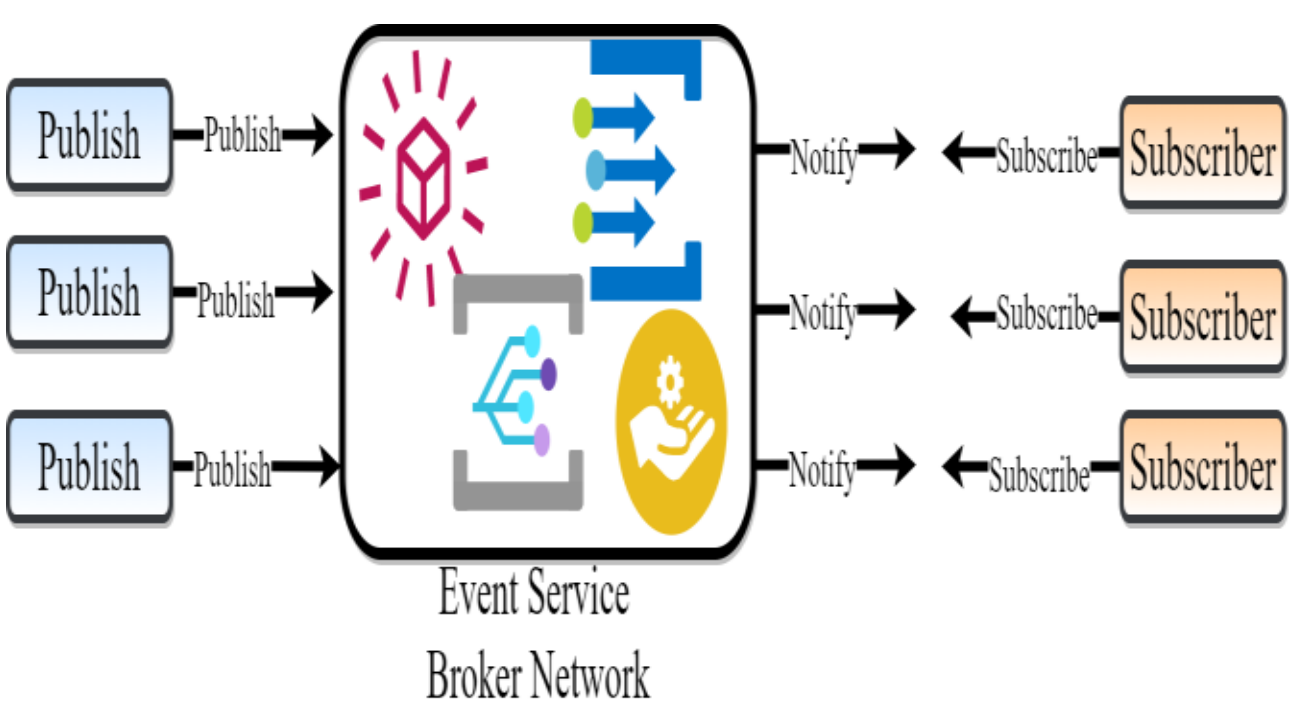

FIGURE 18: Event based IoT Middleware architecture

IoT middleware based on SOA is implemented on servers and in the cloud. Because the middleware can be implemented in all tiers and IoT devices can perform computation where it is most advantageous, actor-based delivers the best latency and scalability for large scale linked IoT devices. While these architectures provide some level of security and privacy, cloud-based architecture requires users to place their trust in the cloud provider to protect the privacy and integrity of their data.

Since a middleware cannot be implemented within the physical device and the data exchanged between physical devices and the middleware can be compromised, there is a weak security link between the physical devices and the middleware in both service and cloud-based architecture. To ensure QoS, the middleware must have a service discovery component that allows new services to be made accessible on demand and failed services to be dynamically replaced. The most widely used event-based middleware is discussed below and summarized in the Table XV.

a) **Hermes:** An event-based, scalable middleware designed to make distributed and large-scale applications easier. To address big size and dynamic situations, Hermes provides self-managed event brokers based on a P2P routing layer. It features an adaptive solution that takes into account failed event-broker events and routing stacks, all while maintaining the event-broker network. Hermes middleware has two versions, both of which share the majority of the codebase and are intended for use in distributed and large-scale systems as well as in communication and implementation among event brokers[155].

b) **Gryphon:** It is a highly scalable publish/subscribe middleware designed to distribute large amounts of real-time data over the network. Gryphon is a Java-based interface that enables the development of web applications and the creation of a robust, redundant, publish/subscribe, and content-based multi-broker. This middleware is completely secured and offers simple yet efficient routing and event handling. It also uses a messaging information flow paradigm (BKS+99) to specify communication between publisher and subscriber[168].

c) **Rebeca**: This middleware is based on publish/subscribe technology and focuses on the design of efficient routing algorithms and use of professional software engineering methodologies to implement large-scale business applications. To avoid network flooding, Rebeca employs advanced routing techniques. It incorporates interoperability and subscription merging capabilities into its services to facilitate location mobility and reduce the size of the routing table. The event scope function abstracts away the implementation details of a service, such as transmission policies, security, data transmission methods, external and internal interfaces, and notification representation[169].

d) **Fi WARE:** It enables distributed IoT devices and applications to communicate in an efficient, flexible, secure, and scalable manner. It was created to facilitate the control and monitoring of a variety of IoT applications, including logistics, retail, and smart cities. This platform is comprised of numerous components, including APIs, reusable modules, and massive code, all of which enable a user to create an IoT application. A collection of sensed data from IoT sensors is captured via REST API and later sent to a dedicated server called the broker. FiWare has developed an API for querying and storing various IoT contents, which enables any application registered as a content consumer to retrieve the necessary data from the broker. This platform has a component called an adapter that is in charge of transferring a certain type of material to subscriber applications[170].

## X. INTERFACING ARCHITECTURE WITH THE EXISTING NETWORK

The architecture in the Fig.19 represents the real time interfacing of the existing 5G/B5G network scenario with the next generation 6G all the way through the network slicing phenomenon. The slicing aspect comprises of the intelligent cloud slicing, the RAN slicing and the application slicing. The cloud slicing intends to facilitate the cloud computing and storage access to the edge devices and users via different slices catering to different users simultaneously. The RAN slicing is beneficial in connecting and routing the end users and devices to their receiving ends, accordingly making use of techniques like spectrum sharing, beam forming, cognitive antenna and radio transmission, tactile support and transmission system through different slices.

All the applications accessed at the user end owing to the slicing phenomenon are furnished by the application slicing technique. The applications accessed may range from robotics, tactile, haptic and touch based communications to the real time online gaming using the AR/VR/XR with the UHD video streaming. Other applications may range from the remote industrial application (IIoT), tele-medicine and tele-surgery, smart classroom with UHD video streaming. The autonomous vehicular and UAV system may be operated by application slice providing the IoT connectivity along with the URLLC of *1ms*.

The complete autonomy of the system and devices connected to the internet with the minimum latency and high data rate facilitate an intelligent and autonomous functioning system. The smart cities comprising of the smart homes, smart traffic monitoring systems and the rest instill another level of intelligence in the existing wireless network system. The subsequent section highlights some of the future aspects concerning the application of intelligent technologies in the wireless 6G communication system followed by some of the collaborative and ongoing projects

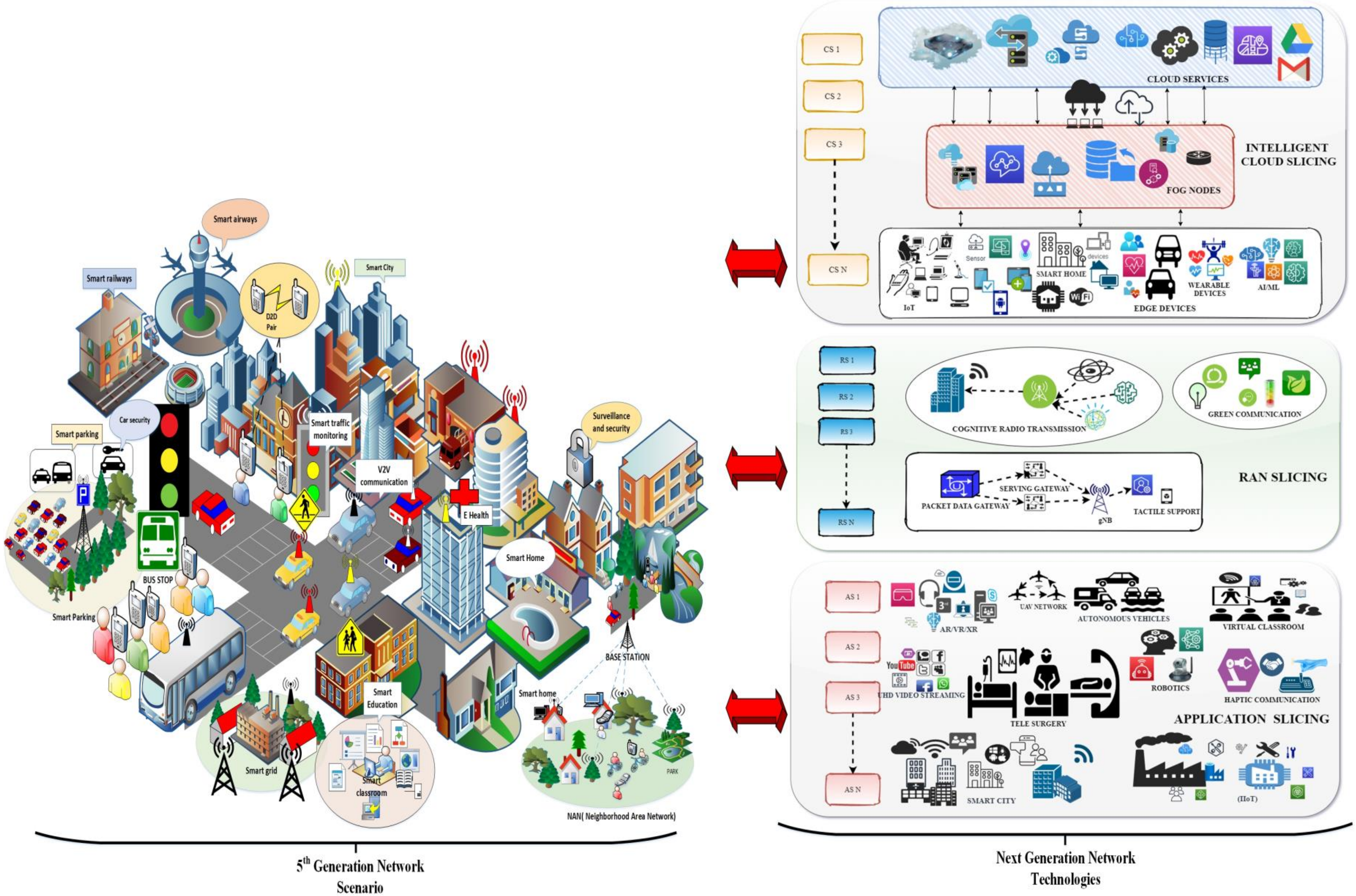

**FIGURE 19: Intelligent interfacing architecture of the existing communication system with the next generation technologies (B5G/6G).**

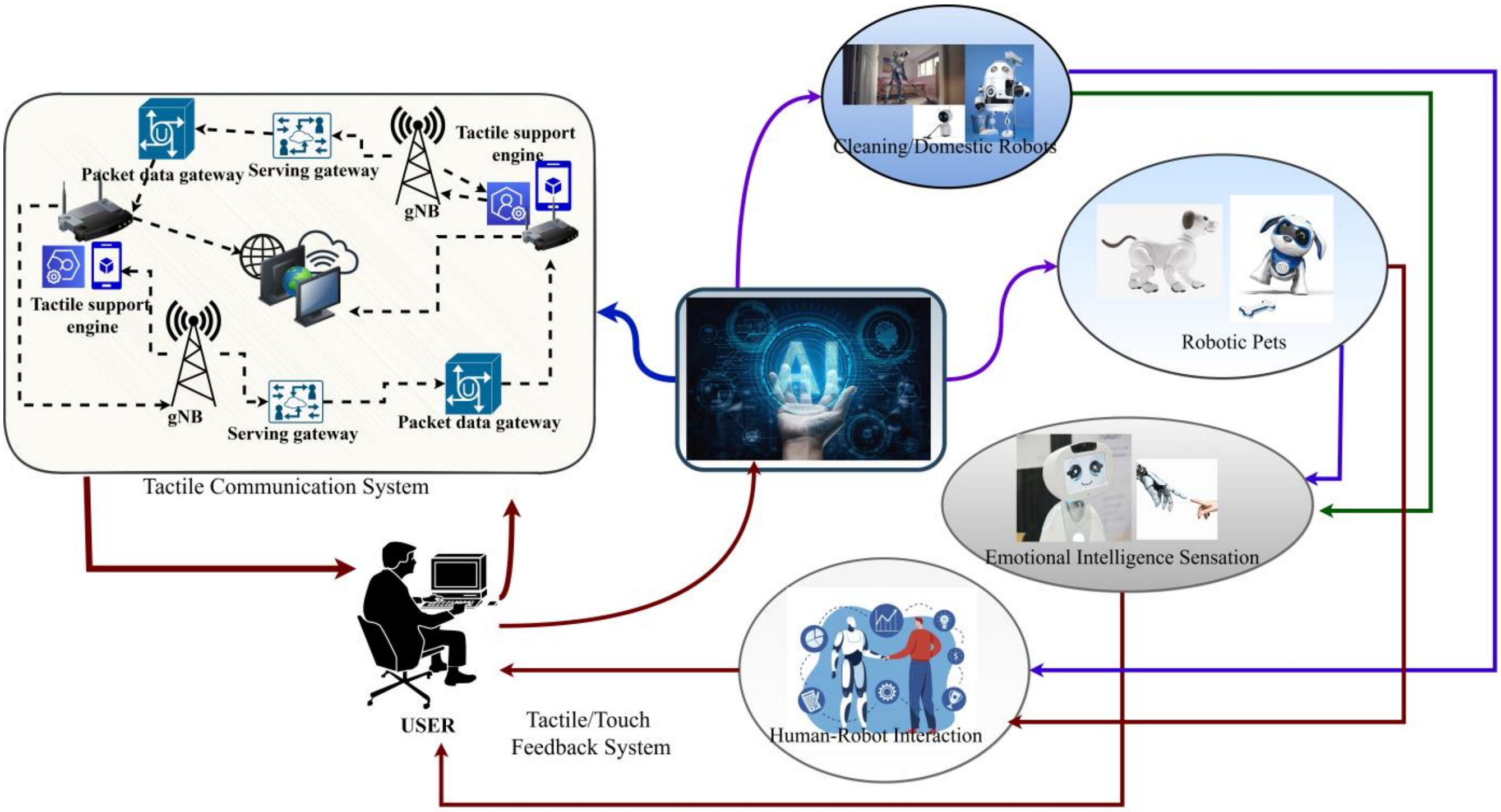

**FIGURE 20:** Touch induced human to robotic interactions with emotional intelligence.

implemented in the B5G/IoT and 6G wireless communication systems.

## XI. APPLICABLE USE CASES OF INTELLIGENT TOUCH TECHNOLOGY

This section explores the touch technology framework and future application elements, with a focus on tactile-based haptic communication integrated with learning approaches to enable efficient network transmission. The touch-enabled tele-diagnosis, robotic interaction, and haptic sensation-based shopping experience are just a few of the potential research aspect areas to look into.

### A. CASE 1- ROBOTIC INTERACTION

Another application of touch technology could be tactile-based robotic interaction between humans and robots, with the bot system being able to recognize and respond to human emotions [171], [172]. Emotions are interfaced into the machinery by creating a database in the system using AI/ML algorithms. The following are some examples of such applications:

1) ROBOTIC PETS:
It entails the integration of AI algorithms and functionality into robotic systems, as well as the introduction of emotional intelligence (EI) into them. The robotic pet dogs [173] is a dog-shaped robot capable of learning and detecting human gestures, face and eye movement, and responding appropriately. By introducing a touch-based interface into the system, these robots are capable of reacting to human touch feelings. Those who are elderly or mentally ill and are looking for companionship to relieve their loneliness will find these useful, as they will aid in their emotional development as a result of the interaction.

2) CLEANING/DOMESTIC ROBOTS:
These robotic systems are capable of following user instructions and performing household chores. These might be driven by popular AI-based home automation systems like Alexa to provide complete automation, allowing them to hear and act on the user's vocal commands rather than their restricted instruction library. The introduction of touch-based sensation/actuation into the framework enables them to automatically respond and act in response to the sensory simulations provided by the environment, thereby saving time that would otherwise be spent on user instructions, programming, and the interface itself.

In this way, the touch-enabled robotic-human interaction system with emotional intelligence is illustrated in Fig.20.

### B. CASE 2-AR/VR BASED ENTERTAINMENT/ SHOPPING EXPERIENCE

AR and VR technology are becoming the two important innovation factors that promote technological progress. Here the AR glasses provide a realistic perspective for viewing augmented reality content, while VR headsets provide an immersive sensory experience. Therefore AR and VR technology are becoming the two important innovation factors that promote technological progress. The virtual reality headsets provide users with an immersive sensory experience by allowing them to view AR content from a realistic perspective. At the same time, AR/VR technology has opened up a slew of new advertising and marketing possibilities.

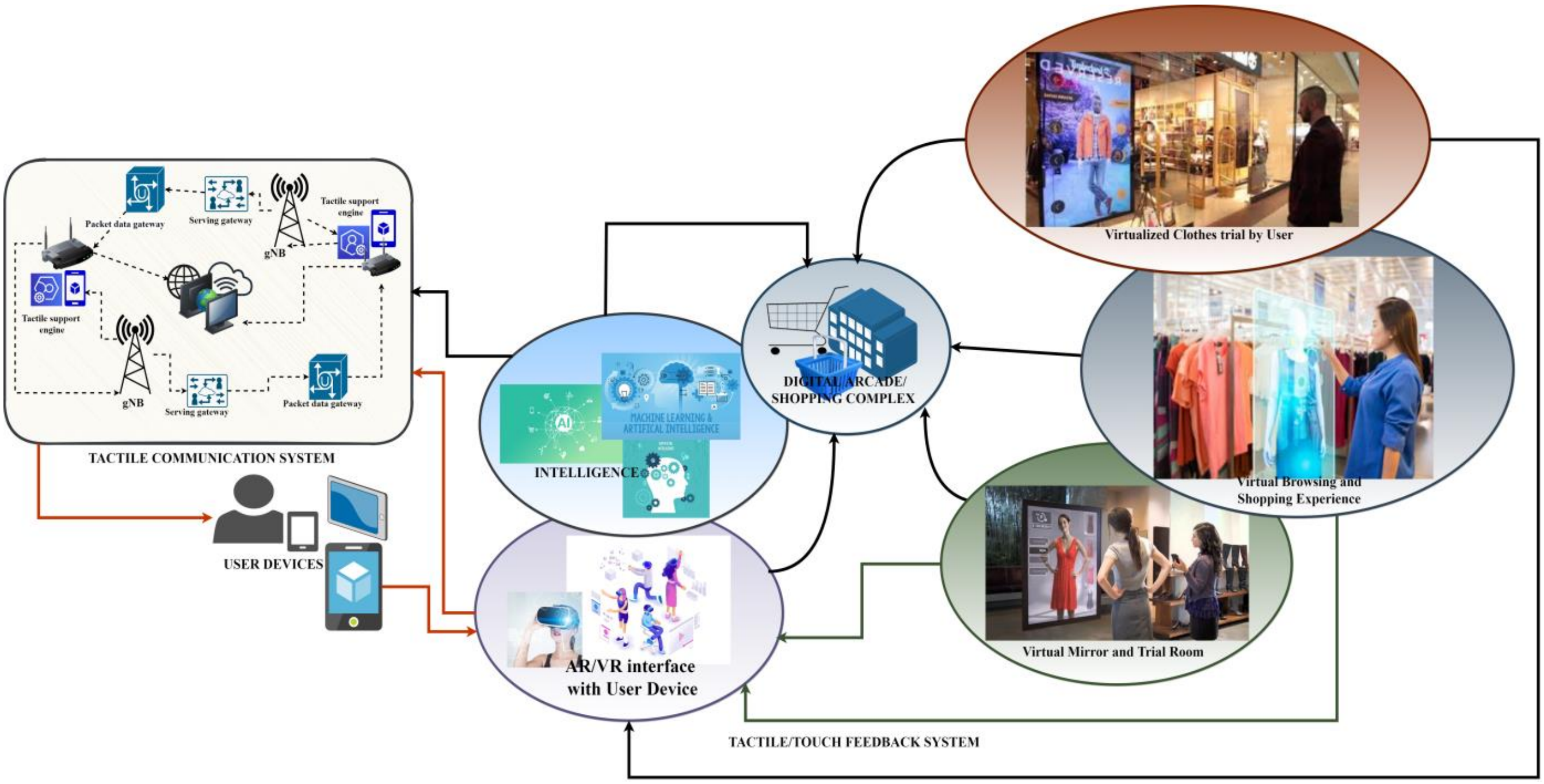

**FIGURE 21: Touch technology enabled virtualized shopping experience.**

Around 75 percent of business owners anticipate adopting AR/VR technology in the next two years, and global AR/VR market spending will be more than double. Moreover, AR/VR users are becoming more common in the world. As consumers increasingly rely on e-commerce and online shopping, there is a high demand for augmented reality content. WIMI plays a significant role in the AR shopping market and also uses 'AI+AR' to support other industries such as advertising, entertainment, and e-commerce[174].

Immersive AR/VR solutions help bridge the gap between online and offline purchasing for consumers, and more and more people are seeing AR/VR as a valuable tool for discovering products and getting brand services. AR/VR can currently improve the shopping experience through virtualization, with technology such as a virtual mirror assisting in a virtual garment trial before purchase being implemented[175]. Hence it acts as a future aspect that complements the integration of tactile/haptic-based intelligence with AR/VR technologies to further enrich our virtual shopping experience using touch technology.

In this regard, the illustration below shows how the proposed methodology can be integrated with existing AR/VR technology to further enhance the real-time user experience. The Fig.21 illustration above depicts a graphical representation of the virtualized shopping experience provided to users via intelligent touch technology.

### *C. CASE 3-TACTILE AND HAPTIC SENSATION BASED TELEDIAGNOSIS FOR CONTACT FREE COVID-19 CASES EXAMINATION.*

Tactile internet-enabled wireless communication systems have been integrated into the conventional healthcare system to create the smart healthcare system. Tactile-enabled healthcare systems, such as telemedicine, tele-surgery, and remote tele-diagnosis, could all benefit from the suggested touch technology. In today's technological environment, let us consider a practical scenario in which a medical surgeon is working from a smart surface or console that is connected to a telecommunications network in Chennai, India, while the patient is lying on an operation table at Fortis Hospital in New Delhi thousands of miles away[176].

The medical surgeon can remotely control the movement of a multi-armed surgical robot to perform gall bladder surgery on the patient by utilizing the smart surface and other communication technologies. Through the use of a tactile communication network, the doctor can communicate with and instruct the robot, while at the patient's end, a multi-machinery robot performs operations on the patient in accordance with the doctor's instructions. Introducing intelligence into the robotic system, which allows the robot to learn on its own while performing operations, can further enhance the technological benefits of the proposed touch technology as illustrated through Fig.22.

The procedure will be made easier if the machine is capable of sensing and responding to the user's touch sensations and interactions. Thus, an attempt is made to incorporate artificial intelligence (AI) into the system while providing instructions through the tactile communication network. Additionally, if a sense of touch or emotion is introduced in the form of emotional intelligence (EI), the robot will be able to comprehend and execute the user's instructions without delay. This method could be used for both remote and local diagnosis.

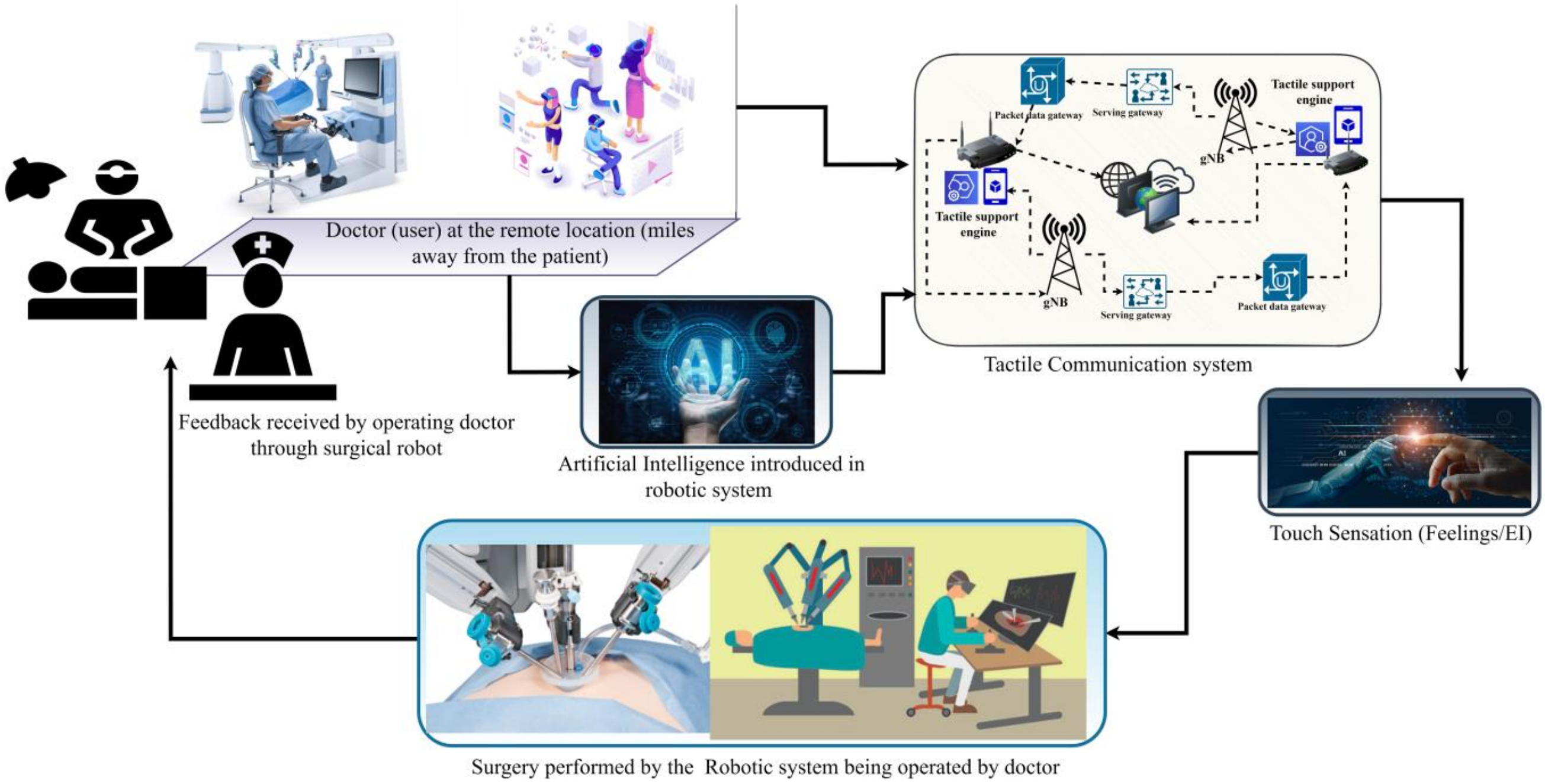

FIGURE 22: **Touch technology enabled contactless tele-diagnosis and surgery.**

In the current covid-19 pandemic, localized tele-diagnosis employing robots could be used to allow on-duty clinicians to undertake contactless diagnosis and testing on infected individuals. That in turn will lessen their chances of contracting the virus from the patients and thus lessen the strain on them during such a pandemic crisis or emergency situation.

## XII. RESEARCH CHALLENGES AND FUTURE ASPECTS

This section comprehends some of the conclusive future aspects concerning the incorporation of the reconfigurable and intelligent technologies like AI/ML/DL in the B5G/6G and IoT governed wireless communication system. In the end, intelligent technology machine learning has become one of the promising tools of artificial intelligence for the intelligent integration of wireless communications in the next generation. Their futuristic scope may be steered towards their relevance in channel estimation and detection, inferring user location and behavior, resource allocation, iterative learning, computational intelligence like neural networks and decision making process. Some of these have been discussed below:

### A. RESEARCH CHALLENGES:

1) INTELLIGENT CHANNEL ESTIMATION: The futuristic scope in the ML based channel estimation technique lies in the fact that it may be put to use in direct scenarios without any need for training. The only way to be able to learn the channel features of various user environments can be that a system is as smart to understand its parametric needs or in other words that such a generalized system requires a vast amount of pre-collected communication data to be used by ML/DL algorithms.
2) NON/SEMI-AUTONOMOUS DEVICE DISCOVERY:
Human intervention in IoT components, such as device discovery, renders these applications non-scalable and prone to error. Because of this limitation, interacting IoT devices like IoT middleware are unsuitable for self-adaptive applications such as M2M communication systems.
3) HETEROGENOUS ENVIRONMENTS: Since most smart IoT devices and middle wares only handle one or two types of heterogeneous components, this is considered a major issue that must be addressed. Due to the fact that non-autonomous and inflexible services and devices restrict the support of IoT applications, it is critical that new approaches address and resolve the heterogeneity of IoT environments, particularly in large scale networks.
4) SERVICE LEVEL AGREEMENT: To provide customers with an agreed level of service, three components should be considered: a model that precisely defines all functional and nonfunctional services required by consumers, automatic service to ensure a high level of QoS and adaptation, and a monitoring tool for SLA services. Human intervention in current autonomous networks will be phased out in favor of the development of intelligent IoT devices.
5) QOS LEVELS: Since there is no mechanism in place to guarantee a specific level of QoS for non-functional IoT services, researchers should develop procedures for optimizing and monitoring QoS levels.

6) PRIVACY AND SECURITY: As a result of the resource-constrained devices in IoT environments, the majority of autonomous and semi-autonomous application services restrict security, authorization and authentication mechanisms, among other things. Hence for the intelligent network to communicate between the cloud, gateway, and sensors securely and efficiently, security and privacy must be both end to end and lightweight.

### B. FUTURE RESEARCH ASPECTS:

1) ALGORITHMIC MODELING: Almost all relevant and physical modeling/construction should be seen as an integral step toward algorithmic implementation using applicable ML tools, and DNN is a key technology component in this regard.
2) ROBOTIC FEELINGS: The implementation of tactile based haptic communication in conjunction with touch enabled gesture recognition is another application that could benefit from this technology. This could be used in the inculcation of feelings and emotions in robotic systems.
3) INTELLIGENT HEALTH MEASURES DURING PANDEMIC LIKE COVID-19: Telemedicine, remote surgery, and treatment using intelligence and haptic sensations may also benefit from it, requiring a suitable algorithm with appropriate ML tools for efficient implementation and risk-free operation. So it may also be useful in the contactless administration to the patients during pandemics such as the one that occurred in COVID-19. In this way, the risk of doctors becoming infected with the virus is reduced because the diagnosis will be done by a robotic system that will use haptic and touch sensations that have been introduced.
4) EFFICIENT NETWORK MANAGEMENT: As a result, a paradigm shift is required for the efficient design of B5G/6G networks in order to leverage AI/ML and take advantage of big data analytics to improve the overall performance of future networks.
5) PROACTIVE NETWORKING: Future networks require a prediction mechanism to help predict and anticipate the future while allocating network resources proactively. As a result, it aids in the prediction and analysis of traffic patterns while determining off peak times on various spectrum bands so that incoming traffic demands can be properly allocated over a given window.
6) BEHAVIORAL LEARNING: Predicting user behaviors will result in better network resource utilization and will allow us to optimally allocate end-to-end network sources in an online fashion, which would be impossible without the assistance of AI and ML techniques.

## XIII. RECENT RESEARCH AND PROJECTS CONCERNING 6G WIRELESS COMMUNICATION SYSTEMS

Numerous 6G research and development activities have already begun on a global scale and this section summarizes the most significant 6G research activities underway at the moment[68].

### A. 6G FLAGSHIP (MAY 2018-APRIL 2026)

The 6G Flagship[177] is an eight-year research initiative that focuses on the wireless smart society and ecosystem enabled by 6G technology. Being funded by the Academy of Finland, this project intends to realize B5G networks from the very outset towards its phase of commercialization, and to develop the new 6G standards for the future digital societies. It aims to develop essential technology components of 6G mobile networks in areas such as wireless connectivity, distributed intelligent computing, security, and privacy. In addition to human-to-human communication, the research focuses on communication between devices, processes, and objects. This, in turn, enables a highly automated, smart society that will permeate all aspects of life. Ultimately, the 6G flagship project is to conduct large-scale pilots with a test network using industry and academic support.

### B. HEXA-X: A FLAGSHIP FOR 6G VISION AND INTELLIGENT FABRIC OF TECHNOLOGY ENABLERS CONNECTING HUMANS, PHYSICAL AND DIGITAL WORLDS (JAN 2021-JUNE 2023)

HEXA-X[178] is the first flagship project of the European Commission for implementing 6G vision and establishing an intelligent fabric of technology enablers for integrating the human, physical, and digital worlds. HEXA-X is a European industry-academic collaboration that intends to prepare the way for the next generation of wireless networks through exploratory research. Its objective is to connect humans, with the physical and digital worlds through a fabric of 6G essential enablers. In order to achieve this goal and vision, the HEXA-X project is concentrating on building critical technical enablers in the following areas:

1) High-frequency and high-resolution long-range access using New Radio access technologies.
2) Connected intelligence to future networks via AI-powered air interface.
3) Network disaggregation and dynamic dependability are enabled by 6G architectural drivers.

### C. TERA FLOW: SECURED AUTONOMIC TRAFFIC MANAGEMENT FOR A TERA OF SDN FLOWS (JAN 2021-JUNE 2023)

Tera flow[179] is working on developing a cloud native SDN controller for B5G/6G networks. This novel SDN controller is compatible with the contemporary NFV and MEC frameworks. It will also provide new features for traffic flow aggregation, service layer management, infrastructure layer network equipment integration, AI/ML-based security and forensic evidence for multi-tenancy networks.

### D. DAEMON: NETWORK INTELLIGENCE FOR ADAPTIVE AND SELF LEARNING MOBILE NETWORKS (JAN 2021- JUNE 2023)

The major goal of the DAEMON project[180] is to enable high-quality Network Intelligence (NI) for 6G systems, which will completely automate network administration. The project includes an end-to-end B5G/6G NI architecture, which can be fully coordinated with the NI-assisted features. DAEMON performs a systematic analysis of each NI task that is solved using AI models and also provides a solid set of guidelines for incorporating machine learning into network functionalities. A major goal of the DAEMON project is to focus on existing B5G network-specific AI methods that go beyond the current trend of integrating AI into network controllers and orchestrators.

### E. 6G BRAINS: BRINGING REINFORCEMENT LEARNING INTO RADIO LIGHTWEIGHT NETWORK FOR MASSIVE CONNECTIONS (JAN 2021-JUNE 2023)

The 6G BRAINS[181] project is focused on implementing multi-agent DRL for 6G radio links using AI. In order to improve massive connection over D2D assisted highly dynamic cell free networks, a novel comprehensive cross-layer DRL driven resource allocation solution will be required to perform resource allocation for Sub 6GHz/mm-wave/THz/optical wireless communication (OWC) medium. This significantly improves the capacity, reliability, and latency of future industrial, intelligent transportation, and e-health networks.

### F. SOUTH KOREA MSIT 6G RESEARCH PROGRAM

The Ministry of Science and ICT (MSIT) [182] in South Korea is working on a bold strategy to be the first country to deploy 6G networks. 6G services are expected to be commercially available in Korea between 2028 and 2030, according to the South Korean government. The initial deployment is expected in 2028, followed by a mass commercial deployment in 2030, with a total investment of $169 million in R&D for 6G technology. The preliminary goal is to deploy a 6G pilot in five important areas, including digital healthcare, immersive content, self-driving cars, smart cities, and smart manufacturing, by 2026[183]. The 6G research program's objectives are as follows.

1) Attain a data rate of 1Tbps.
2) Latency reduction of 0.1ms for wireless networks
3) Increases the range of connectivity coverage to up to 10 km from the ground.
4) AI must be integrated into the network to ensure that everything is covered.
5) By implementing security by design, it is possible to protect the network.

### G. JAPAN 6G PROMOTION STRATEGY

Japan invests approximately 50 billion dollars in the 6G development project. This initiative aims to strengthen collaboration between the public and private sectors in the field of 6G research and development. Furthermore, by 2025, this 6G promotion strategy seeks to establish and exhibit the 6G system's basic technologies, as well as put new technologies into practice by 2030[184].

### H. 6TH GENERATION INNOVATION CENTER

With the continuation of the 5th Generation Innovation Center (5GIC), the University of Surrey in the United Kingdom launched the 6GIC[185] in 2020 to focus on 6G-related research activities across two themes.

1) AMBIENT INFORMATION: There is a use of the advanced wireless technologies, high resolution sensing, and highly accurate geo-location methods to improve the fusion of virtual and physical environments. This will enable a new level of 6G digital services by better connecting of human senses with ambient and remote data.
2) UBIQUITOUS COVERAGE: It has a role in emphasizing on increasing the quality and range of 6G communication network coverage. The research will focus on expanding coverage indoors, utilizing intelligent surfaces, and satellite technology to enable 6G services to be available globally.

### I. INDUSTRIAL 6G PROGRAMS

The Table XVI summarizes several industrial research programs focused on the development and implementation of 6G. While Table XVII and Table XVIII give an overall projection of the ongoing recent projects in B5G/IoT and 6G respectively in the Appendix I

## XIV. CONCLUSION

The internet, with the advances in technology, drastically affects the human lifestyle, in profound ways, transforming various facets of life via interactions between the individuals at virtual level throughout most of the applications. The wireless technologies have thus transformed many elements of life including the business, living standards and the infrastructure. In a never-ending quest for elegant solutions to various problems, the society is always on the lookout for new avenues of progress. Therefore the motivation behind seamless connectivity has resulted in the evolution of wireless communication from 1G to 5G.

TABLE XVI
INDUSTRIAL AVENUES ON 6G DEVELOPMENTS

| Reference | Industries | Research projects/collaborations |
|---|---|---|
| [186] | Sony, NTT and Intel | Cooperative research on 6G technologies to be commercialized by 2030 |
| [187] | Huawei | 6G wireless technology research at the Canadian Research Center and 13 other universities. |
| [188] | SK Telecom | A 6G-based commercial networking project with Nokia and Ericsson. |
| [189] | Samsung | Commercialization is anticipated by 2028, along with 6G services that incorporate XR, a high-fidelity mobile hologram, and a digital replica. |
| [190] | LG and KAIST[32] | The opening of a new Research Center aimed at developing the 6G network standard. |
| [191] | NTT | Demonstrating a 100Gbps communication solution using OAM at 28GHz with MIMO. |
| [192] | Tektronix | Wireless fiber, a 100Gbps communication solution. |

[32] KAIST: Korea Advanced Institute of Science and Technology

Despite the recent development, work in this field is still underway all over the world with the aim of deploying the 6G communication network by 2030, thus making it one of the most in demand research fields, with the potential to revolutionize personal life of individual and society, business and communication systems. The real time applications from the individual user's point of view are perceptible in both professional as well as domestic fields. It may vary from e-health, smart appliances to the smart classroom based enhanced learning.

From the industrial point of view, the evident outcomes are discernible in departments like automation and industrial manufacturing, logistics, business and process management along with the intelligent transportation of people and goods. Therefore the current existing technologies make the IoT concept viable but do not necessarily fits well with the expected scalability and proficiency criteria of the 6G system. Therefore many problems do exist in the technical process and therefore, a societal reform is vital in the technology, universally conceptualizing the IoT connectivity.

A complete interoperability of the integrated devices is required, provided with a high percentage of smartness, maintaining the trust, safety and protection in the system for creation of technologies that emphasize technical requirements. However, the most intriguing aspect of the next generation 6G standard is the incorporation of an intelligent interface linking the existing communication standards with the recently researched ones, so as to competently fulfill the high data rate and the stringent latency requirements.

In other words, all we lack is an intelligent interface that is capable enough to integrate and enable a tactile based haptic (touch based) communication in the existing B5G network right from its source to its destination, complete with its feedback mechanism, altogether incorporating intelligence in the system so as to satisfy the stringent latency requirement of *1ms*.

This paper therefore provides the preliminary insight to and answers the above mentioned challenges by providing a comprehensive survey of the touch based intelligent communication system using network slicing and TI coupled the intelligence like AI and ML, in 6G.

## APPENDIX I:

### A. COLLABORATIVE AND ONGOING PROJECTS IN B5G/IOT

Table XVII illustrates the recent collaborative projects in B5G and IoT based wireless communication networks.

### B. GLOBAL LEVEL ONGOING PROJECTS FOR 6G DEPLOYMENT

Table XVIII summarizes all the ongoing research works and projects pertaining to 6G wireless communication system along with their application.

TABLE XVII
COLLABORATIVE AND ONGOING PROJECTS IN B5G AND IOT WIRELESS COMMUNICATION SYSTEMS.

| Ref | Research Project/ Group | Institution | Research area | Source |
|---|---|---|---|---|
| [74] | 5G Evolution and 6G | NTT-DOCOMO | Performance enhancement of mobile communication through exploring high frequency bands and improvement in wireless technologies. | https://www.nttdocomo.co.jp/english/binary/pdf/corporate/technology/whitepaper_6g/DOCOMO_6G_White_PaperEN_v3.0.pdf |
| [193] | 5G exchange (5GEx) | European Commission | Multiparty collaboration and multi domain orchestration for multiple carriers in 5G/B5G network infrastructure. | https://ieeexplore.ieee.org/document/7901481;doi:10.1109/MCOM.2017.1600197 |
| [194] | MATILDA | European commission | A holistic 5G E2E operational framework with smart and unified orchestration and management supporting edge/cloud computing. | http://www.matilda-5g.eu/ |
| [195] | Strategic Research and Innovation Agenda (SRIA-2021-27) | European Technology Platform Net world 2020 | Integration and management of AI and ML for supporting network and applications like slicing, URLLC networks, satellite communication with seamless fog/edge and cloud orchestration. | https://bscw.5g-ppp.eu/pub/bscw.cgi/d367342/Networld2020%20SRIA%202020%20Final%20Version%202.2%20.pdf |
| [196] | ML for future networks including 5G (ML5G) | ITU | Identification of gaps, issues and standardization concerning the ML for future networks for enhancing the network interfacing algorithms, protocols and architecture. | https://www.itu.int/en/ITU-T/focusgroups/ml5g/pages/default.aspx |
| [197] | AI and applied ML | TIP | Application of AI and ML for network planning, operation and optimization while leveraging the customer behavior driven optimization experience. | https://telecominfraproject.com/artificial-intelligence-and-applied-machine-learning/ |
| [198] | Network data analytics function (NWDAF) | 3GPP | ML function facilitating the monitoring of the network slice status concerning the purpose of the third party involvement. | http://www.tech-invite.com/3m29/tinv-3gpp-29-520.html |
| [199] | ITU-T Study group 13 | ITU | Evaluation of intelligence in the future networks including the IMT-2020. | http://handle.itu.int/11.1002/1000/14133 |
| [200] | 6Genesis Flagship Program | Academy of Finland | Implementation and testing of the key enabling technologies for 6G using the existing state of art 5G test bed. | http://jultika.oulu.fi/files/nbnfi-fe2019081624413.pdf |
| [201] | Zero touch provisioning | CISCO | Device configuration, management and orchestration to the local Ethernet in remote locations via dynamic host configuration protocol (DHCP)-IP. | https://www.cisco.com/c/en/us/td/docs/switches/lan/catalyst3850/software/release/16-5/configuration_guide/prog/b_165_prog_3850_cg/zero_touch_provisioning.pdf |
| [79] | What should 6G be? | KAUST. | Human centric perspective of 6G vision as research channel in post 5G era. | http://hdl.handle.net/10754/661147 |

TABLE XVIII
GLOBAL-LEVEL ONGOING PROJECTS ON 6G ALONGSIDE THEIR ASSOCIATED APPLICATIONS

| S.No | Ongoing Projects | Institute/ Organization | Enabling Technologies | Applications | Source |
|---|---|---|---|---|---|
| 1. | 6G Flagship | Academy of Finland | • To enable THz communication.<br>• AI/ML/FL implementation<br>• Blockchain/DLT<br>• ZSM<br>• NTN/3D networking<br>• VLC<br>• Quantum computing | • Extended reality<br>• Autonomous driving<br>• Intelligent healthcare<br>• Personalized Body area networks<br>• Industry 4.0/5.0 | [177] |
| 2. | HEXA-X | European Commission | • THz communication<br>• AI/ML/FL<br>• Compressive sensing<br>• Swarm networking | • UAV based networking<br>• Internet of everything (IoE)<br>• Industry 4.0/5.0<br>• Collaborative robots | [178] |
| 3. | Tera flow | 5GPPP | • AI/FL<br>• Blockchain/DLT<br>• ZSM | • Autonomous driving<br>• IoE | [179] |
| 4. | DAEMON | European Commission | • AI/FL<br>• ZSM<br>• Compressive sensing | • UAV connectivity<br>• Autonomous driving<br>• IoE<br>• Intelligent Heathcare | [180] |
| 5. | 6G BRAINS | 5GPPP | • THz communication<br>• AI/FL<br>• ZSM<br>• Swarm networking<br>• Compressive sensing<br>• VLC | • UAV navigation and connectivity<br>• Collaborative autonomous driving<br>• IoE | [181] |
| 6. | MSIT | South Korea | • THz communication<br>• AI/FL<br>• Smart surfaces<br>• VLC | • XR<br>• Collaborative robots and autonomous driving<br>• IoE<br>• Smart grid 2.0<br>• Industry 4.0/5.0<br>• Intelligent Healthcare | [182] |
| 7. | JAPAN | JAPAN | • THz communication<br>• AI/FL<br>• Smart surfaces<br>• Swarm networking<br>• VLC<br>• Quantum computing | • UAV mobility<br>• XR<br>• Autonomous driving<br>• IoE<br>• Smart grid 2.0<br>• Industry 4.0/5.0<br>• Intelligent Healthcare | [184] |
| 8. | 6GIC | University of Surrey, UK | • THz communication<br>• AI/FL<br>• Compressive sensing<br>• NTN/3D networking | • IoE<br>• Industry 4.0/5.0<br>• Hyper Intelligent IoT | [185] |

## ACKNOWLEDGMENT

The authors gratefully acknowledge the support provided by 5G and IoT Lab, DoECE, Shri Mata Vaishno Devi University, Katra, Jammu

## REFERENCES

[1] A. Goldsmith, *Wireless communications*. Cambridge university press, 2005.

[2] 'Cisco Annual Internet Report - Cisco Annual Internet Report (2018–2023) White Paper', CISCO, Mar. 2020. Accessed: Apr. 22, 2021.[Online].Available:https://www.cisco.com/c/en/us/solutions/collateral/executive-perspectives/annual-internet-report/white-paper-c11-741490.html

[3] N. Sharma, M. Shamkuwar, and I. Singh, 'The history, present and future with IoT', in *Internet of Things and Big Data Analytics for Smart Generation*, Springer, 2019, pp. 27–51.

[4] J. G. Andrews *et al.*, 'What will 5G be?', *IEEE J. Sel. Areas Commun.*, vol. 32, no. 6, pp. 1065–1082, 2014.

[5] A. Gupta and R. K. Jha, 'A Survey of 5G Network: Architecture and Emerging Technologies', *IEEE Access*, vol. 3, pp. 1206–1232, 2015, doi: 10.1109/ACCESS.2015.2461602.

[6] I. Farris, T. Taleb, H. Flinck, and A. Iera, 'Providing ultra-short latency to user-centric 5G applications at the mobile network edge', *Trans. Emerg. Telecommun. Technol.*, vol. 29, no. 4, p. e3169, 2018, doi: https://doi.org/10.1002/ett.3169.

[7] S. Parkvall, E. Dahlman, A. Furuskar, and M. Frenne, 'NR: The new 5G radio access technology', *IEEE Commun Stand Mag*, vol.1,no.4,pp.24–30,2017,doi:10.1109/MCOMSTD.2017.1700042.

[8] A. Dogra, R. K. Jha, and S. Jain, 'A Survey on beyond 5G network with the advent of 6G: Architecture and Emerging Technologies', *IEEE Access*, pp. 67512–67547, 2020, doi: 10.1109/ACCESS.2020.3031234.

[9] P. Gandotra, R. K. Jha, and S. Jain, 'Sector-Based Radio Resource Allocation ( SBRRA ) Algorithm for Better Quality of Service and Experience in Device-to-Device (D2D) Communication', *IEEE Trans. Veh. Technol.*, vol. 67, no. 7, pp. 5750–5765, 2017, doi: 10.1109/TVT.2017.2787767.

[10] H. Wang, J. Wang, G. Ding, L. Wang, T. A. Tsiftsis, and P. K. Sharma, 'Resource allocation for energy harvesting-powered D2D

communication underlaying UAV-assisted networks', *IEEE Trans. Green Commun. Netw.*, vol. 2, no. 1, pp. 14–24, 2018.

[11] M. Shafi and R. K. Jha, 'Half-Duplex Attack: An Effectual Attack Modelling in D2D Communication', in *2020 International Conference on COMmunication Systems & NETworkS (COMSNETS)*, 2020, pp. 879–881.

[12] S. Akbar, Y. Deng, A. Nallanathan, M. Elkashlan, and G. K. Karagiannidis, 'Massive multiuser MIMO in heterogeneous cellular networks with full duplex small cells', *IEEE Trans. Commun.*, vol. 65, no. 11, pp. 4704–4719, 2017.

[13] F. Al-Turjman, E. Ever, and H. Zahmatkesh, 'Small cells in the forthcoming 5G/IoT: Traffic modelling and deployment overview', *IEEE Commun. Surv. Tutor.*, vol. 21, no. 1, pp. 28–65, 2018.

[14] V. Hassija, V. Chamola, V. Saxena, D. Jain, P. Goyal, and B. Sikdar, 'A survey on IoT security: application areas, security threats, and solution architectures', *IEEE Access*, vol. 7, pp. 82721–82743, 2019.

[15] 'Cloud VR–oriented Bearer Network White Paper', 2017. Accessed: Aug. 25, 2020. [Online]. Available: https://www-file.huawei.com/-/media/corporate/pdf/ilab/cloud_vr_oriented_bearer_network_white_paper_en_v2.pdf

[16] S. Sharma, R. Miller, and A. Francini, 'A cloud-native approach to 5G network slicing', *IEEE Commun. Mag.*, vol. 55, no. 8, pp. 120–127, 2017.

[17] N. Alliance, '5G white paper', *Gener. Mob. Netw. White Pap.*, vol. 1, 2015.

[18] A. Ghosh, A. Maeder, M. Baker, and D. Chandramouli, '5G Evolution: A View on 5G Cellular Technology beyond 3GPP Release 15', *IEEE Access*, vol. 7, no. March, pp. 127639–127651, 2019, doi: 10.1109/ACCESS.2019.2939938.

[19] A. Ali and W. Hamouda, 'Advances on spectrum sensing for cognitive radio networks: Theory and applications', *IEEE Commun. Surv. Tutor.*, vol. 19, no. 2, pp. 1277–1304, 2016.

[20] W. Lu *et al.*, 'SWIPT Cooperative Spectrum Sharing for 6G-Enabled Cognitive IoT Network', *IEEE Internet Things J.*, 2020.

[21] B. Soret, A. De Domenico, S. Bazzi, N. H. Mahmood, and K. I. Pedersen, 'Interference Coordination for 5G New Radio', *IEEE Wirel Commun*, vol. 25, no. 3, pp. 131–137, 2018, doi: 10.1109/MWC.2017.1600441.

[22] G. Chopra, R. K. Jha, and S. Jain, 'Rank-based secrecy rate improvement using NOMA for ultra dense network', *IEEE Trans. Veh. Technol.*, vol. 68, no. 11, pp. 10687–10702, 2019.

[23] M. Giordani, M. Polese, A. Roy, D. Castor, and M. Zorzi, 'A tutorial on beam management for 3GPP NR at mmwave frequencies', *IEEE Commun Surv Tutor.*, vol. 21, no. 1, pp. 173–196, 2019, doi: 10.1109/COMST.2018.2869411.

[24] S. Bakri, P. A. Frangoudis, A. Ksentini, and M. Bouaziz, 'Data-Driven RAN Slicing Mechanisms for 5G and Beyond', *IEEE Trans. Netw. Serv. Manag.*, pp. 1–1, 2021, doi: 10.1109/TNSM.2021.3098193.

[25] N. Nomikos *et al.*, 'A UAV-based moving 5G RAN for massive connectivity of mobile users and IoT devices', *Veh. Commun.*, vol. 25, p. 100250, Oct. 2020, doi: 10.1016/j.vehcom.2020.100250.

[26] A. Manzalini, C. Lin, J. Huang, C. Buyukkoc, and M. Bursell, 'Towards 5G software-defined ecosystems: Technical challenges, business sustainability and policy issues', *IEEE SDN White Pap.*, 2016.

[27] P. P. Ray and N. Kumar, 'SDN/NFV architectures for edge-cloud oriented IoT: A systematic review', *Comput. Commun.*, 2021.

[28] K. K. Karmakar, V. Varadharajan, S. Nepal, and U. Tupakula, 'SDN-Enabled Secure IoT Architecture', *IEEE Internet Things J.*, vol. 8, no. 8, pp. 6549–6564, 2020.

[29] A. A. Barakabitze, A. Ahmad, R. Mijumbi, and A. Hines, '5G network slicing using SDN and NFV: A survey of taxonomy, architectures and future challenges', *Comput. Netw.*, vol. 167, p. 106984, 2020.

[30] A. Cárdenas and D. Fernández, 'Network Slice Lifecycle Management Model for NFV-based 5G Virtual Mobile Network Operators', in *2020 IEEE Conference on Network Function Virtualization and Software Defined Networks (NFV-SDN)*, Nov. 2020, pp. 120–125. doi: 10.1109/NFV-SDN50289.2020.9289883.

[31] A. El-Mekkawi, X. Hesselbach, and J. R. Piney, 'Novel NFV Aware Network Service for Intelligent Network Slicing Based on Squatting-Kicking Model', *IEEE Access*, vol. 8, pp. 223041–223068, 2020, doi: 10.1109/ACCESS.2020.3044951.

[32] G. Nencioni, R. G. Garroppo, A. J. Gonzalez, B. E. Helvik, and G. Procissi, 'Orchestration and Control in Software-Defined 5G Networks: Research Challenges', *Wirel. Commun. Mob. Comput.*, vol. 2018, pp. 1–18, Aug. 2018, doi: 10.1155/2018/6923867.

[33] Z. Hou, C. She, Y. Li, T. Q. Quek, and B. Vucetic, 'Burstiness-aware bandwidth reservation for ultra-reliable and low-latency communications in tactile Internet', *IEEE J. Sel. Areas Commun.*, vol. 36, no. 11, pp. 2401–2410, 2018.

[34] S. Yin, Y. Zhao, and L. Li, 'Resource allocation and basestation placement in cellular networks with wireless powered UAVs', *IEEE Trans. Veh. Technol.*, vol. 68, no. 1, pp. 1050–1055, 2018.

[35] N. Alliance, 'Description of network slicing concept', *NGMN 5G P*, vol. 1, no. 1, 2016.

[36] X. Foukas, G. Patounas, A. Elmokashfi, and M. K. Marina, 'Network slicing in 5G: Survey and challenges', *IEEE Commun. Mag.*, vol. 55, no. 5, pp. 94–100, 2017.

[37] F. Granelli, 'Network slicing', in *Computing in Communication Networks*, Elsevier, 2020, pp. 63–76. doi: 10.1016/B978-0-12-820488-7.00014-1.

[38] J. Prados-Garzon, J. J. Ramos-Munoz, P. Ameigeiras, P. Andres-Maldonado, and J. M. Lopez-Soler, 'Modeling and Dimensioning of a Virtualized MME for 5G Mobile Networks', *IEEE Trans. Veh. Technol.*, vol. 66, no. 5, pp. 4383–4395, May 2017, doi: 10.1109/TVT.2016.2608942.

[39] M. Richart, J. Baliosian, J. Serrat, and J.-L. Gorricho, 'Resource Slicing in Virtual Wireless Networks: A Survey', *IEEE Trans. Netw. Serv. Manag.*, vol. 13, no. 3, pp. 462–476, Sep. 2016, doi: 10.1109/TNSM.2016.2597295.

[40] T. Taleb, A. Ksentini, and A. Kobbane, 'Lightweight Mobile Core Networks for Machine Type Communications', *IEEE Access*, vol. 2, pp. 1128–1137, 2014, doi: 10.1109/ACCESS.2014.2359649.

[41] Q. Zhang, M. Jiang, Z. Feng, W. Li, W. Zhang, and M. Pan, 'IoT Enabled UAV: Network Architecture and Routing Algorithm', *IEEE Internet Things J.*, vol. 6, no. 2, pp. 3727–3742, Apr. 2019, doi: 10.1109/JIOT.2018.2890428.

[42] P. Agyapong, M. Iwamura, D. Staehle, W. Kiess, and A. Benjebbour, 'Design considerations for a 5G network architecture', *IEEE Commun. Mag.*, vol. 52, no. 11, pp. 65–75, Nov. 2014, doi: 10.1109/MCOM.2014.6957145.

[43] A. Imran, A. Zoha, and A. Abu-Dayya, 'Challenges in 5G: how to empower SON with big data for enabling 5G', *IEEE Netw.*, vol. 28, no. 6, pp. 27–33, 2014.

[44] N. H. Mahmood *et al.*, 'White paper on critical and massive machine type communication towards 6G', 2020. [Online]. Available: arXiv preprint arXiv:2004.14146

[45] I. Afolabi, A. Ksentini, M. Bagaa, T. Taleb, M. Corici, and A. Nakao, 'Towards 5G network slicing over multiple-domains', *IEICE Trans. Commun.*, vol. 100, no. 11, pp. 1992–2006, 2017, doi: 10.1587/transcom.2016NNI0002.

[46] G. P. Fettweis, 'The tactile internet: Applications and challenges', *IEEE Veh Technol Mag*, vol. 9, no. 1, pp. 64–70, 2014, doi: 10.1109/MVT.2013.2295069.

[47] M. Simsek, A. Aijaz, M. Dohler, J. Sachs, and G. Fettweis, '5G-enabled tactile internet', *IEEE J. Sel. Areas Commun.*, vol. 34, no. 3, pp. 460–473, 2016.

[48] M. Geller and P. Nair, '5G security innovation with Cisco', *Whitepaper Cisco Public*, pp. 1–29, 2018.

[49] S. J. Nawaz, S. K. Sharma, S. Wyne, M. N. Patwary, and M. Asaduzzaman, 'Quantum machine learning for 6G communication networks: State-of-the-art and vision for the future', *IEEE Access*, vol. 7, pp. 46317–46350, 2019.

[50] K. David and H. Berndt, '6G vision and requirements: Is there any need for beyond 5G?', *IEEE Veh. Technol. Mag.*, vol. 13, no. 3, pp. 72–80, 2018.

[51] M. Latva-aho, K. Leppänen, F. Clazzer, and A. Munari, 'Key drivers and research challenges for 6G ubiquitous wireless intelligence', 2020. [Online]. Available: University of Oulu, http://urn. fi/urn:isbn:9789526223544

[52] M. Maier, M. Chowdhury, B. P. Rimal, and D. P. Van, 'The tactile internet: vision, recent progress, and open challenges', *IEEE Commun. Mag.*, vol. 54, no. 5, pp. 138–145, 2016.

[53] T. Taleb, S. Dutta, A. Ksentini, M. Iqbal, and H. Flinck, 'Mobile edge computing potential in making cities smarter', *IEEE Commun. Mag.*, vol. 55, no. 3, pp. 38–43, 2017.

[54] X. Foukas, G. Patounas, A. Elmokashfi, and M. K. Marina, 'Network Slicing in 5G: Survey and Challenges', *IEEE Commun Mag*, vol. 55, no. 5, pp. 94–100, 2017, doi: 10.1109/MCOM.2017.1600951.

[55] A. Kaloxylos, 'A Survey and an Analysis of Network Slicing in 5G Networks', *IEEE Commun. Stand. Mag.*, vol. 2, no. 1, pp. 60–65, Mar. 2018, doi: 10.1109/MCOMSTD.2018.1700072.

[56] I. Afolabi, T. Taleb, K. Samdanis, A. Ksentini, and H. Flinck, 'Network slicing and softwarization: A survey on principles, enabling technologies, and solutions', *IEEE Commun Surv Tutor.*, vol. 20, no. 3, pp. 2429–2453, 2018, doi: 10.1109/COMST.2018.2815638.

[57] M. E. Morocho-Cayamcela, H. Lee, and W. Lim, 'Machine learning for 5G/B5G mobile and wireless communications: Potential, limitations, and future directions', *IEEE Access*, vol. 7, pp. 137184–137206, 2019.

[58] M. Mohammadi and A. Al-Fuqaha, 'Enabling cognitive smart cities using big data and machine learning: Approaches and challenges', *IEEE Commun. Mag.*, vol. 56, no. 2, pp. 94–101, 2018.

[59] V. P. Kafle, Y. Fukushima, P. Martinez-Julia, and T. Miyazawa, 'Consideration On Automation of 5G Network Slicing with Machine Learning', in *2018 ITU Kaleidoscope: Machine Learning for a 5G Future (ITU K)*, Santa Fe, Nov. 2018, pp. 1–8. doi: 10.23919/ITU-WT.2018.8597639.

[60] A. Aijaz, M. Dohler, A. Hamid Aghvami, V. Friderikos, and M. Frodigh, 'Realizing the Tactile Internet: Haptic Communications over Next Generation 5G Cellular Networks', *IEEE Wirel Commun*, vol. 24, no. 2, pp. 82–89, 2017, doi: 10.1109/MWC.2016.1500157RP.

[61] K. Antonakoglou, X. Xu, E. Steinbach, T. Mahmoodi, and M. Dohler, 'Toward haptic communications over the 5G tactile internet', *IEEE Commun Surv Tutor.*, vol. 20, no. 4, pp. 3034–3039, 2018, doi: 10.1109/COMST.2018.2851452.

[62] A. Aijaz and M. Sooriyabandara, 'The Tactile Internet for Industries: A Review', *Proc IEEE*, vol. 107, no. 2, pp. 414–435, 2019, doi: 10.1109/JPROC.2018.2878265.

[63] Y. Sun, M. Peng, Y. Zhou, Y. Huang, and S. Mao, 'Application of Machine Learning in Wireless Networks: Key Techniques and Open Issues', *IEEE Commun. Surv. Tutor.*, vol. 21, no. 4, pp. 3072–3108, Fourthquarter 2019, doi: 10.1109/COMST.2019.2924243.

[64] C.-X. Wang, M. Di Renzo, S. Stanczak, S. Wang, and E. G. Larsson, 'Artificial intelligence enabled wireless networking for 5G and beyond: Recent advances and future challenges', *IEEE Wirel. Commun.*, vol. 27, no. 1, pp. 16–23, 2020.

[65] D. Bega, M. Gramaglia, A. Banchs, V. Sciancalepore, and X. Costa-Perez, 'A Machine Learning Approach to 5G Infrastructure Market Optimization', *IEEE Trans Mob Comput*, vol. 19, no. 3, pp. 498–512, 2020, doi: 10.1109/TMC.2019.2896950.

[66] M. F. Zhani and H. ElBakoury, 'FlexNGIA: A flexible Internet architecture for the next-generation tactile Internet', *J. Netw. Syst. Manag.*, pp. 1–45, 2020.

[67] X. You *et al.*, 'Towards 6G wireless communication networks: Vision, enabling technologies, and new paradigm shifts', *Sci. China Inf. Sci.*, vol. 64, no. 1, pp. 1–74, 2021.

[68] C. De Alwis *et al.*, 'Survey on 6G frontiers: Trends, applications, requirements, technologies and future research', *IEEE Open J. Commun. Soc.*, vol. 2, pp. 836–886, 2021.

[69] 'WHITE PAPER ON 6G NETWORKING', 6G Research Visions, No. 6, Jun. 2020. Accessed: Oct. 22, 2021. [Online]. Available: http://jultika.oulu.fi/files/isbn9789526226842.pdf

[70] M. Giordani, M. Polese, M. Mezzavilla, S. Rangan, and M. Zorzi, 'Toward 6G networks: Use cases and technologies', *IEEE Commun. Mag.*, vol. 58, no. 3, pp. 55–61, 2020.

[71] A. Al-Dulaimi, X. Wang, and I. Chih-Lin, *5G networks: fundamental requirements, enabling technologies, and operations management*. John Wiley & Sons, 2018.

[72] L. U. Khan, I. Yaqoob, M. Imran, Z. Han, and C. S. Hong, '6G wireless systems: A vision, architectural elements, and future directions', *IEEE Access*, vol. 8, pp. 147029–147044, 2020.

[73] W. Saad, M. Bennis, and M. Chen, 'A vision of 6G wireless systems: Applications, trends, technologies, and open research problems', *IEEE Netw.*, vol. 34, no. 3, pp. 134–142, 2019.

[74] NTT DOCOMO, 'White Paper 5G Evolution and 6G', White Paper, Feb. 2021. Accessed: Jun. 12, 2021. [Online]. Available: https://www.nttdocomo.co.jp/english/binary/pdf/corporate/technology/whitepaper_6g/DOCOMO_6G_White_PaperEN_v3.0.pdf

[75] Z. Zhang *et al.*, '6G wireless networks: Vision, requirements, architecture, and key technologies', *IEEE Veh. Technol. Mag.*, vol. 14, no. 3, pp. 28–41, 2019.

[76] Y. Al-Eryani and E. Hossain, 'The D-OMA method for massive multiple access in 6G: Performance, security, and challenges', *IEEE Veh. Technol. Mag.*, vol. 14, no. 3, pp. 92–99, 2019.

[77] B. Zong, C. Fan, X. Wang, X. Duan, B. Wang, and J. Wang, '6G technologies: Key drivers, core requirements, system architectures, and enabling technologies', *IEEE Veh. Technol. Mag.*, vol. 14, no. 3, pp. 18–27, 2019.

[78] G. Gui, M. Liu, F. Tang, N. Kato, and F. Adachi, '6G: Opening new horizons for integration of comfort, security, and intelligence', *IEEE Wirel. Commun.*, vol. 27, no. 5, pp. 126–132, 2020.

[79] S. Dang, O. Amin, B. Shihada, and M.-S. Alouini, 'What should 6G be?', *Nat. Electron.*, vol. 3, no. 1, pp. 20–29, 2020.

[80] M. S. Elbamby, C. Perfecto, M. Bennis, and K. Doppler, 'Toward low-latency and ultra-reliable virtual reality', *IEEE Netw.*, vol. 32, no. 2, pp. 78–84, 2018.

[81] T. Park and W. Saad, 'Distributed learning for low latency machine type communication in a massive internet of things', *IEEE Internet Things J.*, vol. 6, no. 3, pp. 5562–5576, 2019.

[82] M. Alsenwi, N. H. Tran, M. Bennis, S. R. Pandey, A. K. Bairagi, and C. S. Hong, 'Intelligent resource slicing for eMBB and URLLC coexistence in 5G and beyond: A deep reinforcement learning based approach', *IEEE Trans. Wirel. Commun.*, 2021.

[83] M. A. Siddiqi, H. Yu, and J. Joung, '5G ultra-reliable low-latency communication implementation challenges and operational issues with IoT devices', *Electronics*, vol. 8, no. 9, p. 981, 2019.

[84] H. Yang, A. Alphones, Z. Xiong, D. Niyato, J. Zhao, and K. Wu, 'Artificial-intelligence-enabled intelligent 6G networks', *IEEE Netw.*, vol. 34, no. 6, pp. 272–280, 2020.

[85] C. Huang, A. Zappone, G. C. Alexandropoulos, M. Debbah, and C. Yuen, 'Reconfigurable intelligent surfaces for energy efficiency in wireless communication', *IEEE Trans. Wirel. Commun.*, vol. 18, no. 8, pp. 4157–4170, 2019.

[86] N. Janbi, I. Katib, A. Albeshri, and R. Mehmood, 'Distributed Artificial Intelligence-as-a-Service (DAIaaS) for Smarter IoE and 6G Environments', *Sensors*, vol. 20, no. 20, 2020, doi: 10.3390/s20205796.

[87] L. U. Khan, I. Yaqoob, N. H. Tran, Z. Han, and C. S. Hong, 'Network Slicing: Recent Advances, Taxonomy, Requirements, and Open Research Challenges', *IEEE Access*, vol. 8, pp. 36009–36028, 2020, doi: 10.1109/ACCESS.2020.2975072.

[88] 'Perspectives on Vertical Industries and Implications for 5G', Jun. 2016. Accessed: Aug. 25, 2021. [Online]. Available: https://www.ngmn.org/fileadmin/user_upload/160610_NGMN_Perspectives_on_Vertical_In dustries_and_Implications_for_5G_v1_0.pdf

[89] C. Li *et al.*, '5G-Based Systems Design for Tactile Internet', *Proc. IEEE*, vol. 107, no. 2, pp. 307–324, Feb. 2019, doi: 10.1109/JPROC.2018.2864984.

[90] S. K. Sharma, I. Woungang, A. Anpalagan, and S. Chatzinotas, 'Toward Tactile Internet in Beyond 5G Era: Recent Advances, Current Issues, and Future Directions', *IEEE Access*, vol. 8, pp. 56948–56991, 2020, doi: 10.1109/ACCESS.2020.2980369.

[91] J. Arshad, M. A. Azad, K. Salah, R. Iqbal, M. I. Tariq, and T. Umer, 'Performance analysis of content discovery for ad-hoc tactile networks', *Future Gener. Comput. Syst.*, vol. 94, pp. 726–739, 2019.

[92] I. Budhiraja, S. Tyagi, S. Tanwar, N. Kumar, and J. J. Rodrigues, ‘Tactile internet for smart communities in 5g: An insight for noma-based solutions’, *IEEE Trans. Ind. Inform.*, vol. 15, no. 5, pp. 3104–3112, 2019.
[93] K. Antonakoglou, X. Xu, E. Steinbach, T. Mahmoodi, and M. Dohler, ‘Toward Haptic Communications Over the 5G Tactile Internet’, *IEEE Commun. Surv. Tutor.*, vol. 20, no. 4, pp. 3034–3059, 2018, doi: 10.1109/COMST.2018.2851452.
[94] ‘The Tactile Internet’, *ITU-T Technol. Watch Rep. August 2014*, p. 24, 2014.
[95] S. Bi, R. Zhang, Z. Ding, and S. Cui, ‘Wireless communications in the era of big data’, *IEEE Commun. Mag.*, vol. 53, no. 10, pp. 190–199, 2015.
[96] M. Chiang and T. Zhang, ‘Fog and IoT: An overview of research opportunities’, *IEEE Internet Things J.*, vol. 3, no. 6, pp. 854–864, 2016.
[97] S. Han, I. Chih-Lin, G. Li, S. Wang, and Q. Sun, ‘Big data enabled mobile network design for 5G and beyond’, *IEEE Commun. Mag.*, vol. 55, no. 9, pp. 150–157, 2017.
[98] Y. Sun, H. Song, A. J. Jara, and R. Bie, ‘Internet of things and big data analytics for smart and connected communities’, *IEEE Access*, vol. 4, pp. 766–773, 2016.
[99] X. Wang and Y. He, ‘Learning from uncertainty for big data: future analytical challenges and strategies’, *IEEE Syst. Man Cybern. Mag.*, vol. 2, no. 2, pp. 26–31, 2016.
[100] G. Villarrubia, J. F. De Paz, P. Chamoso, and F. De la Prieta, ‘Artificial neural networks used in optimization problems’, *Neurocomputing*, vol. 272, pp. 10–16, 2018.
[101] M. Chen, U. Challita, W. Saad, C. Yin, and M. Debbah, ‘Artificial Neural Networks-Based Machine Learning for Wireless Networks: A Tutorial’, *IEEE Commun. Surv. Tutor.*, vol. 21, no. 4, pp. 3039–3071, 2019, doi: 10.1109/COMST.2019.2926625.
[102] G. Alnwaimi, S. Vahid, and K. Moessner, ‘Dynamic heterogeneous learning games for opportunistic access in LTE-based macro/femtocell deployments’, *IEEE Trans. Wirel. Commun.*, vol. 14, no. 4, pp. 2294–2308, 2015.
[103] X. Wang, X. Li, and V. C. Leung, ‘Artificial intelligence-based techniques for emerging heterogeneous network: State of the arts, opportunities, and challenges’, *IEEE Access*, vol. 3, pp. 1379–1391, 2015.
[104] R. Li *et al.*, ‘Intelligent 5G: When cellular networks meet artificial intelligence’, *IEEE Wirel. Commun.*, vol. 24, no. 5, pp. 175–183, 2017.
[105] L. Busoniu, R. Babuska, and B. De Schutter, ‘A comprehensive survey of multiagent reinforcement learning’, *IEEE Trans. Syst. Man Cybern. Part C Appl. Rev.*, vol. 38, no. 2, pp. 156–172, 2008.
[106] K. Arulkumaran, M. P. Deisenroth, M. Brundage, and A. A. Bharath, ‘Deep reinforcement learning: A brief survey’, *IEEE Signal Process. Mag.*, vol. 34, no. 6, pp. 26–38, 2017.
[107] C. Jiang, H. Zhang, Y. Ren, Z. Han, K.-C. Chen, and L. Hanzo, ‘Machine learning paradigms for next-generation wireless networks’, *IEEE Wirel. Commun.*, vol. 24, no. 2, pp. 98–105, 2017.
[108] I. Portugal, P. Alencar, and D. Cowan, ‘The use of machine learning algorithms in recommender systems: A systematic review’, *Expert Syst. Appl.*, vol. 97, pp. 205–227, 2018.
[109] T. E. Bogale, X. Wang, and L. B. Le, ‘Machine Intelligence Techniques for Next-Generation Context-Aware Wireless Networks’, *ITU Spec. Issue Impact Artif. Intell. AI Commun. Netw. Serv.*, vol. 1, 2018.
[110] S. M. Aldossari and K.-C. Chen, ‘Machine learning for wireless communication channel modeling: An overview’, *Wirel. Pers. Commun.*, vol. 106, no. 1, pp. 41–70, 2019.
[111] K. Bonawitz *et al.*, ‘Towards federated learning at scale: System design’, *ArXiv Prepr. ArXiv190201046*, 2019.
[112] R. Dong, C. She, W. Hardjawana, Y. Li, and B. Vucetic, ‘Deep Learning for Hybrid 5G Services in Mobile Edge Computing Systems: Learn From a Digital Twin’, *IEEE Trans. Wirel. Commun.*, vol. 18, no. 10, pp. 4692–4707, Oct. 2019, doi: 10.1109/TWC.2019.2927312.
[113] D. Gündüz, P. de Kerret, N. D. Sidiropoulos, D. Gesbert, C. R. Murthy, and M. van der Schaar, ‘Machine learning in the air’, *IEEE J. Sel. Areas Commun.*, vol. 37, no. 10, pp. 2184–2199, 2019.
[114] J. Huang *et al.*, ‘A big data enabled channel model for 5G wireless communication systems’, *IEEE Trans. Big Data*, vol. 6, no. 2, pp. 211–222, 2018.
[115] X. Ma, J. Zhang, Y. Zhang, and Z. Ma, ‘Data scheme-based wireless channel modeling method: motivation, principle and performance’, *J. Commun. Inf. Netw.*, vol. 2, no. 3, pp. 41–51, 2017.
[116] H. Li, Y. Li, S. Zhou, and J. Wang, ‘Wireless channel feature extraction via GMM and CNN in the tomographic channel model’, *J. Commun. Inf. Netw.*, vol. 2, no. 1, pp. 41–51, 2017.
[117] A. Charrada and A. Samet, ‘Joint interpolation for LTE downlink channel estimation in very high-mobility environments with support vector machine regression’, *IET Commun.*, vol. 10, no. 17, pp. 2435–2444, 2016.
[118] H. He, C.-K. Wen, S. Jin, and G. Y. Li, ‘Deep learning-based channel estimation for beamspace mmWave massive MIMO systems’, *IEEE Wirel. Commun. Lett.*, vol. 7, no. 5, pp. 852–855, 2018.
[119] Y. Li, Y. Zhang, X. Huang, H. Zhu, and J. Ma, ‘Large-scale remote sensing image retrieval by deep hashing neural networks’, *IEEE Trans. Geosci. Remote Sens.*, vol. 56, no. 2, pp. 950–965, 2017.
[120] R. El Hattachi, J. Erfanian, ‘NGMN 5G White Paper’, NMGN Alliance, Feb. 2015. Accessed: Aug. 25, 2020. [Online]. Available: https://www.ngmn.org/uploads/media/NGMN_5G_White_Paper_V1_0.pdf
[121] S. Zhang, C. Xiang, and S. Xu, ‘6G: Connecting everything by 1000 times price reduction’, *IEEE Open J. Veh. Technol.*, vol. 1, pp. 107–115, 2020.
[122] S. Antipolis, ‘Feasibility study on new services and markets technology enablers for critical communications, relese 14’, 3GPP, France, Rep. TR 22.862, Jun. 2016.
[123] S. Antipolis, ‘Study on enhancement of 3GPP support for 5G V2X services, release 15’, 3GPP,France, Rep. TR 22.886, Dec. 2016.
[124] M. Series, ‘IMT Vision–Framework and overall objectives of the future development of IMT for 2020 and beyond’, *Recomm. ITU*, vol. 2083, 2015.
[125] ‘White Paper on Machine Learning in 6G Wireless Communication Networks,6G Research Visions, No. 7’, White Paper No. 7, Jun. 2020. Accessed: Mar. 25, 2021. [Online]. Available: https://www.6gchannel.com/items/6g-white-paper-machine-learning/
[126] M. R. Abdmeziem, D. Tandjaoui, and I. Romdhani, ‘Architecting the internet of things: state of the art’, *Robots Sens. Clouds*, pp. 55–75, 2016.
[127] A. A. O. Bahashwan and S. Manickam, ‘A brief review of messaging protocol standards for internet of things (IoT)’, *J. Cyber Secur. Mobil.*, pp. 1–14, 2019.
[128] H. V. Nguyen and L. L. Iacono, ‘RESTful IoT authentication protocols’, in *Mobile Security and Privacy*, Elsevier, 2017, pp. 217–234.
[129] N. Naik, ‘Choice of effective messaging protocols for IoT systems: MQTT, CoAP, AMQP and HTTP’, in *2017 IEEE international systems engineering symposium (ISSE)*, 2017, pp. 1–7.
[130] A. A. Ahmed and W. Ali, ‘A lightweight reliability mechanism proposed for datagram congestion control protocol over wireless multimedia sensor networks’, *Trans. Emerg. Telecommun. Technol.*, vol. 29, no. 3, p. e3296, 2018.
[131] E. Rescorla and T. Dierks, ‘The transport layer security (TLS) protocol version 1.3’, 2018.
[132] C.-S. Park and W.-S. Park, ‘A group-oriented DTLS handshake for secure IoT applications’, *IEEE Trans. Autom. Sci. Eng.*, vol. 15, no. 4, pp. 1920–1929, 2018.
[133] M. Taillon, T. Saad, R. Gandhi, Z. Ali, and M. Bhatia, ‘Updates to the resource reservation protocol for fast reroute of traffic engineering GMPLS Label Switched Paths (LSPs)’, *Internet Req. Comments RFC Ed. RFC*, vol. 8271, 2017.
[134] M. B. Yassien, S. A. Aljawarneh, M. Eyadat, and E. Eaydat, ‘Routing protocol for low power and lossy network–load balancing

time-based', *Int. J. Mach. Learn. Cybern.*, vol. 12, no. 11, pp. 3101–3114, Nov. 2021, doi: 10.1007/s13042-020-01261-w.

[135] S. A. Hashemian and V. V. Tabataba, 'A multigate scheme to improve CORPL under traffic load in cognitive radio based smart grids with mesh topology', 2019.

[136] A. Bello, 'IMPLEMENTATION OF A CHANNEL-AWARE ROUTING PROTOCOL IN THE NETWORK SIMULATOR FOR UNDERWATER ACOUSTIC COMMUNICATION NETWORKING', 2020.

[137] O. Gnawali, R. Fonseca, K. Jamieson, M. Kazandjieva, D. Moss, and P. Levis, 'CTP: An efficient, robust, and reliable collection tree protocol for wireless sensor networks', *ACM Trans. Sens. Netw. TOSN*, vol. 10, no. 1, pp. 1–49, 2013.

[138] T. Clausen, J. Yi, and U. Herberg, 'Lightweight on-demand ad hoc distance-vector routing-next generation (LOADng): Protocol, extension, and applicability', *Comput. Netw.*, vol. 126, pp. 125–140, 2017.

[139] P. Li, L. Guo, and F. Wang, 'A multipath routing protocol with load balancing and energy constraining based on AOMDV in ad hoc network', *Mob. Netw. Appl.*, pp. 1–10, 2019.

[140] V. Coskun, B. Ozdenizci, and K. Ok, 'A survey on near field communication (NFC) technology', *Wirel. Pers. Commun.*, vol. 71, no. 3, pp. 2259–2294, 2013.

[141] Y. Choi, Y.-G. Hong, J.-S. Youn, D. Kim, and J. Choi, 'Transmission of IPv6 Packets over Near Field Communication, draft-ietf-6lo-nfc-12', 6LO Working Group, 2018.

[142] C. L. Devasena, 'Ipv6 low power wireless personal area network (6lowpan) for networking internet of things (iot)–analyzing its suitability for iot', *Indian J. Sci. Technol.*, vol. 9, no. 30, pp. 6–11, 2016.

[143] P. Sethi and S. R. Sarangi, 'Internet of things: architectures, protocols, and applications', *J. Electr. Comput. Eng.*, vol. 2017, 2017.

[144] Z. Yang and C. H. Chang, '6LoWPAN Overview and Implementations.', in *EWSN*, 2019, pp. 357–361.

[145] S. M. Darroudi, C. Gomez, and J. Crowcroft, 'Bluetooth low energy mesh networks: A standards perspective', *IEEE Commun. Mag.*, vol. 58, no. 4, pp. 95–101, 2020.

[146] R. Singh, J. Kaur, and I. S. Gill, 'Evaluation of hybrid topologies under mobility of ZigBee devices using different trajectories', *Int. J. Comput. Appl.*, vol. 122, no. 20, 2015.

[147] M. B. Yassein, W. Mardini, and A. Khalil, 'Smart homes automation using Z-wave protocol', in *2016 International Conference on Engineering & MIS (ICEMIS)*, 2016, pp. 1–6.

[148] S. J. Danbatta and A. Varol, 'Comparison of Zigbee, Z-Wave, Wi-Fi, and bluetooth wireless technologies used in home automation', in *2019 7th International Symposium on Digital Forensics and Security (ISDFS)*, 2019, pp. 1–5.

[149] K. Mekki, E. Bajic, F. Chaxel, and F. Meyer, 'A comparative study of LPWAN technologies for large-scale IoT deployment', *ICT Express*, vol. 5, no. 1, pp. 1–7, 2019.

[150] S. Popli, R. K. Jha, and S. Jain, 'A Survey on Energy Efficient Narrowband Internet of Things (NBIoT): Architecture, Application and Challenges', *IEEE Access*, vol. 7, pp. 16739–16776, 2019, doi: 10.1109/ACCESS.2018.2881533.

[151] L. Atzori, A. Iera, and G. Morabito, 'Siot: Giving a social structure to the internet of things', *IEEE Commun. Lett.*, vol. 15, no. 11, pp. 1193–1195, 2011.

[152] M. A. da Cruz, J. J. Rodrigues, A. K. Sangaiah, J. Al-Muhtadi, and V. Korotaev, 'Performance evaluation of IoT middleware', *J. Netw. Comput. Appl.*, vol. 109, pp. 53–65, 2018.

[153] A. Farahzadi, P. Shams, J. Rezazadeh, and R. Farahbakhsh, 'Middleware technologies for cloud of things: a survey', *Digit. Commun. Netw.*, vol. 4, no. 3, pp. 176–188, 2018.

[154] A. H. Ngu, M. Gutierrez, V. Metsis, S. Nepal, and Q. Z. Sheng, 'IoT middleware: A survey on issues and enabling technologies', *IEEE Internet Things J.*, vol. 4, no. 1, pp. 1–20, 2016.

[155] W. Kassab and K. A. Darabkh, 'A–Z survey of Internet of Things: Architectures, protocols, applications, recent advances, future directions and recommendations', *J. Netw. Comput. Appl.*, vol. 163, p. 102663, 2020.

[156] 'Link Smart Docs'. Accessed: Jul. 24, 2021. [Online]. Available: https://docs.linksmart.eu

[157] 'Global Sensor Network (GSN)'. Accessed: Jul. 24, 2021. [Online]. Available: https://www.epfl.ch/labs/lsir/global-sensor-networks/

[158] 'Understand Your Things The open IoT platform with MATLAB analytics'. Accessed: Jul. 25, 2021. [Online]. Available: https://thingspeak.com

[159] 'Aura Middleware'. Accessed: Jul. 25, 2021. [Online]. Available: https://github.com/AuraMiddleware/aura-middleware

[160] 'Service Discovery'. Accessed: Jul. 24, 2021. [Online]. Available: https://aws.amazon.com

[161] 'Azure IoT Hub'. Accessed: Jul. 24, 2021. [Online]. Available: https://azure.microsoft.com

[162] 'IBM Watson IoT Platform'. Accessed: Jul. 24, 2021. [Online]. Available: https://www.ibm.com/internet-of-things/solutions/iot-platform/watson-iot-platform.

[163] 'Oracle IoT'. Accessed: Jul. 24, 2021. [Online]. Available: https://www.oracle.com/internet-of-things/.

[164] 'Calvin Middleware'. Accessed: Jul. 25, 2021. [Online]. Available: https://github.com/EricssonResearch/calvin-base

[165] 'Node-RED'. Accessed: Jul. 25, 2021. [Online]. Available: https://Nodered.org

[166] 'Ptolemy II'. Accessed: Jul. 25, 2021. [Online]. Available: https://ptolemy.berkeley.edu/ptolemyII/index.htm

[167] 'Akka: Part of Lightbend Platform'. Accessed: Jul. 25, 2021. [Online]. Available: https://www.lightbend.com/akka-part-of-lightbend-platform

[168] 'Gryphon Trading Framework 0.12 Documentation'. Accessed: Jul.25,2021.[Online].Available:https://gryphon.readthedocs.io/en/latest/

[169] 'REBECA-Publish/Subscribe Middleware'. Accessed: Jul. 25, 2021. [Online]. Available: https://www.ava.uni-rostock.de/en/ava-research/projects/rebeca/

[170] 'FIWARE Step by Step'. Accessed: Jul. 25, 2021. [Online]. Available: https://fiware-tutorials.readthedocs.io/en/latest/

[171] 'New "emotional" robots aim to read human feelings'. https://techxplore.com/news/2018-01-emotional-robots-aim-human.html (accessed Jul. 21, 2021).

[172] 'Emotional Robots: Machines that Recognize Human Feelings', *Discovery*. https://www.discovery.com/science/emotional-robots--machines-that-recognize-human-feelings (accessed Jul. 21, 2021).

[173] A. Pardes, 'The Second Coming of the Robot Pet', *Wired*. Accessed:Jul.21,2021.[Online].Available:https://www.wired.com/story/the-second-coming-of-the-robot-pet/

[174] 'AR Shopping Is the Future Growth Point, and WIMI Hologram Cloud Focuses on the 5G Consumer Market - ZEE5 News', *ZEE5*, Jul. 20, 2021. https://www.zee5.com/zee5news/ar-shopping-is-the-future-growth-point-and-wimi-hologram-cloud-focuses-on-the-5g-consumer-market (accessed Jul. 21, 2021).

[175] 'Virtual Mirror Technology - It Will Change the Way You Shop', *Quytech Blog*, Jul. 06, 2018. https://www.quytech.com/blog/how-virtual-mirror-technology-will-change-the-way-you-shop/ (accessed Jul. 21, 2021).

[176] 'Robotic Surgery'. Accessed: Jul. 27, 2021. [Online]. Available: https://www.womencentre.com.au/robotic-surgery.html

[177] '6G Flagship', Univ. Oulu, Oulu, Finland, 2020. Accessed: Jul. 18, 2021. [Online]. Available: https://www.oulu.fi/6gflagship/

[178] 'Hexa-X: A flagship for 6G vision and intelligent fabric of technology'. Accessed: Jul. 18, 2021. [Online]. Available: https://hexa-x.eu/

[179] 'TeraFlow: Secured Autonomic Traffic Management for a Tera of SDN Flows.' Accessed: Jul. 18, 2021. [Online]. Available: https://teraflow-h2020.eu/

[180] 'DAEMON: Network Intelligence for Adaptive and Self-Learning Mobile Networks.' Accessed: Jul. 18, 2021. [Online]. Available: https://h2020daemon.eu/

[181] 'Bringing Reinforcement-Learning Into Radio Light Network for Massive Connections (6G BRAINS)'. Accessed: Jul. 19, 2021. [Online]. Available: https://6g-brains.eu/

[182] 'Ministry of Science and ICT, South Korea.' Accessed: Jul. 19, 2021. [Online]. Available: http://english.msip.go.kr

[183] 'South Korea to launch 6G pilot project in 2026'. Accessed: Jul. 19,2021.[Online].Available:https://www.rcrwireless.com/20200810/asia-pacific/south-korea-launch-6g-pilot-project-2026-report
[184] 'Japan to Earmark 50 Billion for 6G Development.' Accessed: Jul. 19,2021.[Online].Available:https://www.japantimes.co.jp/news/2020/ 12/10/business/japan-earmark%C%A550-billion-6g-development/
[185] '6th Generation Innovation Center (6GIC).' Accessed: Jul. 19, 2021. [Online]. Available: https://www.surrey.ac.uk/news
[186] 'NTT, Intel and Sony Establish New Global Forum Dedicated to Realizing the Communications of the Future.' Accessed: Jul. 19, 2021.[Online].Available:https://www.ntt.co.jp/news2019/1910e/191031a.html
[187] 'Huawei Starts 6G Research at Its Canada Lab.' Accessed: Jul. 19, 2021. [Online]. Available: https://www.techspot.com/news/81457-huawei-starts-6g-research-canada-lab.html
[188] 'SK Telecom to Collaborate on "6G" Tech With Nokia, Ericsson and Samsung.' Accessed: Jul. 19, 2021. [Online]. Available: https://www.rcrwireless.com/20190618/5g/sk-telecom-collaborate-6g-nokia-ericsson-samsung
[189] 'Samsung's 6G White Paper Lays Out the Company's Vision for the Next Generation of Communications Technology'. Accessed: Jul.19,2021.[Online].Available:https://news.samsung.com/global/samsungs-6g-white-paper-lays-out-the-companys-vision-for-the-next-generation-of-communications-technology
[190] 'LG Electronics, KRISS and KAIST Team Up for 6G Development.' Accessed: Jul. 19, 2021. [Online]. Available: http://www. businesskorea.co.kr/news/articleView.html?idxno=50389
[191] 'NTT Successfully Demonstrates 100 Gbps Wireless Transmission Using a New Principle (OAM Multiplexing) as a World's First.' Accessed:Jul.19,2021.[Online].Available:https://www.ntt.co.jp/news2018/1805e/180515a.html
[192] 'Tektronix, IEMN Demonstrate 100 Gb/s Wireless Transmissions Using New IEEE 802.15.3d Standard.' Accessed: Jul. 19, 2021. [Online]. Available: http://news.tektronix.com
[193] G. Biczok, M. Dramitinos, L. Toka, P. E. Heegaard, and H. Lonsethagen, 'Manufactured by software: Sdn-enabled multi-operator composite services with the 5g exchange', *IEEE Commun. Mag.*, vol. 55, no. 4, pp. 80–86, 2017.
[194] MATILDA, Deliverable D1.1, 'MATILDA Reference Architecture, Conceptualization and Use Cases', 2017. Accessed: Jun. 14, 2021. [Online]. Available: http://www.matilda-5g.eu/
[195] 'Smart Networks in Context of NGI', European Technology Platform Networld 2020, Networld2020, SRIA-2020, Sep. 2020. Accessed: Jun. 12, 2021. [Online]. Available: https://bscw.5g-ppp.eu/pub/bscw.cgi/d367342/Networld2020%20SRIA%202020%20Final%20Version%202.2%20.pdf
[196] 'Machine Learning for Future Networks including 5G (ML5G)', ITU Focus groups, 2017. Accessed: Jun. 14, 2021. [Online]. Available:https://www.itu.int/en/ITU-T/focusgroups/ml5g/pages/default.aspx
[197] 'AI and applied machine learning', TIP, 2017. Accessed: Jun. 14, 2021.[Online].Available: https://telecominfraproject.com/artificial-intelligence-and-applied-machine-learning/
[198] 'Network Data Analytics Function (NWDAF)', 3GPP, 2018. Accessed: Jun. 14, 2021. [Online]. Available: http://www.tech-invite.com/3m29/tinv-3gpp-29-520.html
[199] ITU-T Study group 13, 'Framework for evaluating intelligence levels of future networks including IMT-2020', ITU-T Y.3173. Accessed: Jun. 14, 2021. [Online]. Available: http://handle.itu.int/11.1002/1000/14133
[200] M. Katz, M. Matinmikko-Blue, and M. Latva-Aho, '6Genesis flagship program: Building the bridges towards 6G-enabled wireless smart society and ecosystem', in *2018 IEEE 10th Latin-American Conference on Communications (LATINCOM)*, 2018, pp. 1–9.
[201] 'Zero Touch Provisioning', CISCO, 2019. Accessed: Jun. 14, 2021.[Online].Available:https://www.cisco.com/c/en/us/td/docs/switches/lan/catalyst3850/software/release/16-5/configuration_guide/prog/b_165_prog_3850_cg/zero_touch_provisioning.pdf

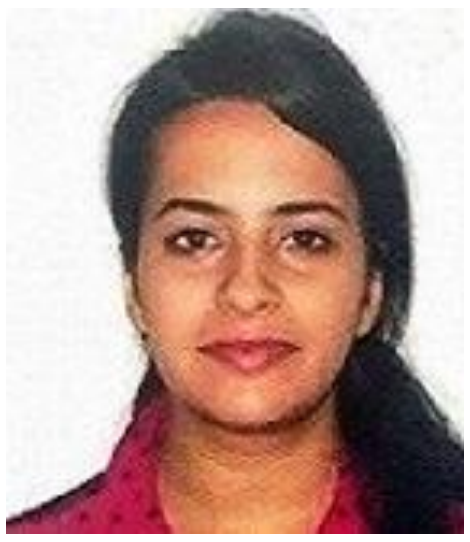

**Mantisha Gupta (S' 21)** received the B.E degree in Electronics and Communication Engineering from Jammu University, Jammu and Kashmir, India in 2017 and the M.Tech Degree in Electronics and Communication Engineering from Shri Mata Vaishno Devi University, Katra, Jammu and Kashmir, India in 2019, where she is pursuing the Ph,D degree in Electronics and Communication Engineering. Her research interest includes the emerging technologies involving the B5G/6G and IoT enabled wireless communication and security network and currently she is doing her research on IoT configured networks in B5G/6G wireless communication systems. She is a student member of Institute of Electrical and Electronics Engineers (IEEE).

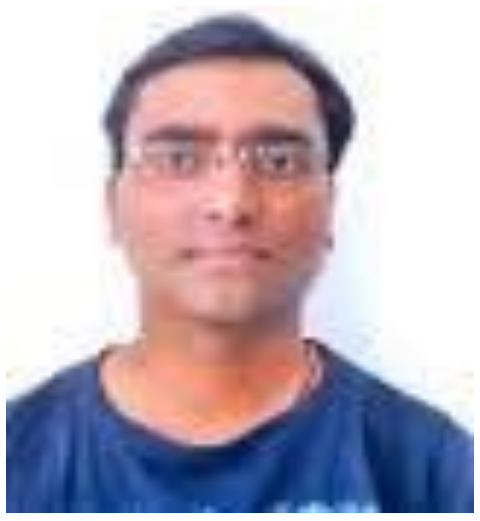

**. Rakesh K Jha (S'10, M'13)** is an Associate Professor in the Department of Electronics and Communication Engineering, Indian Institute of Information Technology, Design and Manufacturing, Jabalpur (IIITDM Jabalpur). He is carrying out his research in wireless communication, power optimizations, wireless security issues, and optical fiber communication. He has done B. Tech (Hon's) in Electronics and Communication Engineering and M.Tech from NIT Jalandhar (Hon's), India in 2008. Received his Ph.D. degree from NIT Surat, India in 2013. He has completed his 10th exam from govt. High school and Class 12th from Science College. He has published more than 101 Journal Papers out of which more than 61 SCI Journal papers including IEEE Transactions, IEEE Journal, Elsevier, Springer, Taylor & Francis, Hindawi, etc. He has published more than 25 Interference including ITU-T, IEEE ANTS, INDICON, IEEE ANTS, and APAN. Dr. Jha's one concept related to the router of Wireless Communication has been accepted by ITU (International Telecommunication Union) in 2010.He has received the young scientist author award by ITU in Dec 2010. He has received an APAN fellowship in 2011, 2012-Srilanka, 2016, and in 2017-China, 2018-Singapore,2018-New Zealand, 2019-South Korea, and a student travel grant from COMSNET 2012. He is a Senior Member of IEEE, GISFI, and SIAM, International Association of Engineers (IAENG), ACCS (Advanced Computing and Communication Society), CSI, etc. He has filed 8 Patents out of which 4 are published. Dr. Jha had 10 years of rich academic, Industrial, and research experience in various institutes/University including NIT-Surat, Capgemini India Pvt. Ltd and SMVD University. He has also served as an organizing member and TPC member for several national and international conferences. He has organized many workshops and has also been invited as a resource person in many workshops organized by prestigious research institutes. He has guided 05 Ph.D. students, 01 Submitted the thesis, 03 Defended Pre-Ph.D. Synopsis and 03 students are presently pursuing. He has guided more than 15 M.Tech and more than 41 B.Tech students for various projects. More than 4001 citations in his credit in the area of wireless communication.

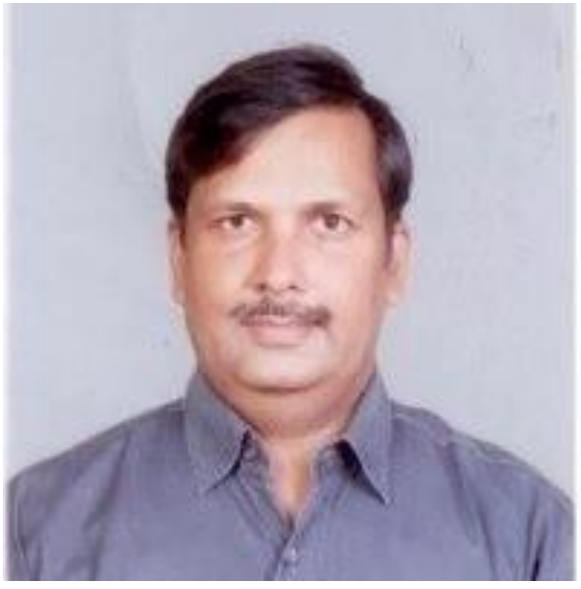

**PROF. SANJEEV JAIN**, born at Vidisha in Madhya Pradesh in 1967, obtained his Post Graduate Degree in Computer Science and Engineering from Indian Institute of Technology, Delhi, in 1992. He later received his Doctorate Degree in Computer Science & Engineering and has over 24 years' experience in teaching and research. He has served as Director, Madhav Institute of Technology and Science (MITS), Gwalior. He has worked as a vice chancellor at Shri Mata Vaishno Devi University, Katra. Presently he is a Professor in the Computer Science Department, Central University Jammu, Jammu and Kashmir. Besides teaching at Post Graduate level Professor Jain has the credit of making significant contribution to R & D in the area of Image Processing and Mobile Adhoc Network. He has guided Ph.D. Scholars and has undertaken a number of major R & D projects sponsored by the Government and Private Agencies. His work on Digital Watermarking for Image Authentication is highly valued in the research field. He is also a member of Association for Computing Machinery (ACM).